\documentclass{article}
\usepackage[utf8]{inputenc}
\usepackage{graphicx}
\usepackage{color}
\usepackage{float}
\usepackage{slashed}
\usepackage{amsthm,amsmath,amssymb,mathrsfs}
\usepackage{setspace}
\pdfoutput=1 
\allowdisplaybreaks
\usepackage{jheppub}
\usepackage{xcolor} 
\usepackage{babel}
\begin{document}
\title{CP violation and flavor invariants in the seesaw effective field theory}
\author[a,b]{Bingrong Yu}
\emailAdd{yubr@ihep.ac.cn}
\author{and} 
\author[a,b]{Shun Zhou}
\emailAdd{zhoush@ihep.ac.cn}
\affiliation[a]{Institute of High Energy Physics, Chinese Academy of Sciences, Beijing 100049, China}
\affiliation[b]{School of Physical Sciences, University of Chinese Academy of Sciences, Beijing 100049, China}	
\abstract{
In this paper, we systematically study the algebraic structure of the ring of the flavor invariants and the sources of CP violation in the seesaw effective field theory (SEFT), which is obtained by integrating out heavy Majorana neutrinos in the type-I seesaw model at the tree level and thus includes the dimension-five Weinberg operator and one dimension-six operator. For the first time, we calculate the Hilbert series and explicitly construct all the primary flavor invariants in the SEFT. We show that all the physical parameters can be extracted using the primary invariants and any CP-violating observable can be expressed as the linear combination of CP-odd flavor invariants. The calculation of the Hilbert series shows that there is an equal number of primary flavor invariants in the SEFT and in the full seesaw model, which reveals the intimate connection between the flavor space of the SEFT and that of its ultraviolet theory. A proper matching procedure of the flavor invariants is accomplished between the SEFT and the full seesaw model, through which one can establish a direct link between the CP asymmetries in leptogenesis and those in low-energy neutrino oscillations. 
}

\maketitle


























\section{Introduction}
The violation of charge-parity (CP) symmetry is a crucial concept in particle physics and serves as an indispensable ingredient to dynamically generate the matter-antimatter asymmetry in our universe~\cite{Sakharov:1967dj,Bodeker:2020ghk}. In the Standard Model (SM), CP violation has been observed in the quark sector~\cite{Christenson:1964fg,KTeV:1999kad,BaBar:2001pki}. Moreover, since the neutrino oscillation experiments have firmly established that neutrinos are massive and lepton flavors are significantly mixed~\cite{Xing:2020ijf}, CP violation is also expected in the leptonic sector~\cite{Branco:2011zb}, which is the primary goal of the future long-baseline accelerator neutrino oscillation experiments~\cite{Hyper-KamiokandeProto-:2015xww,DUNE:2015lol,Hyper-Kamiokande:2016srs,T2K:2018rhz}.

From the theoretical point of view, the violation of CP symmetry in the fermionic sector in a specific model comes from the complex couplings in the Lagrangian.\footnote{For the bosonic sources of CP violation, such as the $\theta$-term in quantum chromodynamics (QCD), the condition for CP conservation is trivially the vanishing of all couplings of CP-violating terms. In this paper, we only consider the fermionic sources of CP violation.} However, one should keep in mind that the couplings are \emph{not} invariant under the basis transformation in the flavor space. Thus the sufficient and necessary condition for CP conservation should be: \emph{It is possible to find a specific flavor basis such that in this basis every coupling parameter in the Lagrangian is real.} This criterion suffers from the flavor-basis dependence that one has to change the values of parameters from one flavor basis to another. Therefore, it is well motivated to introduce some quantities composed of the coupling parameters in the Lagrangian that are invariant under the flavor-basis transformations and thus one only needs to calculate these basis-independent quantities to judge whether there is CP violation in a given model. This is exactly the reason why flavor invariants are physically interesting. Furthermore, since any physical observables calculated from the parameters in the Lagrangian must be independent of the flavor basis, it will be helpful to express them as some functions of flavor invariants.

The first flavor invariant was constructed by Jarlskog~\cite{Jarlskog:1985ht,Jarlskog:1985cw,Jarlskog:1986mm} to characterize the CP violation in the quark sector. As is well known, the CP-violating phase in the Cabibbo-Kobayashi-Maskawa (CKM) matrix~\cite{Kobayashi:1973fv} appears in the quark charged-current interaction, leading to the CP violation in the neutral meson systems. Although the Yukawa coupling matrices $Y_{\rm u}^{}$ and $Y_{\rm d}^{}$ of up- and down-type quarks are not invariant under the flavor-basis transformations, it is possible to define the following basis-independent quantity~\cite{Jarlskog:1985ht, Jarlskog:1985cw, Jarlskog:1986mm} 
\begin{eqnarray}
	\label{eq:Jarlskog}
J\equiv {\rm Det}\left(\left[Y_{\rm u}^{}Y_{\rm u}^\dagger,Y_{\rm d}^{}Y_{\rm d}^\dagger\right]\right)=\frac{1}{3}{\rm Tr}\left(\left[Y_{\rm u}^{}Y_{\rm u}^\dagger,Y_{\rm d}^{}Y_{\rm d}^\dagger\right]_{}^3\right) \;.
\end{eqnarray}
It can be checked in general that any CP-violating observable in the quark sector is proportional to $J$, so the vanishing of $J$ is equivalent to the absence of CP violation in the quark sector. 

The application of flavor invariants to studying CP violation was later generalized to arbitrary generations of quarks~\cite{Branco:1986quark} and to the leptonic sector~\cite{Branco:1986lepton}. The situation becomes more complicated in the leptonic sector if neutrinos are Majorana particles~\cite{Majorana:1937vz,Racah:1937qq}, because there are two extra Majorana-type CP phases entering the lepton flavor mixing matrix, i.e., the Pontecorvo-Maki-Nakagawa-Sakata (PMNS) matrix~\cite{Pontecorvo:1957cp, Maki:1962mu}. In this case, the minimal sufficient and necessary conditions for CP conservati-\\on in the leptonic sector are the vanishing of three flavor invariants~\cite{Dreiner:2007yz, Yu:2019ihs, Yu:2020xyy}, which are analogous to $J$ in Eq.~(\ref{eq:Jarlskog}). Moreover, if neutrino masses were degenerate, there would be redundant degrees of freedom and the number of conditions to guarantee CP conservation would accordingly be reduced~\cite{  Branco:1986lepton,Branco:1998bw,Mei:2003gu,Yu:2020gre}. 

It is always possible to construct an infinite number of invariants in the flavor space though not all of them are independent. In fact, since the addition or multiplication of any two flavor invariants is also a flavor invariant, all of them form a \emph{ring} in the flavor space in the sense of algebraic structure. It was noticed in Refs.~\cite{Manohar:2009dy,Manohar:2010vu} that the Hilbert series (HS) in the invariant theory~\cite{sturmfels2008algorithms,derksen2015computational} provides a powerful tool in investigating the algebraic structure of the invariants in the flavor space and establishing the relations between flavor invariants and physical parameters.\footnote{The HS has also been widely applied to counting the number of independent gauge- and Lorentz-invariant effective operators of a certain mass dimension in the effective theories~\cite{Lehman:2015via, Henning:2015daa, Lehman:2015coa, Henning:2015alf, Henning:2017fpj, Graf:2020yxt}. In addition, the construction of the flavor invariants in the scalar sector of multi-Higgs-doublet models were thoroughly studied and the corresponding HS were calculated in the literature~\cite{Trautner:2018ipq, Trautner:2020qqo, Bento:2020jei, Bento:2021hyo}. The explicit and spontaneous breaking of a flavor group to its subgroups were also studied using the HS, see Ref.~\cite{Merle:2011vy}.} The maximum number of the algebraically-independent invariants in the flavor space, namely, the \emph{primary} invariants, is equal to the number of independent physical parameters in the theory~\cite{Manohar:2009dy,Manohar:2010vu}. Furthermore, as long as the symmetry group of the flavor space is reductive, the ring of invariants can be finitely generated~\cite{sturmfels2008algorithms, derksen2015computational}, which means there exist a finite number of \emph{basic} invariants such that any invariant in the flavor space can be expressed as the polynomial of the basic ones. The plethystic program~\cite{Hanany:2006qr} provides a convenient method to find out all the basic invariants as well as the polynomial identities (known as syzygies) among them. In the type-I seesaw model~\cite{Minkowski:1977sc, Yanagida:1979as, Gell-Mann:1979vob, Glashow:1979nm, Mohapatra:1979ia} and its low-energy effective theory with only the dimension-five Weinberg operator~\cite{Weinberg:1979sa}, the HS in the flavor space have been calculated and the basic flavor invariants are explicitly constructed~\cite{Manohar:2009dy, Manohar:2010vu, Wang:2021wdq, Yu:2021cco}. Moreover, the renormalization-group equations of all the basic flavor invariants in the effective theory have been derived as well~\cite{Wang:2021wdq}. 

Recently, the number of independent CP-violating phases in the Standard Model effective field theory (SMEFT) with operators of mass dimension $d\leqslant6$ ~\cite{Buchmuller:1985jz,Grzadkowski:2010es,Trott:2017vri} has been systematically counted in Ref.~\cite{Bonnefoy:2021tbt}. Furthermore, an equal number of CP-odd flavor invariants are explicitly constructed and the vanishing of these CP-old invariants guarantees the CP conservation in the SMEFT up to dimension six. However, the $d=5$ Weinberg operator~\cite{Weinberg:1979sa} is not included, so the flavor mixing and CP violation in the leptonic sector are ignored in Ref.~\cite{Bonnefoy:2021tbt}. In this paper, we aim to explore the leptonic CP violation using the language of invariant theory in the framework of seesaw effective field theory (SEFT) at the tree-level matching,\footnote{The full one-loop matching of the type-I seesaw model onto the SMEFT has been accomplished recently~\cite{Zhang:2021jdf, Ohlsson:2022hfl, Du:2022vso}, where 31 dimension-six operators in the Warsaw basis of the SMEFT are involved. However, as we shall demonstrate later, the Wilson coefficients of ${\cal O}_5^{\alpha\beta}$ and ${\cal O}_6^{\alpha\beta}$ are already adequate to incorporate all physical information about the full seesaw model. Therefore, we will restrict ourselves within the SEFT at the tree-level matching.} which includes the $d=5$ Weinberg operator ${\cal O}^{\alpha \beta}_5 = \overline{\ell^{}_{\alpha \rm L}} \widetilde{H} \widetilde{H}^{\rm T} \ell^{\rm C}_{\beta \rm L}$ and one $d=6$ operator ${\cal O}^{\alpha \beta}_6 =\left(\overline{\ell^{}_{\alpha \rm L}} \widetilde{H}\right)i\slashed{\partial}\left(\widetilde{H}^\dagger \ell^{}_{\beta \rm L}\right)$~\cite{Yu:2022nxj}, where $\alpha, \beta = e, \mu, \tau$ are lepton flavor indices. The main motivation for such an exploration is three-fold. 

First, as has been mentioned before, the flavor mixing and CP violation in the leptonic sector are switched off in Ref.~\cite{Bonnefoy:2021tbt}. However, neutrino oscillation experiments have revealed that neutrinos are indeed massive. In the spirit of the SMEFT, the most natural way to explain the nonzero neutrino masses is to introduce the $d=5$ Weinberg operator ${\cal O}_5^{}$~\cite{Weinberg:1979sa}, which accounts for Majorana neutrino masses after the spontaneous gauge symmetry breaking. Once ${\cal O}_5^{}$ is included, there will be extra sources of CP violation from the leptonic sector, which are observable in the low-energy oscillation experiments~\cite{Xing:2013ty, Xing:2013woa, Wang:2021rsi}. Therefore, it is \emph{necessary} to take into account the source of CP violation from ${\cal O}_5^{}$ apart from the $d=6$ operators.

Second, it has been shown that there are totally 699+1+1+6 independent CP phases in the SMEFT with operators of $d\leqslant6$~\cite{Bonnefoy:2021tbt}, the number of which is too large to receive restrictive constraints from low-energy experiments. If the Weinberg operator is included and the sources of CP violation from the leptonic sector are considered, the number of CP-violating phases will be further increased. Instead, we take in this paper the ultraviolet (UV) theory to be the type-I seesaw model, which is one of the most natural models to simultaneously explain nonzero neutrino masses and generate the cosmological matter-antimatter asymmetry via leptogenesis~\cite{Fukugita:1986hr}, and explore the sources of CP violation at the low-energy scale. 

Finally, it is well known that there are six independent CP phases in type-I seesaw model in the three-generation case that appear in the CP asymmetries in the decays of right-handed (RH) neutrinos. The inclusion of ${\cal O}_5^{}$ at the low-energy scale contains only three CP phases in the PMNS matrix, which is obviously not enough to incorporate all the CP-violating sources in the full theory. It was firstly mentioned in Refs.~\cite{Broncano:2002rw,Broncano:2003fq} that the simultaneous inclusion of ${\cal O}_5^{}$ and ${\cal O}_6^{}$ reproduces the same number of independent physical parameters as in the full seesaw model if the number of RH neutrinos matches that of active neutrinos. This observation indicates that both ${\cal O}_5^{}$ and ${\cal O}_6^{}$ have already been \emph{adequate} to cover all physical information about the full theory. In this paper, we will show that this is indeed the case through the language of invariant theory. In particular, we shall establish the intimate connection between the invariant ring of the flavor space in the effective theory and that in the full theory. The matching between the flavor invariants in the SEFT and those in the full seesaw model will be accomplished by a proper procedure. Through the matching of the flavor invariants, one can directly link the CP asymmetries in leptogenesis to those in neutrino-neutrino and neutrino-antineutrino oscillations at low energies in a basis-independent way.

The remaining part of this paper is structured as follows. In Sec.~\ref{sec:framework}, we first define the flavor-basis transformation in the SEFT and set up our formalism and notations. In Sec.~\ref{sec:construction2g} and Sec.~\ref{sec:construction3g} we systematically study the algebraic structure of the invariant ring in the SEFT and construct the flavor invariants using the tool of invariant theory for the two- and three-generation case, respectively. Moreover, we will explain how to extract all physical parameters in the SEFT from the primary flavor invariants. The minimal sufficient and necessary conditions for CP conservation are also given in terms of CP-odd flavor invariants. Phenomenological applications of the flavor invariants are discussed in Sec.~\ref{sec:observables}, where we illustrate how to express a general CP-violating observable as the linear combination of CP-odd flavor invariants. In Sec.~\ref{sec:matching} we explore the connection between the SEFT and the full seesaw model through the matching between flavor invariants at low- and high-energy scales. The intimate relationship between two sets of basic flavor invariants will be revealed. Our main conclusions are summarized in Sec.~\ref{sec:summary}. Finally, some indispensable details are presented in two appendices. In Appendix~\ref{app:HS} we calculate the HS in the SEFT while in Appendix~\ref{app:matching} the matching procedure between flavor invariants in the SEFT and those in the full theory is given.

\section{Flavor-basis transformation in the SEFT}
\label{sec:framework}
The type-I seesaw model extends the SM by introducing three RH neutrinos $N_{\rm R}^{}$, which are singlets under the SM gauge symmetry. The Lagrangian of the type-I seesaw reads
\begin{eqnarray}
	\label{eq:full lagrangian}
	{\cal L}_{\rm seesaw}^{}={\cal L}_{\rm SM}^{}+\overline{N_{\rm R}^{}}{\rm i}\slashed{\partial}N_{\rm R}^{}-\left[\overline{\ell_{\rm L}^{}}Y_\nu^{}\widetilde{H}N_{\rm R}^{}+\frac{1}{2}\overline{N_{\rm R}^{\rm C}}M_{\rm R}^{}N_{\rm R}^{}+{\rm h.c.}\right]\;,
\end{eqnarray}
where ${\cal L}_{\rm SM}$ is the SM Lagrangian, ${\ell }_{\rm L}^{}\equiv \left(\nu_{\rm L}^{},l_{\rm L}^{}\right)_{}^{\rm T}$ and $\widetilde{H}\equiv {\rm i}\sigma^2 H_{}^{*}$ stand for the left-handed lepton doublet and the Higgs doublet, respectively. In addition, $Y_\nu^{}$ denotes the Dirac neutrino Yukawa coupling matrix and $M_{\rm R}^{}$ is the Majorana mass matrix of RH neutrinos. Note that $N_{\rm R}^{\rm C}\equiv {\cal C} \overline{N_{\rm R}^{}}^{\rm T}$ has been defined with ${\cal C}\equiv {\rm i}\gamma_{}^2\gamma_{}^0$ being the charge-conjugation matrix.

If the mass scale of RH neutrinos $\Lambda={\cal O}\left(M_{\rm R}^{}\right)$ is much higher than the electroweak scale characterized by the vacuum expectation value $v \approx 246\, {\rm GeV}$ of the Higgs field, one can integrate $N_{\rm R}^{}$ out at the tree level to obtain the low-energy effective theory. The effective Lagrangian to the order of ${\cal O}\left(1/\Lambda^2\right)$ turns out to be
\begin{eqnarray}
	\label{eq:effective lagrangian}
	{\cal L}_{\rm SEFT}^{}={\cal L}_{\rm SM}^{}-\left(\frac{C_5^{}}{2\Lambda} {\cal O}_5^{}+{\rm h.c.}\right)+\frac{C_6^{}}{\Lambda^2}{\cal O}_6^{}\;,
\end{eqnarray}
with
\begin{eqnarray}
{\cal O}_5^{}=\overline{\ell_{\rm L}^{}}\widetilde{H}\widetilde{H}_{}^{\rm T}\ell_{\rm L}^{\rm C}\;,\quad
{\cal O}_6^{}=\left(\overline{\ell_{\rm L}^{}}\widetilde{H}\right)i\slashed{\partial}\left(\widetilde{H}_{}^\dagger\ell_{\rm L}^{}\right)\;.	
\end{eqnarray}
Note that the lepton flavor indices have been suppressed. At the tree-level matching, the Wilson coefficients are related to the Yukawa couplings in the full theory as
\begin{eqnarray}
	\label{eq:wilson coe}
	C_5^{}=-Y_\nu^{}Y_{\rm R}^{-1}Y_\nu^{\rm T}\;, \quad
	C_6^{}=Y_\nu^{} \left(Y_{\rm R}^{\dagger}Y_{\rm R}^{}\right)_{}^{-1}Y_\nu^\dagger\;,
\end{eqnarray}
where we have defined the dimensionless quantity $Y_{\rm R}^{}\equiv M_{\rm R}^{}/\Lambda$. Given the full Lagrangian in Eq.~(\ref{eq:full lagrangian}), one can perform the general unitary transformation in the flavor space in the leptonic sector
\begin{eqnarray}
	\label{eq:field trans}
	\ell_{\rm L}^{}\to \ell_{\rm L}^{\prime}=U_{\rm L}^{}\ell_L^{}\;,\quad
	l_{\rm R}^{}\to l_{\rm R}^{\prime}=V_{\rm R}l_{\rm R}^{}\;,\quad
	N_{\rm R}^{}\to N_{\rm R}^{\prime}=U_{\rm R}^{}N_{\rm R}^{}\;,
\end{eqnarray}
where $l_{\rm R}^{}$ is the RH charged-lepton field, $U_{\rm L}^{}, V_{\rm R} \in {\rm U}(m)$ and $U_{\rm R}^{} \in {\rm U}(n)$ (for $m$ generations of lepton doublets and $n$ generations of RH neutrinos) are three arbitrary unitary matrices. Then Eq.~(\ref{eq:full lagrangian}) is unchanged if we treat the Yukawa coupling matrices as spurions, namely, taking them as spurious fields that transform as
\begin{eqnarray}
	\label{eq:Yukawa trans}
	Y_l^{} \to Y_l^\prime=U_{\rm L}^{}Y_l^{}V_{\rm R}^\dagger\;,\quad
	Y_\nu^{} \to Y_\nu^\prime=U_{\rm L}^{}Y_\nu^{}U_{\rm R}^\dagger\;,\quad
	Y_{\rm R}^{} \to Y_{\rm R}^\prime=U_{\rm R}^* Y_{\rm R}^{}U_{\rm R}^\dagger\;,
\end{eqnarray} 
where $Y_l^{}$ denotes the charged-lepton Yukawa coupling matrix. At the matching scale, the transformati-\\on in Eq.~(\ref{eq:Yukawa trans}) together with Eq.~(\ref{eq:wilson coe}) induces the transformation of Wilson coefficients in the flavor space of the SEFT, i.e.,
\begin{eqnarray}
	\label{eq:wilson coe trans}
	C_5^{}\to C_5^\prime=U_{\rm L}^{}C_5^{}U_{\rm L}^{\rm T}\;,\quad
	C_6^{}\to C_6^\prime=U_{\rm L}^{}C_6^{}U_{\rm L}^\dagger\;.
\end{eqnarray}
It is easy to verify that the SEFT Lagrangian in Eq.~(\ref{eq:effective lagrangian}) is also unchanged under the  transformation in Eqs.~(\ref{eq:field trans})-(\ref{eq:wilson coe trans}), as it should be. 

Now it is obvious that the building blocks for the construction of flavor invariants in the SEFT are $\left\{X_l^{}\equiv Y_l^{}Y_l^\dagger,C_5^{},C_6^{}\right\}$ with the symmetry group ${\rm U}(m)$, while the building blocks in the full seesaw model  are $\left\{Y_l^{},Y_\nu^{},Y_{\rm R}^{}\right\}$ with the symmetry group ${\rm U}(m)\otimes{\rm U}(n)$.\footnote{Notice that the flavor transformation of RH charged-lepton fields [i.e., $V_{\rm R}^{}$ in Eq.~(\ref{eq:field trans})] is unphysical. Therefore one may use $Y_l^{}Y_l^\dagger$ as a building block instead of $Y_l^{}$ when constructing flavor invariants and calculating the HS.} Note that by flavor invariants, we refer to the \emph{polynomial} matrix invariants constructed from the building blocks that are unchanged under the unitary transformation in the flavor space~\cite{procesi1976invariant,procesi2017invariant}.

\underline{\emph{Notations for flavor invariants}}: Throughout this paper, we use ${\cal I}_{abc}^{}$ (or ${\cal J}_{abc}$) to label the flavor invariant with the degrees $\left\{a,b,c\right\}$ of the building blocks $\left\{X_l^{},C_5^{},C_6^{}\right\}$ for the two- (or three-) generation case in the SEFT. On the other hand, $I_{abc}^{}$ (or $J_{abc}^{}$) will be used to label the flavor invariant with the degrees $\left\{a,b,c\right\}$ of the building blocks $\left\{Y_l^{},Y_\nu^{},Y_{\rm R}^{}\right\}$ for the two- (or three-) generation case in the full seesaw model. Here $a,b,c$ are all non-negative integers.

\section{Algebraic structure of the SEFT flavor invariants: Two-generation case}
\label{sec:construction2g}
Let us start with the case of only two-generation leptons, which is unrealistic but very instructive for the study of the three-generation case.

The first step is to compute the HS in the flavor space, which encodes all information about the ring of invariants. Given the transformation rules of the building blocks in Eq.~(\ref{eq:wilson coe trans}) with $m=2$, one can calculate the HS using the Molien-Weyl (MW) formula~\cite{molien1897invarianten,weyl1926darstellungstheorie} (see Appendix~\ref{app:HS} for details)
\begin{eqnarray}
	\label{eq:HS eff 2g main}
	{\mathscr H}_{\rm SEFT}^{(2\rm g)}(q)=\frac{1+3q^4+2q^5+3q^6+q^{10}}{\left(1-q\right)^2\left(1-q^2\right)^4\left(1-q^3\right)^2\left(1-q^4\right)^2}\;.
\end{eqnarray}
The numerator of the HS in Eq.~(\ref{eq:HS eff 2g main}) exhibits the palindromic structure as expected while the denominator has 10 factors, which implies the maximum number of the algebraically-independent invariants (i.e., primary invariants) in the flavor space is 10. This number, as a highly nontrivial result, is also equal to the number of independent physical parameters in the two-generation SEFT.
  
In order to find out the generators of the invariant ring, one can substitute Eq.~(\ref{eq:HS eff 2g main}) into Eq.~(\ref{eq:PL def}) to calculate the plethystic logarithm (PL) function, 
\begin{eqnarray}
	\label{eq:PL eff 2g main}
	{\rm PL}\left[{\mathscr H}_{\rm SEFT}^{(2\rm g)}(q)\right]=2q+4q^2+2q^3+5q^4+2q^5+3q^6-6q^8-{\cal O}\left(q^9\right)\;,
\end{eqnarray}
whose leading positive terms correspond to the number and degrees of the basic invariants~\cite{Hanany:2006qr}. More explicitly, there are in total 18 basic invariants in the ring, two of degree 1, four of degree 2, two of degree 3, five of degree 4, two of degree 5 and three of degree 6. With the help of Eq.~(\ref{eq:PL eff 2g main}) we explicitly construct all the basic flavor invariants in the SEFT, as summarized in Table~\ref{table:2g eff}. The CP parities listed in the last column of Table~\ref{table:2g eff} describe the behaviors of the flavor invariants under the CP transformation: CP-even invariants are unchanged under the CP-conjugate transformation while CP-odd invariants bring about an extra minus sign. The 18 basic invariants (12 CP-even and 6 CP-odd) in Table~\ref{table:2g eff} serve as the generators of the invariant ring, in the sense that any flavor invariant can be written as the polynomial of them. For example, the CP-even counterparts of the 6 CP-odd basic invariants in Table~\ref{table:2g eff} can be decomposed into the polynomials of the CP-even basic invariants in Table~\ref{table:2g eff} as follows:\footnote{See Appendix C of Ref.~\cite{Wang:2021wdq} for a general algorithm to decompose an arbitrary invariant into the polynomial of the basic ones and to find out all the syzygies among the basic invariants at a certain degree.}
\begin{eqnarray}
	&&{\cal I}_{121}^{(+)}\equiv {\rm Re}\, {\rm Tr}\left(X_l^{}X_5^{}C_6^{}\right)=\frac{1}{2}\left[{\cal I}_{020}^{}\left({\cal I}_{101}^{}-{\cal I}_{100}^{}{\cal I}_{001}^{}\right)+{\cal I}_{100}^{}{\cal I}_{021}^{}+{\cal I}_{001}^{}{\cal I}_{120}^{}\right]\;,\\
	&&{\cal I}_{141}^{(+)}\equiv {\rm Re}\, {\rm Tr}\left(X_5^{}C_6^{}G_{l5}^{}\right)=\frac{1}{4}\left[{\cal I}_{100}^{}{\cal I}_{001}^{}\left({\cal I}_{040}^{}-{\cal I}_{020}^2\right)+2\left({\cal I}_{020}^{}{\cal I}_{121}^{(1)}+{\cal I}_{120}^{}{\cal I}_{021}^{}\right)\right]\;,\\
	&&{\cal I}_{221}^{(+)}\equiv {\rm Re}\, {\rm Tr}\left(X_l^{}G_{l5}^{}C_6^{}\right)=\frac{1}{2}\left[{\cal I}_{101}^{}{\cal I}_{120}^{}+{\cal I}_{001}^{}{\cal I}_{220}^{}+{\cal I}_{100}^{}\left({\cal I}_{121}^{(1)}-{\cal I}_{001}^{}{\cal I}_{120}^{}\right)\right]\;,\\
	&&{\cal I}_{122}^{(+)}\equiv {\rm Re}\, {\rm Tr}\left(C_6^{}G_{56}^{}X_l^{}\right)=\frac{1}{2}\left[{\cal I}_{101}^{}{\cal I}_{021}^{}+{\cal I}_{100}^{}{\cal I}_{022}^{}+{\cal I}_{001}^{}\left({\cal I}_{121}^{(1)}-{\cal I}_{100}^{}{\cal I}_{021}^{}\right)\right]\;,\\
	&&{\cal I}_{240}^{(+)}\equiv {\rm Re}\, {\rm Tr}\left(X_l^{}X_5^{}G_{l5}^{}\right)=\frac{1}{4}\left[{\cal I}_{100}^2\left({\cal I}_{040}^{}-{\cal I}_{020}^2\right)+2\left({\cal I}_{120}^2+{\cal I}_{020}^{}{\cal I}_{220}^{}\right)\right]\;,\\
	&&{\cal I}_{042}^{(+)}\equiv {\rm Re}\, {\rm Tr}\left(C_6^{}X_5^{}G_{56}^{}\right)=\frac{1}{4}\left[{\cal I}_{001}^2\left({\cal I}_{040}^{}-{\cal I}_{020}^2\right)+2\left({\cal I}_{021}^2+{\cal I}_{020}^{}{\cal I}_{022}^{}\right)\right]\;.
\end{eqnarray}

Although there are 18 basic invariants in the SEFT, not all of them are algebraically independent. There exist nontrivial polynomial identities among them that are identically equal to zero, known as syzygies in the invariant theory.\footnote{However, note that none of the basic invariants in the ring can be written as the \emph{polynomial} of the other 17 basic ones. This is different from the case of vector space, where the statement that a set of vectors are not linearly independent means any vector of this set can be written as the linear combination of the others.} For example, the six lowest-degree syzygies begin to appear at degree 8 (in the sense that the total degree of each term in the syzygies is 8):
\begin{eqnarray}
	\label{eq:syzygy1}
	&{\cal I}_{121}^{(2)}&\left(2{\cal I}_{220}^{}-{\cal I}_{100}^{}{\cal I}_{120}^{}\right)+{\cal I}_{221}^{}\left({\cal I}_{100}^{}{\cal I}_{020}^{}-2{\cal I}_{120}^{}\right)+{\cal I}_{240}^{}\left({\cal I}_{001}^{}{\cal I}_{100}^{}-2{\cal I}_{101}^{}\right)\nonumber\\
	&&+{\cal I}_{141}^{}\left({\cal I}_{100}^2-2{\cal I}_{200}^{}\right)=0\;,\\
	\label{eq:syzygy2}
	&{\cal I}_{121}^{(2)}&\left(2{\cal I}_{022}^{}-{\cal I}_{001}^{}{\cal I}_{021}^{}\right)-{\cal I}_{122}^{}\left({\cal I}_{001}^{}{\cal I}_{020}^{}-2{\cal I}_{021}^{}\right)-{\cal I}_{042}^{}\left({\cal I}_{001}^{}{\cal I}_{100}^{}-2{\cal I}_{101}^{}\right)\nonumber\\
	&&-{\cal I}_{141}^{}\left({\cal I}_{001}^2-2{\cal I}_{002}^{}\right)= 0\;,\\
	\label{eq:syzygy3}
	&{\cal I}_{121}^{(2)}&\left(2{\cal I}_{121}^{(1)}-{\cal I}_{001}^{}{\cal I}_{120}^{}\right)+{\cal I}_{221}^{}\left({\cal I}_{001}^{}{\cal I}_{020}^{}-2{\cal I}_{021}^{}\right)+{\cal I}_{240}^{}\left({\cal I}_{001}^2-2{\cal I}_{002}^{}\right)\nonumber\\
	&&+{\cal I}_{141}^{}\left({\cal I}_{001}^{}{\cal I}_{100}^{}-2{\cal I}_{101}^{}\right) = 0\;,\\
	\label{eq:syzygy4}
	&{\cal I}_{121}^{(2)}&\left(2{\cal I}_{121}^{(1)} -{\cal I}_{100}^{}{\cal I}_{021}^{}\right)-{\cal I}_{122}^{}\left({\cal I}_{100}^{}{\cal I}_{020}^{}-2{\cal I}_{120}^{}\right)-{\cal I}_{042}^{}\left({\cal I}_{100}^2-2{\cal I}_{200}^{}\right)\nonumber\\
	&&-{\cal I}_{141}^{}\left({\cal I}_{001}^{}{\cal I}_{100}^{}-2{\cal I}_{101}^{}\right) = 0\;,
\end{eqnarray}
together with another two syzygies involving only CP-even invariants. These six syzygies correspond to the first negative term $-6q_{}^8$ in Eq.~(\ref{eq:PL eff 2g main}). Eqs.~(\ref{eq:syzygy1})-(\ref{eq:syzygy4}) establish 4 linear relations among the 6 CP-odd basic invariants, which is consistent with the fact that there are only $6-4=2$ independent phases in the SEFT for the two-generation case.

Among the 18 basic invariants in Table~\ref{table:2g eff}, the 10 
primary ones that are algebraically independent are labeled with ``$(*)$" in the first column. Note that the choice of primary invariants is by no means unique. Later on we will show that from the 10 primary flavor invariants one can explicitly extract all the physical parameters in the two-generation SEFT. In this sense, the set of primary invariants is \emph{equivalent to} the set of independent physical parameters in the theory. 

\subsection{Physical parameters in terms of primary invariants}
\label{subsec:extract2g}
\renewcommand\arraystretch{1.2}
\begin{table}[t!]
	\centering
	\begin{tabular}{l|c|c}
		\hline \hline
		flavor invariants &  degree & CP parity \\
		\hline \hline
		${\cal I}_{100}^{}\equiv {\rm Tr}\left(X_l^{}\right)\quad (*)$ &  1 & $+$ \\
		\hline
		${\cal I}_{001}^{}\equiv {\rm Tr}\left(C_6^{}\right)\quad (*)$ &  1 & $+$\\
		\hline
		${\cal I}_{200}^{}\equiv {\rm Tr}\left(X_l^2\right)\quad (*)$ &  2 & $+$\\
		\hline
		${\cal I}_{101}^{}\equiv {\rm Tr}\left(X_l^{}C_6^{}\right)$ & 2 & $+$\\
		\hline
		${\cal I}_{020}^{}\equiv {\rm Tr}\left(X_5^{}\right)\quad (*)$ & 2 & $+$\\
		\hline
		${\cal I}_{002}^{}\equiv {\rm Tr}\left(C_6^2\right)\quad (*)$ &  2 & $+$\\
		\hline
		${\cal I}_{120}^{}\equiv {\rm Tr}\left(X_l^{}X_5^{}\right)\quad (*)$ &  3 & $+$\\
		\hline
		${\cal I}_{021}^{}\equiv {\rm Tr}\left(C_6^{}X_5{}\right)\quad (*)$ & 3 & $+$\\
		\hline
		${\cal I}_{220}^{}\equiv {\rm Tr}\left(X_l{}G_{l5}^{}\right)\quad (*)$ & 4 & $+$\\
		\hline
		${\cal I}_{121}^{(1)}\equiv {\rm Tr}\left(G_{l5}^{}C_6^{}\right)$ & 4 & $+$\\
		\hline
		${\cal I}_{121}^{(2)}\equiv {\rm Im}\,{\rm Tr}\left(X_l^{}X_5{}C_6^{}\right)$ & 4 & $-$\\
		\hline
		${\cal I}_{040}^{}\equiv {\rm Tr}\left(X_5^{2}\right)\quad (*)$ & 4 & $+$\\
		\hline
		${\cal I}_{022}^{}\equiv {\rm Tr}\left(C_6^{}G_{56}^{}\right)\quad (*)$ & 4 & $+$\\
		\hline
		${\cal I}_{221}^{}\equiv {\rm Im}\,{\rm Tr}\left(X_l^{}G_{l5}^{}C_6^{}\right)$ & 5 & $-$\\
		\hline
		${\cal I}_{122}^{}\equiv {\rm Im}\,{\rm Tr}\left(C_6^{}G_{56}^{}X_l^{}\right)$ & 5 & $-$\\
		\hline
		${\cal I}_{240}^{}\equiv {\rm Im}\,{\rm Tr}\left(X_l^{}X_5^{}G_{l5}^{}\right)$ & 6 & $-$\\
		\hline
		${\cal I}_{141}^{}\equiv {\rm Im}\,{\rm Tr}\left(X_5^{}C_6^{}G_{l5}^{}\right)$ & 6 & $-$\\
		\hline
		${\cal I}_{042}^{}\equiv {\rm Im}\,{\rm Tr}\left(C_6^{}X_5^{}G_{56}^{}\right)$ & 6 & $-$\\
		\hline
		\hline
	\end{tabular}
	\vspace{0.5cm}
	\caption{\label{table:2g eff}Summary of the basic flavor invariants along with their degrees and CP parities in the SEFT for two-generation leptons, where the subscripts of the invariants denote the degrees of $X_l^{}\equiv Y_l^{}Y_l^\dagger$, $C_5^{}$ and $C_6^{}$, respectively. Note that we have defined $X_5^{}\equiv C_5^{}C_5^\dagger$, $G_{l5}^{}\equiv C_5^{}X_l^*C_5^\dagger$ and $G_{56}^{}\equiv C_5^{}C_6^*C_5^\dagger$ that transform adjointly under the flavor-basis transformation. There are totally 12 CP-even and 6 CP-odd basic invariants in the invariant ring of the flavor space. The 10 primary invariants that are algebraically independent have been labeled with ``$(*)$" in the first column.}
\end{table}
\renewcommand\arraystretch{1}
Without loss of generality, we start with the flavor basis where $C_5^{}$ is diagonal with real and non-negative eigenvalues, namely, $C_5^{}={\rm Diag}\{c_1^{},c_2^{}\}$, while $X_l^{}\equiv Y_l^{} Y_l^\dagger$ and $C_6^{}$ are two general $2\times 2$ Hermitian matrices
\begin{eqnarray}
	\label{eq:parametrization of C6 2g}
	X_l^{}=\left(
	\begin{matrix}
		a_{11}^{}&a_{12}^{}e_{}^{{\rm i}\alpha}\\
		a_{12}^{}e_{}^{-{\rm i}\alpha}&a_{22}^{}
	\end{matrix}
	\right)\;,\quad
	C_6^{}=\left(
	\begin{matrix}
		b_{11}^{}&b_{12}^{}e_{}^{{\rm i}\beta}\\
		b_{12}^{}e_{}^{-{\rm i}\beta}&b_{22}^{}
	\end{matrix}
	\right)\;,
\end{eqnarray}
where $a_{ij}^{}$ and $b_{ij}^{}$ are real numbers while $\alpha$ and $\beta$ are two phases. In this basis, the 10 independent physical parameters are $\{c_1^{}, c_2^{}, a_{11}^{}, a_{12}^{}, a_{22}^{}, b_{11}^{}, b_{12}^{}, b_{22}^{}, \alpha, \beta\}$.\footnote{Here we assume that these parameters are not equal or vanishing and $c_1^{}<c_2^{}$, which are true in general.} First, the eigenvalues of $C_5^{}$ can be extracted from ${\cal I}_{020}^{}\equiv {\rm Tr}\left(X_5^{}\right)=c_1^2+c_2^2$ and ${\cal I}_{040}^{}\equiv {\rm Tr}\left(X_5^2\right)=c_1^4+c_2^4$ with $X_5^{}\equiv C_5^{} C_5^\dagger$, i.e.,
\begin{eqnarray}
	c_{1,2}^{}=\frac{1}{\sqrt{2}}\sqrt{{\cal I}_{020}^{}\mp\sqrt{2{\cal I}_{040}^{}-{\cal I}_{020}^2}}\;.
\end{eqnarray}
Second, from ${\cal I}_{100}^{}\equiv {\rm Tr}\left(X_l^{}\right)=a_{11}^{}+a_{22}^{}$ and ${\cal I}_{120}^{}\equiv {\rm Tr}\left(X_l^{}X_5^{}\right)=c_1^2a_{11}^{}+c_2^2a_{22}^{}$ one can extract $a_{11}^{}$ and $a_{22}^{}$ as follows
\begin{eqnarray}
	a_{11,22}^{}=\frac{1}{2}\left({\cal I}_{100}^{}\pm\frac{{\cal I}_{100}{\cal I}_{020}-2{\cal I}_{120}}{\sqrt{2{\cal I}_{040}-{\cal I}_{020}^2}}\right)\;.
\end{eqnarray}
Then from ${\cal I}_{200}^{}\equiv {\rm Tr}\left(X_l^2\right)=a_{11}^2+2a_{12}^2+a_{22}^2$, we obtain\footnote{It is always possible to choose the phase $\alpha$ to make $a_{12}^{}$ positive, and likewise for $b_{12}^{}$ via the phase $\beta$.}
\begin{eqnarray}
	a_{12}^{}=\frac{1}{\sqrt{2}}\sqrt{\frac{{\cal I}_{100}\left({\cal I}_{100}{\cal I}_{040}-2{\cal I}_{020}{\cal I}_{120}\right)+{\cal I}_{200}\left({\cal I}_{020}^2-2{\cal I}_{040}\right)+2{\cal I}_{120}^2}{{\cal I}_{020}^2-2{\cal I}_{040}}}\;.
\end{eqnarray}
Finally, using ${\cal I}_{220}^{}\equiv {\rm Tr}\left(X_l^{}G_{l5}^{}\right)=c_1^2 a_{11}^2+c_2^2 a_{22}^2+2a_{12}^2c_1^{}c_2^{}\cos2\alpha$, one can get
\begin{eqnarray}
	\cos2\alpha=\frac{\left({\cal I}_{100}^2{\cal I}_{020}-4{\cal I}_{100}{\cal I}_{120}+2{\cal I}_{220}\right)\left({\cal I}_{020}^2-{\cal I}_{040}\right)+2\left({\cal I}_{020}{\cal I}_{120}^2-{\cal I}_{040}{\cal I}_{220}\right)}{\sqrt{2}\sqrt{{\cal I}_{020}^2-{\cal I}_{040}}\left[{\cal I}_{200}^{}\left({\cal I}_{020}^2-{\cal I}_{040}\right)+{\cal I}_{040}\left({\cal I}_{100}^2-{\cal I}_{200}\right)-2{\cal I}_{120}\left({\cal I}_{100}{\cal I}_{020}-{\cal I}_{120}\right)\right]}\;. \quad
\end{eqnarray}
Similarly, the parameters in $C_6^{}$ can be extracted in a parallel manner. The final results are given by
\begin{eqnarray}
	\label{eq:extract C6 2g 1}
	b_{11,22}^{}&=&\frac{1}{2}\left({\cal I}_{001}^{}\pm\frac{{\cal I}_{001}{\cal I}_{020}-2{\cal I}_{021}}{\sqrt{2{\cal I}_{040}-{\cal I}_{020}^2}}\right)\;,\\
	\label{eq:extract C6 2g 2}
	b_{12}^{}&=&\frac{1}{\sqrt{2}}\sqrt{\frac{{\cal I}_{001}\left({\cal I}_{001}{\cal I}_{040}-2{\cal I}_{020}{\cal I}_{021}\right)+{\cal I}_{002}\left({\cal I}_{020}^2-2{\cal I}_{040}\right)+2{\cal I}_{021}^2}{{\cal I}_{020}^2-2{\cal I}_{040}}}\;,\\
	\label{eq:extract C6 2g 3}
	\cos2\beta&=&\frac{\left({\cal I}_{001}^2{\cal I}_{020}-4{\cal I}_{001}{\cal I}_{021}+2{\cal I}_{022}\right)\left({\cal I}_{020}^2-{\cal I}_{040}\right)+2\left({\cal I}_{020}{\cal I}_{021}^2-{\cal I}_{040}{\cal I}_{022}\right)}{\sqrt{2}\sqrt{{\cal I}_{020}^2-{\cal I}_{040}}\left[{\cal I}_{002}^{}\left({\cal I}_{020}^2-{\cal I}_{040}\right)+{\cal I}_{040}\left({\cal I}_{001}^2-{\cal I}_{002}\right)-2{\cal I}_{021}\left({\cal I}_{001}{\cal I}_{020}-{\cal I}_{021}\right)\right]}\;.
\end{eqnarray}
When calculating the observables in the SEFT, one usually turns to the basis where the Higgs field acquires its vacuum expectation value and the electroweak gauge symmetry is spontaneously broken down. In this case, two neutrino masses are related to the eigenvalues of $C_5^{}$ via
\begin{eqnarray}
	\label{eq:extract neutrino mass 2g}
	m_{1,2}^{}=\frac{v^2}{2\Lambda}c_{1,2}^{}=\frac{v^2}{2\sqrt{2}\Lambda}\sqrt{{\cal I}_{020}^{}\mp\sqrt{2{\cal I}_{040}^{}-{\cal I}_{020}^2}}\;.
\end{eqnarray}
On the other hand, the masses of charged leptons can be obtained via the diagonalization of $X_l^{}$: $2\,{\rm Diag}\left\{m_e^2,m_\mu^2\right\}/v^2=V_2^{}X_l^{}V_2^\dagger$,
where
\begin{eqnarray}
	\label{eq:parametrization of V 2g}
	V_2^{}=\left(
	\begin{matrix}
		\cos\theta&\sin\theta\\
		-\sin\theta&\cos\theta
	\end{matrix}
	\right)\cdot
	\left(
	\begin{matrix}
		e_{}^{{\rm i}\phi}&0\\
		0&1
	\end{matrix}
	\right)
\end{eqnarray}
is the flavor mixing matrix in the two-generation case, with $\theta$ the flavor mixing angle and $\phi$ the Majorana-type CP phase. Thus the charged-lepton masses, flavor mixing angle and CP phase can be related to the elements in $X_l^{}$ by
\begin{eqnarray}
	m_{e,\mu^{}}=\frac{v}{2}\sqrt{a_{11}^{}+a_{22}^{}\pm\frac{2a_{12}^{}}{\sin2\theta}}\;,\quad
	\tan2\theta=\frac{2a_{12}}{a_{11}-a_{22}}\;,\quad
	\phi=-\alpha\;.
\end{eqnarray}
More explicitly, we express these physical parameters in terms of the flavor invariants as below
\begin{eqnarray}
	\label{eq:extract chargd-lepton mass 2g}
	m_{e,\mu}&=&\frac{v}{2}\sqrt{{\cal I}_{100}^{}\mp\sqrt{2{\cal I}_{200}^{}-{\cal I}_{100}^2}}\;,\\
	\label{eq:extract theta}
	\cos2\theta&=&\frac{2{\cal I}_{120}-{\cal I}_{100}{\cal I}_{020}}{\sqrt{2{\cal I}_{040}-{\cal I}_{020}^2}\sqrt{2{\cal I}_{200}-{\cal I}_{100}^2}}\;,\\
	\label{eq:extract phi}
	\cos2\phi&=&\frac{\left({\cal I}_{100}^2{\cal I}_{020}-4{\cal I}_{100}{\cal I}_{120}+2{\cal I}_{220}\right)\left({\cal I}_{020}^2-{\cal I}_{040}\right)+2\left({\cal I}_{020}{\cal I}_{120}^2-{\cal I}_{040}{\cal I}_{220}\right)}{\sqrt{2}\sqrt{{\cal I}_{020}^2-{\cal I}_{040}}\left[{\cal I}_{200}^{}\left({\cal I}_{020}^2-{\cal I}_{040}\right)+{\cal I}_{040}\left({\cal I}_{100}^2-{\cal I}_{200}\right)-2{\cal I}_{120}\left({\cal I}_{100}{\cal I}_{020}-{\cal I}_{120}\right)\right]}\;. \quad
\end{eqnarray}
To summarize, we have shown that the 10 physical parameters after the gauge symmetry breaking 
$$\left\{m_1^{},m_2^{},m_e^{},m_\mu^{},\theta,\phi,b_{11}^{},b_{12}^{},b_{22}^{},\beta\right\}$$ 
can be extracted from the 10 primary flavor invariants 
$$\left\{{\cal I}_{020}^{},{\cal I}_{040}^{}, {\cal I}_{100}^{}, {\cal I}_{001}^{}, {\cal I}_{200}, {\cal I}_{002}^{}, {\cal I}_{120}^{}, {\cal I}_{021}^{}, {\cal I}_{220}^{}, {\cal I}_{022}^{} \right\}$$
by Eqs.~(\ref{eq:extract C6 2g 1})-(\ref{eq:extract C6 2g 3}), Eq.~(\ref{eq:extract neutrino mass 2g}) and Eqs.~(\ref{eq:extract chargd-lepton mass 2g})-(\ref{eq:extract phi}).
\subsection{Conditions for CP conservation}
\label{subsec:conditions2g}
Although there are 6 CP-odd basic invariants in the ring, they are not algebraically independent but related to each other by the syzygies in Eqs.~(\ref{eq:syzygy1})-(\ref{eq:syzygy4}). On the other hand, there are only 2 independent physical phases in the leptonic sector (i.e., $\phi$ and $\beta$ in the symmetry-breaking basis). Therefore, the \emph{minimal} conditions to guarantee CP conservation in the leptonic sector is the vanishing of only 2 CP-odd invariants. In the symmetry-breaking basis chosen above, it is straightforward to explicitly calculate the following two CP-odd flavor invariants
\begin{eqnarray}
	\label{eq:i1212}
	{\cal I}_{121}^{(2)}&\equiv&{\rm Im}\,{\rm Tr}\left(X_l^{}X_5^{}C_6^{}\right)=\frac{1}{v^2}\left(m_\mu^2-m_e^2\right)\left(c_2^2-c_1^2\right)b_{12}^{}\sin2\theta\sin\left(\beta+\phi\right)\;,\\
	\label{eq:i240}
	{\cal I}_{240}^{}&\equiv&{\rm Im}\,{\rm Tr}\left(X_l^{}X_5^{}G_{l5}^{}\right)=-\frac{1}{v^4}\left(m_\mu^2-m_e^2\right)_{}^2\left(c_2^2-c_1^2\right)c_1^{}c_2^{}\sin_{}^22\theta\sin2\phi\;.
\end{eqnarray} 
We have assumed that there is neither degeneracy for the lepton masses nor texture zero in the matrix elements of $C_6^{}$ and $X_l^{}$ in general. Thus the vanishing of ${\cal I}_{121}^{(2)}$ and ${\cal I}_{240}^{}$ enforces the phases to take only trivial values
\begin{eqnarray}
	\beta, \; \phi=k\pi\qquad {\rm or} \qquad
	\beta, \; \phi=\frac{2k+1}{2}\pi\;,
\end{eqnarray}
where $k$ is an arbitrary integer. In this basis, one can verify that all the CP-violating observables are proportional to $\sin2\phi$, $\sin2\beta$ or $\sin\left(\beta\pm\phi\right)$, so ${\cal I}_{121}^{(2)}={\cal I}_{240}^{}=0$ serves as the minimal sufficient and necessary conditions for CP conservation in the leptonic sector. As will be shown in Sec.~\ref{sec:observables}, ${\cal I}_{121}^{(2)}$ and ${\cal I}_{240}^{}$ are responsible for the CP violation in  neutrino-neutrino and neutrino-antineutrino oscillations, respectively. Moreover, any CP-violating observable in the two-generation SEFT can be expressed as the linear combination of ${\cal I}_{121}^{(2)}$ and ${\cal I}_{240}^{}$ with the combination coefficients being functions of only CP-even invariants. As a consequence, the vanishing of ${\cal I}_{121}^{(2)}$ and ${\cal I}_{240}^{}$ implies the vanishing of any CP-violating observable.

\section{Algebraic structure of the SEFT flavor invariants: Three-generation case}
\label{sec:construction3g}
Now we proceed with the realistic case of three-generation leptons. As we shall see, the algebraic structure of the invariant ring in the three-generation SEFT is much more complicated than that in the two-generation case.

From the transformation behaviors of the building blocks under ${\rm U}(3)$ group in Eq.~(\ref{eq:wilson coe trans}), one can calculate the HS using the MW formula (see Appendix~\ref{app:HS} for more details). Though very tedious, the result can be recast into the standard form
\begin{eqnarray}
	\label{eq:HS eff 3g main}
	{\mathscr H}_{\rm SEFT}^{(3\rm g)}(q)=\frac{{\mathscr N}_{\rm SEFT}^{(3\rm g)}(q)}{{\mathscr D}_{\rm SEFT}^{(3\rm g)}(q)}\;,
\end{eqnarray}
where 
\begin{eqnarray}
	\label{eq:numerator eff 3g main}
	{\mathscr N}_{\rm SEFT}^{(3\rm g)}(q)&=&q^{65}+2 q^{64}+4 q^{63}+11 q^{62}+23 q^{61}+48 q^{60}+120 q^{59}+269 q^{58}+587 q^{57}+1258 q^{56}\nonumber\\
	&&+2543 q^{55}+4895 q^{54}+9124 q^{53}+16281 q^{52}+27963 q^{51}+46490 q^{50}+74644 q^{49}\nonumber\\
	&&+115871q^{48}+174433 q^{47}+254494 q^{46}+360055 q^{45}+494873 q^{44}+660820 q^{43}\nonumber\\
	&&+857677 q^{42}+1083226 q^{41}+1331628 q^{40}+1593650 q^{39}+1858178 q^{38}+2111158 q^{37}\nonumber\\
	&&+2337226 q^{36}+2522435
	q^{35}+2654026 q^{34}+2721987 q^{33}+2721987 q^{32}+2654026q^{31}\nonumber\\
	&&+2522435 q^{30}+2337226 q^{29}+2111158 q^{28}+1858178 q^{27}+1593650 q^{26}+1331628 q^{25}\nonumber\\
	&&+1083226 q^{24}+857677 q^{23}+660820
	q^{22}+494873 q^{21}+360055 q^{20}+254494 q^{19}\nonumber\\
	&&+174433 q^{18}+115871 q^{17}+74644 q^{16}+46490 q^{15}+27963 q^{14}+16281 q^{13}+9124 q^{12}\nonumber\\
	&&+4895 q^{11}+2543 q^{10}+1258 q^9+587 q^8+269 q^7+120
	q^6+48 q^5+23 q^4+11 q^3+4 q^2\nonumber\\
	&&+2 q+1\;,
\end{eqnarray}
and
\begin{eqnarray}
	\label{eq:denominator eff 3g main}
	{\mathscr D}_{\rm SEFT}^{(3\rm g)}(q)=\left(1-q^2\right)^3 \left(1-q^3\right) \left(1-q^4\right)^5 \left(1-q^5\right)^6 \left(1-q^6\right)^6\;.
\end{eqnarray}

As a nontrivial cross-check, the denominator of the HS in Eq.~(\ref{eq:denominator eff 3g main}) do have 21 factors, exactly matching the number of independent physical parameters in the three-generation SEFT. Moreover, the awesomely large numbers in the numerator of the HS in Eq.~(\ref{eq:numerator eff 3g main})
indicate that the richness of the flavor structure and the complexity of the invariant ring grow very quickly with the generation of leptons when there exist Majorana-type building blocks.\footnote{This is very different from the case in the quark sector, where all the fermions are Dirac particles and all the building blocks transform adjointly under the unitary group. The Majorana character of $C_5^{}$ enforces it to transform as the symmetric rank-2 tensor representation, thus leading to a much more complicated algebraic structure of the invariant ring in the leptonic sector.} The complexity of the algebraic structure of the invariant ring for the three-generation case can be better understood by substituting Eq.~(\ref{eq:HS eff 3g main}) into Eq.~(\ref{eq:PL def}) to calculate the PL function of the HS, which is an infinite series of $q$ and encodes all information about the basic invariants and the polynomial relations among them (i.e., syzygies)
\begin{eqnarray}
	\label{eq:PL eff 3g main}
	{\rm PL}\left[{\mathscr H}_{\rm SEFT}^{(3\rm g)}(q)\right]&=&2q+4q^2+6q^3+9q^4+14q^5+33q^6+44q^7+72q^8+74q^9+21q^{10}\nonumber\\
	&&-208q^{11}-708q^{12}-1904q^{13}-3806q^{14}-{\cal O}\left(q_{}^{15}\right)\;.
\end{eqnarray}
From Eq.~(\ref{eq:PL eff 3g main}) one can observe that there are nearly 300 basic invariants in the ring, more than 200 syzygies at degree 11, more than 700 syzygies at degree 12, and so on. Therefore, it is very difficult to explicitly construct all the basic invariants due to the complexity of the invariant ring. However, we have known that there are only 21 independent physical parameters in the theory, which is also the maximum number of the algebraically-independent invariants (i.e., primary invariants) in the ring. Hence, we shall construct only the primary flavor invariants for the three-generation case, with which one can extract all the physical parameters so that any physical observable can be written as the function of flavor invariants.

\subsection{Physical parameters in terms of primary invariants}
\label{subsec:extract3g}
\renewcommand\arraystretch{1.2}
\begin{table}[t!]
	\centering
	\begin{tabular}{l|c|c}
		\hline \hline
		flavor invariants &  degree & CP parity \\
		\hline \hline
		${\cal J}_{100}^{}\equiv {\rm Tr}\left(X_l^{}\right) $ &  1 & $+$ \\
		\hline
		${\cal J}_{001}^{}\equiv {\rm Tr}\left(C_6^{}\right) $ &  1 & $+$\\
		\hline
		${\cal J}_{200}^{}\equiv {\rm Tr}\left(X_l^2\right)$ &  2 & $+$\\
		\hline
		${\cal J}_{020}^{}\equiv {\rm Tr}\left(X_5^{}\right) $ & 2 & $+$\\
		\hline
		${\cal J}_{002}^{}\equiv {\rm Tr}\left(C_6^2\right)$ &  2 & $+$\\
		\hline
		${\cal J}_{120}^{}\equiv {\rm Tr}\left(X_l^{}X_5^{}\right) $ &  3 & $+$\\
		\hline
		${\cal J}_{021}^{}\equiv {\rm Tr}\left(C_6^{}X_5{}\right) $ & 3 & $+$\\
		\hline
		${\cal J}_{220}^{}\equiv {\rm Tr}\left(X_l{}G_{l5}^{}\right) $ & 4 & $+$\\
		\hline
		${\cal J}_{040}^{}\equiv {\rm Tr}\left(X_5^{2}\right) $ & 4 & $+$\\
		\hline
		${\cal J}_{022}^{}\equiv {\rm Tr}\left(C_6^{}G_{56}^{}\right) $ & 4 & $+$\\
		\hline
		${\cal J}_{140}^{}\equiv {\rm Tr}\left(X_l^{}X_5^{2}\right)$ & 5 & $+$\\
		\hline
		${\cal J}_{041}^{}\equiv {\rm Tr}\left(C_6^{}X_5^{2}\right)$ & 5 & $+$\\
		\hline
		${\cal J}_{240}^{}\equiv {\rm Tr}\left(X_l^{2}X_5^{2}\right)$ & 6 & $+$\\
		\hline
		${\cal J}_{240}^{(2)}\equiv {\rm Im}\,{\rm Tr}\left(X_l^{}X_5^{}G_{l5}^{}\right)$ & 6 & $-$\\
		\hline
		${\cal J}_{060}^{}\equiv {\rm Tr}\left(X_5^{3}\right)$ & 6 & $+$\\
		\hline
		${\cal J}_{042}^{}\equiv {\rm Tr}\left(C_6^{2}X_5^{2}\right)$ & 6 & $+$\\
		\hline
		${\cal J}_{042}^{(2)}\equiv {\rm Im}\,{\rm Tr}\left(C_6^{}X_5^{}G_{56}^{}\right)$ & 6 & $-$\\
		\hline
		${\cal J}_{260}^{}\equiv {\rm Im}\,{\rm Tr}\left(X_l^{}X_5^{2}G_{l5}^{}\right)$ & 8 & $-$\\
		\hline
		${\cal J}_{062}^{}\equiv {\rm Im}\,{\rm Tr}\left(C_6^{}X_5^{2}G_{56}^{}\right)$ & 8 & $-$\\
		\hline
		${\cal J}_{280}^{}\equiv {\rm Im}\,{\rm Tr}\left(X_l^{}X_5^{2}G_{l5}^{}X_5^{}\right)$ & 10 & $-$\\
		\hline
		${\cal J}_{082}^{}\equiv {\rm Im}\,{\rm Tr}\left(C_6^{}X_5^{2}G_{56}^{}X_5^{}\right)$ & 10 & $-$\\
		\hline
		\hline
	\end{tabular}
	\vspace{0.5cm}
	\caption{\label{table:3g eff} Summary of the primary flavor invariants along with their degrees and CP parities in the SEFT for three-generation leptons, where the subscripts of the invariants denote the degrees of $X_l^{}\equiv Y_l^{}Y_l^\dagger$, $C_5^{}$ and $C_6^{}$, respectively. We have also defined $X_5^{}\equiv C_5^{}C_5^\dagger$, $G_{l5}^{}\equiv C_5^{}X_l^*C_5^\dagger$ and $G_{56}^{}\equiv C_5^{}C_6^*C_5^\dagger$. There are in total 21 primary invariants in the invariant ring of the flavor space and 6 of them are CP-odd, corresponding to the 6 independent phases in the three-generation SEFT.}
\end{table}
\renewcommand\arraystretch{1}

The 21 primary invariants have been explicitly constructed in Table~\ref{table:3g eff}. Then we show how to extract all physical parameters in the SEFT from the primary invariants.
For convenience, we again start with the basis where $C_5^{}$ is diagonal with real and non-negative eigenvalues, i.e., $C_5^{}={\rm Diag}\left\{c_1^{},c_2^{},c_3^{}\right\}$, while $X_l^{}$ and $C_6^{}$ are two arbitrary $3\times 3$ Hermitian matrices\footnote{As in the two-generation case, we assume that $c_1^{}\neq c_2^{}\neq c_3^{}$ and there are in general no texture zeros in the matrix elements of $X_l^{}$ and $C_6^{}$.}
\begin{eqnarray}
	X_l^{}=\left(
	\begin{matrix}
		a_{11}^{}&a_{12}^{}e_{}^{{\rm i}\alpha_{12}}&a_{31}e_{}^{-{\rm i}\alpha_{31}}\\
		a_{12}e_{}^{-{\rm i}\alpha_{12}}&a_{22}^{}&a_{23}^{}e_{}^{{\rm i}\alpha_{23}}\\
		a_{31}e_{}^{{\rm i}\alpha_{31}}&a_{23}^{}e_{}^{-{\rm i}\alpha_{23}}&a_{33}^{}
	\end{matrix}
	\right)\;,\quad
	C_6^{}=\left(
	\begin{matrix}
		b_{11}^{}&b_{12}^{}e_{}^{{\rm i}\beta_{12}}&b_{31}e_{}^{-{\rm i}\beta_{31}}\\
		b_{12}e_{}^{-{\rm i}\beta_{12}}&b_{22}^{}&b_{23}^{}e_{}^{{\rm i}\beta_{23}}\\
		b_{31}e_{}^{{\rm i}\beta_{31}}&b_{23}^{}e_{}^{-{\rm i}\beta_{23}}&b_{33}^{}
	\end{matrix}
	\right)\;,
\end{eqnarray}
where $a_{ij}^{}$ and $b_{ij}^{}$ are real numbers while $\alpha_{ij}^{}$ and $\beta_{ij}^{}$ are phases. First, the eigenvalues of $C_5^{}$, corresponding to the masses of light Majorana neutrinos, can be extracted from ${\cal J}_{020}^{}$, ${\cal J}_{040}^{}$ and ${\cal J}_{060}^{}$ as follows
\begin{eqnarray}
	\label{eq:extract C5 3g 1}
	{\cal J}_{020}^{}&\equiv& {\rm Tr}\left(X_5^{}\right)=c_1^2+c_2^2+c_3^2\;,\\
	\label{eq:extract C5 3g 2}
	{\cal J}_{040}^{}&\equiv& {\rm Tr}\left(X_5^{2}\right)=c_1^4+c_2^4+c_3^4\;,\\
	\label{eq:extract C5 3g 3}
	{\cal J}_{060}^{}&\equiv& {\rm Tr}\left(X_5^{3}\right)=c_1^6+c_2^6+c_3^6\;.
\end{eqnarray}
In principle $c_1^{}$, $c_2^{}$ and $c_3^{}$ can be solved from Eqs.~(\ref{eq:extract C5 3g 1})-(\ref{eq:extract C5 3g 3}) in term of ${\cal J}_{020}^{}$, ${\cal J}_{040}^{}$ and ${\cal J}_{060}^{}$. However, the most general solution is too lengthy to be listed here. For illustration, we only consider two hierarchical scenarios. For the normal hierarchical mass ordering with $0 \leqslant c_1^{}<c_2^{}\ll c_3^{}$, we have
\begin{eqnarray}
	c_{1,2}^{}=\sqrt{\frac{{\cal J}_{020}^2-{\cal J}_{040}}{4{\cal J}_{060}^{1/3}}\mp\sqrt{\left(\frac{{\cal J}_{020}^2-{\cal J}_{040}}{4{\cal J}_{060}^{1/3}}\right)_{}^2-\frac{{\cal J}_{020}^3-3{\cal J}_{020}{\cal J}_{040}+2{\cal J}_{060}}{6{\cal J}_{060}^{1/3}}}}\;,\quad
	c_3^{}={\cal J}_{060}^{1/6}\;;
\end{eqnarray}
while for the inverted hierarchical mass ordering with $0 \leqslant c_3^{}\ll c_1^{}<c_2^{}$, we get
\begin{eqnarray}
	c_{1,2}^{}=\frac{1}{\sqrt{2}}\sqrt{{\cal J}_{020}^{}-c_3^2\mp\sqrt{2{\cal J}_{040}^{}-{\cal J}_{020}^2+2{\cal J}_{020}^{}c_3^2}}\;,\quad
	c_3^{}=\sqrt{\frac{{\cal J}_{020}^3-3{\cal J}_{020}{\cal J}_{040}+2{\cal J}_{060}}{3\left({\cal J}_{020}^2-{\cal J}_{040}\right)}}\;.
\end{eqnarray}
Then the invariants ${\cal J}_{100}^{}$, ${\cal J}_{120}^{}$ and ${\cal J}_{140}$ can be used to extract the diagonal elements of $X_l^{}$, i.e.,
\begin{eqnarray*}
	{\cal J}_{100}^{}&\equiv&{\rm Tr}\left(X_l^{}\right)=a_{11}^{}+a_{22}^{}+a_{33}^{}\;,\\
	{\cal J}_{120}^{}&\equiv&{\rm Tr}\left(X_l^{}X_5^{}\right)=c_1^2a_{11}^{}+c_2^2a_{22}^{}+c_3^2a_{33}^{}\;,\\
	{\cal J}_{140}^{}&\equiv&{\rm Tr}\left(X_l^{}X_5^{2}\right)=c_1^4a_{11}^{}+c_2^4a_{22}^{}+c_3^4a_{33}^{}\;,
\end{eqnarray*}
from which we obtain
\begin{eqnarray}
	\label{eq:extract aii}
	a_{ii}^{}=\frac{\left(c_j^2+c_k^2\right){\cal J}_{120}-c_j^2c_k^2{\cal J}_{100}-{\cal J}_{140}}{\left(c_k^2-c_i^2\right)\left(c_i^2-c_j^2\right)}\;,
\end{eqnarray}
where $\left(i,j,k\right)=\left(1,2,3\right)$ or $\left(2,3,1\right)$ or $\left(3,1,2\right)$. As for the off-diagonal elements of $X_l^{}$, we can use
\begin{eqnarray*}
	{\cal J}_{200}^{}&\equiv& {\rm Tr}\left(X_l^2\right)=a_{11}^2+a_{22}^2+a_{33}^2+2\left(a_{12}^{2}+a_{23}^{2}+a_{31}^{2}\right)\;,\\
	{\cal J}_{220}^{}&\equiv& {\rm Tr}\left(X_l^2X_5^{}\right)=c_1^2a_{11}^2+c_2^2a_{22}^2+c_3^2a_{33}^2+\left(c_1^2+c_2^2\right)a_{12}^2+\left(c_2^2+c_3^2\right)a_{23}^2+\left(c_3^2+c_1^2\right)a_{31}^2\;,\\
	{\cal J}_{240}^{}&\equiv& {\rm Tr}\left(X_l^2X_5^{2}\right)=c_1^4a_{11}^2+c_2^4a_{22}^2+c_3^4a_{33}^2+\left(c_1^4+c_2^4\right)a_{12}^2+\left(c_2^4+c_3^4\right)a_{23}^2+\left(c_3^4+c_1^4\right)a_{31}^2\;,
\end{eqnarray*}
to derive\footnote{Without loss of generality, the phases $\alpha_{ij}^{}$ and $\beta_{ij}^{}$ can be chosen to ensure $a_{ij}^{}>0$ and $b_{ij}^{} > 0$ for $i\neq j$.}
\begin{eqnarray}
	\label{eq:extract aij}
	a_{ij}^{}=\sqrt{\frac{\left(c_1^2c_2^2+c_2^2c_3^2+c_3^2c_1^2-c_k^4\right){\cal J}_{200}^{\prime}-\left(c_i^2+c_j^2\right){\cal J}_{220}^{\prime}+{\cal J}_{240}^{\prime}}{\left(c_j^2-c_k^2\right)\left(c_k^2-c_i^2\right)}}\;,
\end{eqnarray}
where $\left(i,j,k\right)=\left(1,2,3\right)$ or $\left(2,3,1\right)$ or $\left(3,1,2\right)$ and
\begin{eqnarray*}
	{\cal J}_{200}^\prime &\equiv& \frac{1}{2}\left({\cal J}_{200}^{}-a_{11}^2-a_{22}^2-a_{33}^2\right)\;,\\
	{\cal J}_{220}^\prime &\equiv& {\cal J}_{220}^{}-\left(c_1^2a_{11}^2+c_2^2a_{22}^2+c_3^2a_{33}^2\right)\;,\\
	{\cal J}_{240}^\prime &\equiv& {\cal J}_{240}^{}-\left(c_1^4a_{11}^2+c_2^4a_{22}^2+c_3^4a_{33}^2\right)\;.
\end{eqnarray*}
Finally, the phases in $X_l^{}$ can be conveniently determined by using CP-odd invariants
\begin{eqnarray*}
	{\cal J}_{240}^{(2)}&\equiv&{\rm Im}\,{\rm Tr}\left(X_l^{}X_5^{}G_{l5}^{}\right)\\
	&=&-a_{12}^2c_1^{}c_2^{}\left(c_1^2-c_2^2\right)\sin2\alpha_{12}^{}-a_{23}^2c_2^{}c_3^{}\left(c_2^2-c_3^2\right)\sin2\alpha_{23}^{}-a_{31}^2c_3^{}c_1^{}\left(c_3^2-c_1^2\right)\sin2\alpha_{31}^{}\;,\\
	{\cal J}_{260}^{}&\equiv&{\rm Im}\,{\rm Tr}\left(X_l^{}X_5^{2}G_{l5}^{}\right)\\
	&=&-a_{12}^2c_1^{}c_2^{}\left(c_1^4-c_2^4\right)\sin2\alpha_{12}^{}-a_{23}^2c_2^{}c_3^{}\left(c_2^4-c_3^4\right)\sin2\alpha_{23}^{}-a_{31}^2c_3^{}c_1^{}\left(c_3^4-c_1^4\right)\sin2\alpha_{31}^{}\;,\\
	{\cal J}_{280}^{}&\equiv&{\rm Im}\,{\rm Tr}\left(X_l^{}X_5^{2}G_{l5}^{}\right)\\
	&=&-a_{12}^2c_1^{3}c_2^{3}\left(c_1^2-c_2^2\right)\sin2\alpha_{12}^{}-a_{23}^2c_2^{3}c_3^{3}\left(c_2^2-c_3^2\right)\sin2\alpha_{23}^{}-a_{31}^2c_3^{3}c_1^{3}\left(c_3^2-c_1^2\right)\sin2\alpha_{31}^{}\;,
\end{eqnarray*}
from which we have
\begin{eqnarray}
	\label{eq:extract alphaij}
	\sin2\alpha_{ij}^{}=\frac{c_k^4{\cal J}_{240}^{(2)}-c_k^2{\cal J}_{260}+{\cal J}_{280}}{a_{ij}^{2}c_i c_j\left(c_1^2-c_2^2\right)\left(c_2^2-c_3^2\right)\left(c_3^2-c_1^2\right)}\;,
\end{eqnarray}
where $\left(i,j,k\right)=\left(1,2,3\right)$ or $\left(2,3,1\right)$ or $\left(3,1,2\right)$. Similarly, the elements in $C_6^{}$ can be extracted in a parallel manner
\begin{eqnarray}
	\label{eq:extract bii}
	b_{ii}^{}&=&\frac{\left(c_j^2+c_k^2\right){\cal J}_{021}-c_j^2c_k^2{\cal J}_{001}-{\cal J}_{041}}{\left(c_k^2-c_i^2\right)\left(c_i^2-c_j^2\right)}\;,\\
	\label{eq:extract bij}
	b_{ij}^{}&=&\sqrt{\frac{\left(c_1^2c_2^2+c_2^2c_3^2+c_3^2c_1^2-c_k^4\right){\cal J}_{002}^{\prime}-\left(c_i^2+c_j^2\right){\cal J}_{022}^{\prime}+{\cal J}_{042}^{\prime}}{\left(c_j^2-c_k^2\right)\left(c_k^2-c_i^2\right)}}\;,\\
	\label{eq:extract betaij}
	\sin2\beta_{ij}^{}&=&\frac{c_k^4{\cal J}_{042}^{(2)}-c_k^2{\cal J}_{062}+{\cal J}_{082}}{b_{ij}^{2}c_i c_j\left(c_1^2-c_2^2\right)\left(c_2^2-c_3^2\right)\left(c_3^2-c_1^2\right)}\;,
\end{eqnarray}
where $\left(i,j,k\right)=\left(1,2,3\right)$ or $\left(2,3,1\right)$ or $\left(3,1,2\right)$ and
\begin{eqnarray*}
	{\cal J}_{002}^\prime &\equiv& \frac{1}{2}\left({\cal J}_{002}^{}-b_{11}^2-b_{22}^2-b_{33}^2\right)\;,\\
	{\cal J}_{022}^\prime &\equiv& {\cal J}_{022}^{}-\left(c_1^2b_{11}^2+c_2^2b_{22}^2+c_3^2b_{33}^2\right)\;,\\
	{\cal J}_{042}^\prime &\equiv& {\cal J}_{042}^{}-\left(c_1^4b_{11}^2+c_2^4b_{22}^2+c_3^4b_{33}^2\right)\;.
\end{eqnarray*}
In summary, in the basis where $C_5^{}$ is real and diagonal, the 21 physical parameters in the leptonic sector for the three-generation case
$$\left\{c_1^{},c_2^{},c_3^{},a_{11}^{},a_{22}^{},a_{33}^{},a_{12}^{},a_{23}^{},a_{31}^{},\alpha_{12}^{},\alpha_{23}^{},\alpha_{31}^{},b_{11}^{},b_{22}^{},b_{33}^{},b_{12}^{},b_{23}^{},b_{31}^{},\beta_{12}^{},\beta_{23}^{},\beta_{31}^{}\right\}$$
can be extracted from the 21 primary invariants in Table~\ref{table:3g eff} by Eqs.~(\ref{eq:extract C5 3g 1})-(\ref{eq:extract C5 3g 3}) and Eqs.~(\ref{eq:extract aii})-(\ref{eq:extract betaij}). Note that among the primary invariants we choose 6 of them to be CP-odd and the others to be CP-even, in accordance with the fact that there are 6 independent CP-violating phases in the three-generation SEFT.

After the spontaneous breaking of the gauge symmetry, the masses of neutrinos are given by
\begin{eqnarray}
	m_i^{}=\frac{v^2}{2\Lambda}c_i^{}\;,\qquad
	i=1,2,3\;,
\end{eqnarray}
while the masses of charged leptons are determined by the eigenvalue of $X_l^{}$
\begin{eqnarray}
	m_\alpha^{}=\frac{v}{\sqrt{2}}\sqrt{l_\alpha^{}}\;,\qquad
	\alpha=e,\mu,\tau\;,
\end{eqnarray} 
with $l_\alpha^{}$ the eigenvalues of $X_l^{}$. Furthermore, the flavor mixing matrix $V_3^{}$ is related to $X_l^{}$ by $X_l^{}=V_3^\dagger{\rm Diag}\left\{l_e^{},l_\mu^{},l_\tau^{}\right\}V_3^{}$ and $X_l^{2}=V_3^{\dagger}{\rm Diag}\left\{l_e^{2},l_\mu^{2},l_\tau^{2}\right\}V_3^{}$, from which one can extract the matrix elements of $V_3^{}$ as below
\begin{eqnarray}
	\label{eq:extract Vii}
	\left|V_{\alpha i}\right|^2&=&\frac{\left(X_l\right)_{ii}\left(l_\beta+l_\gamma\right)-\left(X_l^2\right)_{ii}-l_\beta l_\gamma}{\left(l_\gamma-l_\alpha\right)\left(l_\alpha-l_\beta\right)}\;,\qquad
	i=1,2,3\;,\\
	\label{eq:extract Vij}
	V_{\alpha i}^*V_{\alpha j}^{}&=&\frac{\left(X_l^{}\right)_{ij}\left(l_\beta+l_\gamma\right)-\left(X_l^2\right)_{ij}}{\left(l_\gamma-l_\alpha\right)\left(l_\alpha-l_\beta\right)}\;,\qquad
	i,j=1,2,3\; (i\neq j)\;,
\end{eqnarray} 
with $\left(\alpha,\beta,\gamma\right)=\left(e,\mu,\tau\right)$ or $\left(\mu,\tau,e\right)$ or $\left(\tau,e,\mu\right)$, and $V_{\alpha i}^{}$ denoting the $\left(\alpha,i\right)$-element of $V_3^{}$. In the standard parametrization of the PMNS matrix~\cite{ParticleDataGroup:2020ssz}, $V_3^{}$ is written as
\begin{eqnarray}
	\label{eq:standard para}
	V_3^{}=\left( \begin{matrix} c^{}_{13} c^{}_{12} & c^{}_{13} s^{}_{12} & s^{}_{13} e_{}^{-{\rm i}\delta} \cr -s_{12}^{} c_{23}^{} - c_{12}^{} s_{13}^{} s_{23}^{} e^{ {\rm i}\delta}_{} & + c_{12}^{} c_{23}^{} - s_{12}^{} s_{13}^{} s_{23}^{} e^{{\rm i}\delta}_{} & c_{13}^{} s_{23}^{} \cr + s_{12}^{} s_{23}^{} - c_{12}^{} s_{13}^{} c_{23}^{} e^{ {\rm i}\delta}_{} & - c_{12}^{} s_{23}^{} - s_{12}^{} s_{13}^{} c_{23}^{} e^{{\rm i}\delta}_{} & c_{13}^{} c_{23}^{} \end{matrix} \right) \cdot \left(\begin{matrix} e^{{\rm i}\rho}_{} & 0 & 0 \\ 0 & e^{{\rm i}\sigma}_{} & 0 \\ 0 & 0 & 1\end{matrix}\right) \; ,
\end{eqnarray}
where $c_{ij}^{}\equiv\cos\theta_{ij}^{}$ and $s_{ij}^{}\equiv\sin\theta_{ij}^{}$ (for $ij=12,13,23$). Therefore, the flavor mixing angles $\left\{\theta_{12}^{},\theta_{13}^{},\theta_{23}^{}\right\}$, the Dirac-type phase $\delta$ and two Majorana-type phases $\left\{\rho,\sigma\right\}$ can also be extracted from primary flavor invariants through
\begin{eqnarray}
	&&s_{13}^{} = |V_{e3}|\;,\quad s_{12}^{} = \frac{|V_{e2}|}{\sqrt{1-|V_{e3}|^2}}\;,\quad s_{23}^{} = \frac{|V_{\mu 3}|}{\sqrt{1-|V_{e3}|^2}}\;,\quad
	\sin \delta =\frac{{\rm Im}\left(V_{e2}^{}V_{e3}^{*}V_{\mu 2}^{*}V_{\mu 3}\right)}{s_{12}^{} c_{12}^{} s_{23}^{} c_{23}^{} s_{13}^{} c_{13}^2}\;,\nonumber\\
	&&\rho = - \delta - {\rm Arg}\left(\frac{V_{e1}^* V_{e3}^{}}{c_{12}^{} c_{13}^{} s_{13}^{}}\right)\;,\quad
	\sigma = - \delta - {\rm Arg}\left(\frac{V_{e2}^* V_{e3}^{}}{s_{12}^{} c_{13}^{} s_{13}^{}}\right)\;,
\end{eqnarray}
and Eqs.~(\ref{eq:extract Vii})-(\ref{eq:extract Vij}).

\subsection{Conditions for CP conservation}
\label{subsec:conditions3g}
In this subsection we investigate the conditions for CP conservation in the three-generation SEFT. As in the two-generation case, one would expect the minimal conditions to guarantee CP conservation in the leptonic sector require the vanishing of 6 CP-odd invariants, which is also the number of the independent phases in the theory. An immediate choice is the 6 CP-odd primary invariants in Table~\ref{table:3g eff}. However, it can be shown that the vanishing of all the 6 CP-odd invariants in Table~\ref{table:3g eff} is \emph{not} sufficient to guarantee CP conservation in the leptonic sector. An explicit counter example is $\alpha_{ij}^{}=\beta_{ij}^{}=\pi/2$ (for $ij=12,23,31$), which is of course a solution to ${\cal J}_{240}^{(2)}={\cal J}_{042}^{(2)}={\cal J}_{260}^{}={\cal J}_{062}^{}={\cal J}_{280}^{}={\cal J}_{082}^{}=0$. But, in this case, the following Jarlskog-like flavor invariant
\begin{eqnarray}
	\label{eq:j360}
	{\cal J}_{360}^{} &\equiv& {\rm Im}\,{\rm Tr}\left(X_l^2X_5^2X_l^{}X_5^{}\right)=\frac{1}{2{\rm i}}{\rm Det}\left(\left[X_l^{},X_5^{}\right]\right)\nonumber\\
	&=&-a_{12}^{}a_{23}^{}a_{31}^{}\left(c_1^2-c_2^2\right)\left(c_2^2-c_3^2\right)\left(c_3^2-c_1^2\right)\sin\left(\alpha_{12}^{}+\alpha_{23}^{}+\alpha_{31}^{}\right)\;,
\end{eqnarray}
is nonzero if the neutrino masses are not degenerate and there are no texture zeros in $X_l^{}$. The invariant ${\cal J}_{360}^{}$ will appear in the CP asymmetries of neutrino oscillations and cause CP violation in the leptonic sector. In Ref.~\cite{Yu:2019ihs} we have actually proved that in the presence of only the charged-lepton mass matrix and the Majorana neutrino mass matrix [i.e., the effective theory up to ${\cal O}\left(1/\Lambda\right)$], the minimal sufficient and necessary conditions for CP conservation in the leptonic sector are give by
\begin{eqnarray}
	\label{eq:cp conservation condition 1}
	{\cal J}_{360}^{}&\equiv& {\rm Im}\,{\rm Tr}\left(X_l^2X_5^2X_l^{}X_5^{}\right)=0\;,\\
	\label{eq:cp conservation condition 2}
	{\cal J}_{240}^{(2)}&\equiv&{\rm Im}\,{\rm Tr}\left(X_l^{}X_5^{}G_{l5}^{}\right)=0\;,\\
	\label{eq:cp conservation condition 3}
	{\cal J}_{260}^{}&\equiv&{\rm Im}\,{\rm Tr}\left(X_l^{}X_5^{2}G_{l5}^{}\right)=0\;.
\end{eqnarray}
Eqs.~(\ref{eq:cp conservation condition 1})-(\ref{eq:cp conservation condition 3}) enforce the three CP phases in the flavor mixing matrix to take only trivial values: 
\begin{eqnarray*}
	&&({\rm i})\; \delta=\rho=\sigma=0 \iff \alpha_{12}^{}, \; \alpha_{23}^{}, \; \alpha_{31}^{}=k\pi\\
	&&({\rm ii})\; \delta=\rho=0\,,\;\sigma=\pi/2 \iff \alpha_{12}^{}, \; \alpha_{23}^{}=\frac{2k+1}{2}\pi\,,\; \alpha_{31}^{}=k\pi\\
	&&({\rm iii})\; \delta=\sigma=0\,,\;\rho=\pi/2 \iff \alpha_{12}^{}, \; \alpha_{31}^{}=\frac{2k+1}{2}\pi\,,\; \alpha_{23}^{}=k\pi\\
	&&({\rm iv})\; \delta=0\,,\;\rho=\sigma=\pi/2 \iff \alpha_{23}^{}, \; \alpha_{31}^{}=\frac{2k+1}{2}\pi\,,\; \alpha_{12}^{}=k\pi
\end{eqnarray*}
where $k$ is an arbitrary integer. It is easy to check that in any of the four scenarios (i)-(iv), CP is conserved in the leptonic sector up to ${\cal O}\left(1/\Lambda\right)$. This is because in the standard parametrization in Eq.~(\ref{eq:standard para}), any CP-violating observable is proportional to $\sin\left(l\delta+2m\rho+2n\sigma\right)$ with $l,m,n$ being arbitrary integers and would vanish in any of the scenarios (i)-(iv). Particularly, the Jarlskog-like invariant ${\cal J}_{360}^{}$ in Eq.~(\ref{eq:j360}) also vanishes. When the dimension-six operator ${\cal O}^{}_6$ is included, three additional phases $\beta_{ij}^{}$ appear. It is evident that in the basis where $C_5^{}$ is real and diagonal, $\alpha_{ij}^{}$ and $\beta_{ij}^{}$ play the parallel role in describing CP violation. This inspires us to construct the following three CP-odd invariants
\begin{eqnarray}
	{\cal J}_{121}^{}&\equiv&{\rm Im}\,{\rm Tr}\left(X_l^{}X_5^{}C_6^{}\right)\;,\\
	{\cal J}_{141}^{}&\equiv&{\rm Im}\,{\rm Tr}\left(X_l^{}X_5^{2}C_6^{}\right)\;,\\
	{\cal J}_{161}^{}&\equiv&{\rm Im}\,{\rm Tr}\left(X_5^{}X_l^{}X_5^{2}C_6^{}\right)\;,
\end{eqnarray}
and the vanishing of them gives three homogeneous linear equations of $\sin\left(\alpha_{ij}^{}-\beta_{ij}^{}\right)$, namely,
\begin{eqnarray}
	\label{eq:j121}
	{\cal J}_{121}^{}&=&-a_{12}^{}b_{12}^{}\left(c_1^2-c_2^2\right)\sin\left(\alpha_{12}^{}-\beta_{12}^{}\right)-a_{23}^{}b_{23}^{}\left(c_2^2-c_3^2\right)\sin\left(\alpha_{23}^{}-\beta_{23}^{}\right)\nonumber\\
	&&-a_{31}^{}b_{31}^{}\left(c_3^2-c_1^2\right)\sin\left(\alpha_{31}^{}-\beta_{31}^{}\right)=0\;,\\
	\label{eq:j141}
	{\cal J}_{141}^{}&=&-a_{12}^{}b_{12}^{}\left(c_1^4-c_2^4\right)\sin\left(\alpha_{12}^{}-\beta_{12}^{}\right)-a_{23}^{}b_{23}^{}\left(c_2^4-c_3^4\right)\sin\left(\alpha_{23}^{}-\beta_{23}^{}\right)\nonumber\\
	&&-a_{31}^{}b_{31}^{}\left(c_3^4-c_1^4\right)\sin\left(\alpha_{31}^{}-\beta_{31}^{}\right)=0\;,\\
	\label{eq:j161}
	{\cal J}_{161}^{}&=&-a_{12}^{}b_{12}^{}c_1^2c_2^2\left(c_1^2-c_2^2\right)\sin\left(\alpha_{12}^{}-\beta_{12}^{}\right)-a_{23}^{}b_{23}^{}c_2^2c_3^2\left(c_2^2-c_3^2\right)\sin\left(\alpha_{23}^{}-\beta_{23}^{}\right)\nonumber\\
	&&-c_3^2c_1^2a_{31}^{}b_{31}^{}\left(c_3^2-c_1^2\right)\sin\left(\alpha_{31}^{}-\beta_{31}^{}\right)=0\;.
\end{eqnarray}
The determinant of the coefficient matrix in Eqs.~(\ref{eq:j121})-(\ref{eq:j161}) is simply
$$a_{12}^{}a_{23}^{}a_{31}^{}b_{12}^{}b_{23}^{}b_{31}^{}\left(c_1^2-c_2^2\right)_{}^2\left(c_2^2-c_3^2\right)_{}^2\left(c_3^2-c_1^2\right)_{}^2\neq 0\;,$$
thus Eqs.~(\ref{eq:j121})-(\ref{eq:j161}) lead to
\begin{eqnarray}
	\label{eq:integer condition}
	\alpha_{ij}^{}-\beta_{ij}^{}=k\pi\;,\quad
	k\; {\rm integer}
\end{eqnarray}
for $ij=12,23,31$. One can verify that if the phases in the SEFT satisfy Eq.~(\ref{eq:integer condition}) along with the conditions in any one of four scenarios (i)-(iv), then all the CP-violating observables would vanish. Therefore, the vanishing of $\left\{{\cal J}_{360}^{},{\cal J}_{240}^{(2)},{\cal J}_{260}^{},{\cal J}_{121}^{},{\cal J}_{141}^{},{\cal J}_{161}^{}\right\}$ serves as the minimal sufficient and necessary conditions for leptonic CP conservation in the three-generation SEFT.

We close this section by giving some comments on the equivalence between the set of physical observables and the set of primary invariants. Strictly speaking, there are indeed discrete leftover degeneracies when extracting physical parameters by using primary invariants. This is also the origin of the counter example in the beginning of Sec.~\ref{subsec:conditions3g}: It is $\sin(2\alpha_{ij}^{})$ rather than $\sin\alpha_{ij}^{}$ that has been extracted from the primary invariants $\left\{{\cal J}_{240}^{(2)}, {\cal J}_{260}^{}, {\cal J}_{280}^{}\right\}$,
so ${\cal J}_{360}^{}$ cannot be expanded in terms of ${\cal J}_{240}^{(2)}$, ${\cal J}_{260}^{}$, ${\cal J}_{280}^{}$, and ${\cal J}_{240}^{(2)}{\cal J}_{260}^{}{\cal J}_{280}^{}$ when $\alpha_{ij}=\pi/2^{}$. Therefore, it is more strict to state that the set of physical parameters is equivalent to the set of primary invariants in the ring \emph{up to some discrete degeneracies}. The existence of such degeneracies can be ascribed to the complicated structure of the ring: There are high-degree primary invariants composed of high powers of building blocks, implying that they are functions of high multiples of CP-violating phases. The degeneracies can be eliminated by including more basic invariants other than the primary ones. For example, in Sec.~\ref{subsec:extract2g}, the 10 primary invariants can only determine the values of $\cos2\alpha$ and $\cos2\beta$, which remain unchanged under $\alpha\to -\alpha$ and $\beta\to -\beta$. Then one can introduce two more CP-odd invariants ${\cal I}_{240}^{}\propto \sin2\alpha$ and ${\cal I}_{042}^{}\propto \sin2\beta$, whose signs can be used to eliminate the $Z_2^{}$ degeneracy. However, one should keep in mind that ${\cal I}_{240}^{}$ and ${\cal I}_{042}$ are by no means algebraically independent of the primary invariants because their squares can be decomposed into the polynomials of the primary ones due to the syzygies at degree 12.

\section{CP-violating observables and flavor invariants}
\label{sec:observables}
In the last two sections we have shown that all the physical parameters can be extracted from the primary flavor invariants. Therefore, any physical observable can be completely written as the function of a finite number of invariants, which is explicitly independent of the parametrization schemes and the flavor basis that has been chosen. In particular, for any CP-violating processes, one will be able to define a working observable ${\cal A}_{\rm CP}^{}$ that changes its sign under the CP-conjugate transformation and can be expressed as\footnote{Here it is not claimed that any CP-violating observable can be written into the form of Eq.~(\ref{eq:observable decomposition}). But for any CP-violating processes, there exists a working observable that measures the CP asymmetry for the processes and can be written into the form of Eq.~(\ref{eq:observable decomposition}).}
\begin{eqnarray}
	\label{eq:observable decomposition}
	{\cal A}_{\rm CP}^{}=\sum_{j=1}^{j_{\rm max}^{}}{\cal F}_j^{}\left[{\cal I}_k^{\rm even}\right]{\cal I}_j^{\rm odd}\;,
\end{eqnarray}
where ${\cal I}_{j}^{\rm odd}$ and ${\cal F}_j^{}\left[{\cal I}_k^{\rm even}\right]$ (for $j=1,2,...,j_{\rm max}^{}$) are respectively CP-odd basic flavor invariants and functions of only CP-even basic flavor invariants,\footnote{If all the primary invariants happen to be CP-even, then ${\cal I}_k^{\rm even}$ can be restricted to be only primary invariants.} with $j_{\rm max}^{}$ being some positive integer. 

The statement in Eq.~(\ref{eq:observable decomposition}) can be proved as follows. Suppose that there are totally $n$ independent phases coming from the complex couplings in the theory denoted by $\left\{\alpha_1^{},...,\alpha_n^{}\right\}$. First, noticing that both ${\cal A}_{\rm CP}^{}$ and CP-odd flavor invariants change the sign under the CP-conjugate transformation while CP-even flavor invariants remain unchanged, we can generally write ${\cal A}_{\rm CP}^{}$ as
\begin{eqnarray}
	{\cal A}_{\rm CP}^{}=\sum_{j}^{}{\cal G}_j^{}\left(\vec{y}\right)\cos\left(\sum_{i=1}^{n}z_j^i\alpha_i\right)\sin\left(\sum_{i'=1}^{n}\tilde{z}_j^{i'}\alpha_{i'}^{}\right)\;,
\end{eqnarray}
where $z_j^i$ and $\tilde{z}_j^{i'}$ are integers and ${\cal G}_j^{}\left(\vec{y}\right)$ are functions of a set of parameters, denoted by $\vec{y}$, other than the phases. A key observation is that one can convert all the trigonometric functions into the rational functions of $x_i^{}$ by setting $x_i^{}\equiv \tan\left(\alpha_i^{}/2\right)$ (for $i=1,2,...,n$):
\begin{eqnarray}
	\label{eq:tangent half-angle substitution}
	&&\sin\alpha_i^{}=\frac{2x_i}{1+x_i^2}\;,\quad
	\cos\alpha_i^{}=\frac{1-x_i^2}{1+x_i^2}\;,\quad\nonumber\\ &&\sin\left(\alpha_i^{}+\alpha_j^{}\right)=\frac{2x_i\left(1-x_j^2\right)+2x_j\left(1-x_i^2\right)}{\left(1+x_i^2\right)\left(1+x_j^2\right)}\;,\quad
	\cos\left(\alpha_i^{}+\alpha_j^{}\right)=\frac{\left(1-x_i^2\right)\left(1-x_j^2\right)-4x_ix_j}{\left(1+x_i^2\right)\left(1+x_j^2\right)}\;,
	\nonumber\\
	&&\sin\left(\alpha_i^{}+\alpha_j^{}+\alpha_k^{}\right)\nonumber\\
	&&=\frac{2x_i\left(1-x_j^2\right)\left(1-x_k^2\right)+2x_j\left(1-x_k^2\right)\left(1-x_i^2\right)+2x_k\left(1-x_i^2\right)\left(1-x_j^2\right)-8x_ix_jx_k}{\left(1+x_i^2\right)\left(1+x_j^2\right)\left(1+x_k^2\right)}\;,
\end{eqnarray}
and so on. Using Eq.~(\ref{eq:tangent half-angle substitution}), ${\cal A}_{\rm CP}^{}$ can be rewritten as\footnote{This is because under the CP-conjugate transformation, only the odd powers of $x_i^{}$ change signs, but both $\vec{y}$ and even powers of $x_i^{}$ remain unchanged.}
\begin{eqnarray}
	\label{eq:observable decomposition2}
	{\cal A}_{\rm CP}^{}&=&\sum_{i=1}^{n}{\cal G}_i^{}\left(\vec{y}\right){\cal R}_i^{}\left(x_1^2,...,x_n^2\right)x_i^{}+\sum_{1=i_1<i_2<i_3}^{n}{\cal G}_{i_1i_2i_3}^{}\left(\vec{y}\right){\cal R}_{i_1i_2i_3}\left(x_1^2,...,x_n^2\right)x_{i_1}^{}x_{i_2^{}}x_{i_3^{}}\nonumber\\
	&&+\cdots+\sum_{1=i_1<\cdots<i_{\rm max}}^{n}{\cal G}_{i_1\cdots i_{\rm max}}^{}\left(\vec{y}\right){\cal R}_{i_1\cdots i_{\rm max}}\left(x_1^2,...,x_n^2\right)x_{i_1}^{}\cdots x_{i_{\rm max}^{}}\nonumber\\
	&=&\sum_{i=1}^{n}{\cal H}_{i}^{}\left[{\cal I}_k^{\rm even}\right]x_i^{}+\sum_{1=i_1<i_2<i_3}^{n}{\cal H}_{i_1i_2i_3}^{}\left[{\cal I}_k^{\rm even}\right]x_{i_1}^{}x_{i_2}^{}x_{i_3}\nonumber\\
	&&+\cdots+\sum_{1=i_1<\cdots<i_{\rm max}}^{n}{\cal H}_{i_1\cdots i_{\rm max}}^{}\left[{\cal I}_k^{\rm even}\right]x_{i_1}^{}\cdots x_{i_{\rm max}^{}}\;,
\end{eqnarray}
where $i_{\rm max}^{}=n$ (or $n-1$) if $n$ is odd (or even), ${\cal G}$ are functions of $\vec{y}$, ${\cal R}$ are rational functions of $\left\{x_1^2,\cdots,x_n^2\right\}$ and ${\cal H}\left[{\cal I}_k^{\rm even}\right]$ are functions of CP-even basic invariants. The second equality in Eq.~(\ref{eq:observable decomposition2}) has been derived by noticing the fact that $\vec{y}$, the physical parameters other than phases, as well as $\left\{x_1^2,\cdots,x_n^2\right\}$, can all be extracted from only CP-even basic invariants. Note that in Eq.~(\ref{eq:observable decomposition2}) there are totally
\begin{eqnarray*}
	\left(
	\begin{matrix}
		n\\
		1
	\end{matrix}
	\right)+\left(
	\begin{matrix}
		n\\
		3
	\end{matrix}
	\right)+\left(
	\begin{matrix}
		n\\
		5
	\end{matrix}
	\right)+\cdots+
	\left(
	\begin{matrix}
		n\\
		i_{\rm max}^{}
	\end{matrix}
	\right)=2_{}^{n-1}
\end{eqnarray*}
linearly-independent odd-power monomials of $\left\{x_1^{},\cdots,x_n^{}\right\}$. Therefore, Eq.~(\ref{eq:observable decomposition2}) can be written as the linear combination of the monomials
\begin{eqnarray}
	\label{eq:observable decomposition3}
	{\cal A}_{\rm CP}^{}=\sum_{i=1}^{2^{n-1}}{\cal H}_i^{}\left[{\cal I}_k^{\rm even}\right]{\cal M}_i^{}\;,
\end{eqnarray}
where ${\cal M}_i^{}$ (for $i=1,2,...,2_{}^{n-1}$) range over all the odd-power monomials of $\left\{x_1^{},\cdots,x_n^{}\right\}$.

Since CP-odd flavor invariants change their signs under the CP-conjugate transformation as ${\cal A}_{\rm CP}^{}$ does, they can also be decomposed into the form of Eq.~(\ref{eq:observable decomposition3}). In order to \emph{linearly} extract all the ${\cal M}_i^{}$, one must utilize $2_{}^{n-1}$ linearly-independent CP-odd flavor invariants. If the total number of the CP-odd basic invariants in the ring is no smaller than $2^{n-1}_{}$, then it is possible to choose ${\cal I}_j^{\rm odd}$ (for $j=1,2,...,2_{}^{n-1}$) to linearly extract all the odd-power monomials
\begin{eqnarray}
	\label{eq:linear combination}
	{\cal M}_i^{}=\sum_{j=1}^{2^{n-1}}\widetilde{\cal H}_{ij}^{}\left[{\cal I}_k^{\rm even}\right]{\cal I}_j^{\rm odd}\;,\qquad
	i=1,2,...,2^{n-1}_{}\;,
\end{eqnarray}
where $\widetilde{\cal H}_{ij}^{}\left[{\cal I}_k^{\rm even}\right]$ are functions of only CP-even basic invariants. Substituting Eq.~(\ref{eq:linear combination}) back into Eq.~(\ref{eq:observable decomposition3}) one obtains
\begin{eqnarray}
	{\cal A}_{\rm CP}^{}=\sum_{i=1}^{2^{n-1}}\sum_{j=1}^{2^{n-1}}{\cal H}_i^{}\left[{\cal I}_k^{\rm even}\right]\widetilde{\cal H}_{ij}^{}\left[{\cal I}_k^{\rm even}\right]{\cal I}_j^{\rm odd}=\sum_{j=1}^{2^{n-1}}{\cal F}_j^{}\left[{\cal I}_k^{\rm even}\right]{\cal I}_j^{\rm odd}\;,
\end{eqnarray}
where ${\cal F}_j^{}\left[{\cal I}_k^{\rm even}\right]\equiv \sum_{i=1}^{2^{n-1}}{\cal H}_i^{}\left[{\cal I}_k^{\rm even}\right]\widetilde{\cal H}_{ij}^{}\left[{\cal I}_k^{\rm even}\right]$. This completes the proof of Eq.~(\ref{eq:observable decomposition}) with $j_{\rm max}^{}=2_{}^{n-1}$. However, if the total number of the CP-odd basic invariants in the ring is smaller than $2_{}^{n-1}$, then it is in general \emph{impossible} to express an arbitrary CP-violating observable as the \emph{linear} combination of CP-odd basic flavor invariants as in Eq.~(\ref{eq:observable decomposition}).

It should be noted that the number of the CP-odd basic invariants to linearly expand any CP-violating observable needs not to match the minimal number of CP-odd invariants that one requires to be vanishing to guarantee CP conservation in the theory. For example, for a theory with $n=6$ independent phases (such as the SEFT for the three-generation case), one needs $2^{6-1}=32$ linearly-independent CP-odd invariants to linearly expand any CP-violating observable in the most general case. However, as we have shown in Sec.~\ref{subsec:conditions3g}, the vanishing of only 6 CP-odd invariants is sufficient to guarantee the absence of CP violation, which is equivalent to the vanishing of all CP-violating observables. The point is that the vanishing of \emph{some} CP-odd flavor invariants can reduce the number of independent phases. In our case, there are 6 independent phases $\left\{\alpha_{12}^{}, \alpha_{23}^{}, \alpha_{31}^{}, \beta_{12}^{}, \beta_{23}^{}, \beta_{31}^{}\right\}$ at the beginning. The vanishing of $\left\{{\cal J}_{121}^{}, {\cal J}_{141}^{}, {\cal J}_{161}^{}\right\}$ in Eqs.~(\ref{eq:j121})-(\ref{eq:j161}) enforces $\alpha_{ij}^{}=\beta_{ij}^{}+k\pi$ and thus reduces the number of independent phases to 3. In addition, the vanishing of ${\cal J}_{360}^{}$ in Eq.~(\ref{eq:j360}) leads to $\alpha_{12}^{}+\alpha_{23}^{}+\alpha_{31}^{}=k\pi$ and further eliminates one more phase. Therefore, if ${\cal J}_{121}^{}={\cal J}_{141}^{}={\cal J}_{161}^{}={\cal J}_{360}^{}=0$ is satisfied, there remain only 2 independent phases in the theory. Under this condition, any nonzero CP-violating observable can be written as the linear combination of ${\cal J}_{240}^{(2)}$ and ${\cal J}_{260}^{}$ and the further vanishing of ${\cal J}_{240}^{(2)}$ and ${\cal J}_{260}^{}$ will lead to the vanishing of all CP-violating observables in the theory. This explains from another point of view why the vanishing of $2^{6-1-4}_{}+4=6$ CP-odd flavor invariants can serve as the minimal sufficient and necessary conditions for leptonic CP conservation in the SEFT for the three-generation case.

Below we will discuss some concrete CP-violating processes and explain how to write the CP-violating observables into the form of Eq.~(\ref{eq:observable decomposition}).

\subsection{Neutrino-neutrino oscillations}
After the SM gauge symmetry is spontaneously broken, the Wilson coefficient matrix $C_5^{}$ gives the Majorana mass term of light neutrinos while the Wilson coefficient matrix $C_6^{}$ contributes to the unitarity violation of the flavor mixing matrix in the leptonic sector. The effective Lagrangian governing the lepton mass spectra, flavor mixing, and charged-current interaction together with the kinetic term of neutrinos after the gauge symmetry breaking reads
\begin{eqnarray}
	\label{eq:Lagrangian eff after SSB}
	{\cal L}_{\rm eff}^{}=\overline{\nu_{\rm L}^{}}\,{\rm i}\slashed{\partial}\,{\cal K}\nu_{\rm L}^{}- \left[\frac{1}{2} \overline{\nu_{\rm L}^{}}M_\nu^{}\nu_{\rm L}^{\rm C} + \overline{l_{\rm L}^{}}M_l^{}l_{\rm R}^{} - \frac{g}{\sqrt{2}}\overline{l_{\rm L}^{}}\gamma_{}^\mu\nu_{\rm L}^{}W_\mu^{-}+{\rm h.c.} \right]\;,
\end{eqnarray}
where ${\cal K}=1+v_{}^2 C_6^{}/\left(2\Lambda_{}^2\right)$, $M_\nu^{}=v_{}^2 C_5^{}/\left(2\Lambda\right)$, $M_l^{}=vY_l^{}/\sqrt{2}$ and $g$ is the gauge coupling constant of ${\rm SU}(2)_{\rm L}^{}$ group. Recalling that we work in the basis where $C_5^{}$ is real and diagonal, thus $M_\nu^{}=\widehat{M}_\nu^{}={\rm Diag}\left\{m_1^{},m_2^{},m_3^{}\right\}$. In order to normalize the kinetic term of neutrinos, one should rescale the neutrino fields as
$\nu_{\rm L}\to \nu_{\rm L}^{\prime}={\cal K}_{}^{1/2}\nu_{\rm L}^{}$, which will modify the neutrino mass matrix as 
\begin{eqnarray}
	M_\nu^{}\to M_\nu^\prime={\cal K}_{}^{-1/2}\widehat{M}_\nu^{}\left({\cal K}_{}^{\rm T}\right)_{}^{-1/2}=\widehat{M}_\nu^{}+\frac{v^2}{4\Lambda^2}\left(C_6^{}\widehat{M}_\nu^{}+\widehat{M}_\nu^{}C_6^{\rm T}\right)\;.
\end{eqnarray}
However, since $\widehat{M}_\nu^{}$ itself is of order ${\cal O}\left(1/\Lambda\right)$, the difference between $M_\nu^{}$ and $M_\nu^\prime$ is on the order of ${\cal O}\left(1/\Lambda_{}^3\right)$ and thus can be neglected to the order of ${\cal O}\left(1/\Lambda^2_{}\right)$. Hence we have $M_\nu^\prime=\widehat{M}_\nu^{}$. The next step is to diagonalize the charged-lepton mass matrix via $V M_l^{}V_{}^{\prime \dagger}=\widehat{M}_l^{}={\rm Diag}\left\{m_e^{},m_\mu^{},m_\tau^{}\right\}$ with $V$ and $V_{}^\prime$ being unitary matrices and change the basis of charged-lepton fields $l_{\rm L}^{}\to l_{\rm L}^\prime=Vl_{\rm L}^{}$,  $l_{\rm R}^{}\to l_{\rm R}^\prime=V_{}^{\prime}l_{\rm R}^{}$, then the Lagrangian in the mass basis is given by
\begin{eqnarray}
	\label{eq:Lagrangian eff after SSB2}
	{\cal L}_{\rm eff}^{}=\overline{\nu_{\rm L}^{\prime}}\,{\rm i}\slashed{\partial}\nu_{\rm L}^{\prime} - \left[ \frac{1}{2}\overline{\nu_{\rm L}^{\prime}}\widehat{M}_\nu^{}\nu_{\rm L}^{\prime \rm C} + \overline{l_{\rm L}^{\prime}}\widehat{M}_l^{}l_{\rm R}^{\prime} - \frac{g}{\sqrt{2}}\overline{l_{\rm L}^{\prime}}\gamma_{}^\mu V {\cal K}_{}^{-1/2}\nu_{\rm L}^{\prime}W_\mu^{-}+{\rm h.c.} \right]\;.
\end{eqnarray}
From Eq.~(\ref{eq:Lagrangian eff after SSB2}) we obtain the non-unitary flavor mixing matrix
\begin{eqnarray}
	V_{\rm eff}^{}=V {\cal K}_{}^{-1/2}=V\left(1-\frac{v^2}{4\Lambda^2}C_6^{}\right)\;,
\end{eqnarray}
which violates the unitarity to the order of ${\cal O}\left(1/\Lambda_{}^2\right)$. The unitarity violation will contribute to the CP asymmetries in neutrino-neutrino oscillations
\begin{eqnarray}
	\label{eq:CP asymmetry in neutrino oscillation def}
	{\cal A}_{\nu\nu}^{\alpha\beta}\equiv \frac{{\rm P}\left(\nu_\alpha\to\nu_\beta\right)-{\rm P}\left(\bar{\nu}_\alpha\to\bar{\nu}_\beta\right)}{{\rm P}\left(\nu_\alpha\to\nu_\beta\right)+{\rm P}\left(\bar{\nu}_\alpha\to\bar{\nu}_\beta\right)}\;,
\end{eqnarray}
where ${\rm P}(\nu_\alpha^{}\to\nu_\beta^{})$ denotes the transition probability from $\nu_\alpha^{}$ to $\nu_\beta^{}$ while ${\rm P}(\bar{\nu}_\alpha^{}\to\bar{\nu}_\beta^{})$ denotes that of its CP-conjugate process. The CP asymmetries in Eq.~(\ref{eq:CP asymmetry in neutrino oscillation def}) are found to be~\cite{Fernandez-Martinez:2007iaa, Xing:2007zj, Goswami:2008mi, Antusch:2009pm}
\begin{eqnarray}
	\label{eq:CP asymmetry in neutrino oscillation}
	{\cal A}_{\nu\nu}^{\alpha\beta}=\frac{2\sum_{i<j}{\rm Im}\left(Q_{\alpha\beta}^{ij}\right)\sin2\Delta_{ji}^{}}{\delta_{\alpha\beta}-4\sum_{i<j}{\rm Re}\left(Q_{\alpha\beta}^{ij}\right)\sin^2\Delta_{ji}}\;,
\end{eqnarray}
where $Q_{\alpha\beta}^{ij}\equiv \left(V_{\rm eff}^{}\right)_{\alpha i}^{}\left(V_{\rm eff}^{}\right)_{\beta j}^{}\left(V_{\rm eff}^{}\right)_{\alpha j}^{*}\left(V_{\rm eff}^{}\right)_{\beta i}^{*}$ and $\Delta_{ji}^{}\equiv \left(m_j^2-m_i^2\right)L/(4E)$ have been defined with $L$ and $E$ being respectively the propagation distance and neutrino beam energy. It is clear that Eq.~(\ref{eq:CP asymmetry in neutrino oscillation}) is only nonvanishing for $\alpha\neq \beta$, as a consequence of the CPT theorem. Particularly for the two-generation case, ${\cal A}_{\nu\nu}^{\alpha\beta}$ is nonzero for $\alpha\neq \beta$. This is contrary to the result in the unitary limit (i.e., $\Lambda\to \infty$), where CP violation is absent in neutrino oscillations with only two flavors. In fact, we have
\begin{eqnarray}
	\label{eq:CP asymmetry in neutrino oscillation2}
	A_{\nu\nu}^{\mu e}=-A_{\nu\nu}^{e\mu}=\frac{v^2}{\Lambda^2}\frac{b_{12}}{\sin2\theta}\cot\left(\Delta^{}_{21}\right)\sin\left(\beta+\phi\right)\;,
\end{eqnarray}
which is nonvanishing though suppressed by $v_{}^2/\Lambda_{}^2$. Note that we have used the parametrization of $C_6^{}$ and $V$ in Eqs.~(\ref{eq:parametrization of C6 2g}) and (\ref{eq:parametrization of V 2g}). It is then interesting to rewrite the result in Eq.~(\ref{eq:CP asymmetry in neutrino oscillation2}) into a complete form of flavor invariants, which is independent of the parametrization and flavor basis. This can be achieved by using Eqs.~(\ref{eq:extract C6 2g 2})-(\ref{eq:extract neutrino mass 2g}), Eqs.~(\ref{eq:extract chargd-lepton mass 2g})-(\ref{eq:extract theta}) and recalling Eq.~(\ref{eq:i1212}). Finally one arrives at
\begin{eqnarray}
	\label{eq:neutrino oscillation}
	A_{\nu\nu}^{e\mu}=-\frac{v^2}{\Lambda^2}\cot\left(\Delta^{}_{21}\right){\cal F}_{\nu\nu}^{e\mu}\left[{\cal I}_{100}^{},{\cal I}_{200}^{},{\cal I}_{020}^{},{\cal I}_{120}^{},{\cal I}_{040}^{}\right]\,{\cal I}_{121}^{(2)}\;,
\end{eqnarray}
where 
\begin{eqnarray}
	{\cal F}_{\nu\nu}^{e\mu}\left[{\cal I}_{100}^{},{\cal I}_{200}^{},{\cal I}_{020}^{},{\cal I}_{120}^{},{\cal I}_{040}^{}\right]=\frac{\left(2{\cal I}_{040}-{\cal I}_{020}^2\right)^{1/2}\left(2{\cal I}_{200}-{\cal I}_{100}^2\right)^{1/2}}{{\cal I}_{040}\left(2{\cal I}_{200}-{\cal I}_{100}^2\right)-2{\cal I}_{120}^{}\left({\cal I}_{120}-{\cal I}_{020}{\cal I}_{100}\right)-{\cal I}_{020}^2{\cal I}_{200}}\;,
\end{eqnarray}
and
\begin{eqnarray}
	\Delta_{21}^{}= \frac{L}{4E}\left(m_2^2-m_1^2\right)=\frac{Lv^4}{16E\Lambda^2}\left(2{\cal I}_{040}^{}-{\cal I}_{020}^2\right)_{}^{1/2}\;.
\end{eqnarray}
Thus we have successfully recast the CP-asymmetries in two-flavor neutrino oscillations into the form of Eq.~(\ref{eq:observable decomposition}), which is linearly proportional to the unique CP-odd flavor invariant ${\cal I}_{121}^{(2)}$ with the coefficient being function of only  CP-even primary flavor invariants.
\subsection{Neutrino-antineutrino oscillations}
Next we take the example of neutrino-antineutrino oscillations. The CP-asymmetries are defined by
\begin{eqnarray}
	\label{eq:CP asymmetry in neutrino-antineutrino oscillation def}
	{\cal A}_{\nu\bar{\nu}}^{\alpha\beta}\equiv \frac{{\rm P}\left(\nu_\alpha\to \bar{\nu}_\beta\right)-{\rm P}\left(\bar{\nu}_\alpha \to \nu_\beta\right)}{{\rm P}\left(\nu_\alpha\to \bar{\nu}_{\beta}\right)+{\rm P}\left(\bar{\nu}_\alpha\to \nu_\beta\right)}\;,
\end{eqnarray}
and can be calculated immediately~\cite{Xing:2013ty, Xing:2013woa, Wang:2021rsi}
\begin{eqnarray}
	{\cal A}_{\nu\bar{\nu}}^{\alpha\beta}=\frac{2\sum_{i<j}m_im_j\,{\rm Im}\left(\widetilde{Q}_{\alpha\beta}^{ij}\right)\sin2\Delta_{ji}^{}}{\left|\langle m \rangle_{\alpha\beta}\right|^2-4\sum_{i<j}m_im_j\,{\rm Re}\left(\widetilde{Q}_{\alpha\beta}^{ij}\right)\sin^2\Delta_{ji}^{}}\;,
\end{eqnarray}
where $\widetilde Q_{\alpha\beta}^{ij}\equiv \left(V_{\rm eff}^{}\right)_{\alpha i}^{}\left(V_{\rm eff}^{}\right)_{\beta i}^{}\left(V_{\rm eff}^{}\right)_{\alpha j}^{*}\left(V_{\rm eff}^{}\right)_{\beta j}^{*}$ and $\langle m \rangle_{\alpha\beta}\equiv \sum_i^{} m_i^{} \left(V_{\rm eff}\right)_{\alpha i}^{}\left(V_{\rm eff}\right)_{\beta i}^{}$ have been defined. For illustration, we consider the two-generation case and take $\alpha\neq\beta$. We will express the corresponding CP asymmetries into the complete form of flavor invariants. It is easy to show that
\begin{eqnarray}
	{\cal A}_{\nu\bar{\nu}}^{e\mu}={\cal A}_{\nu\bar{\nu}}^{\mu e}=-\frac{2 m_1 m_2 \sin2\phi \sin 2\Delta_{21}}{m_1^2+m_2^2-2m_1 m_2 \cos2\phi\cos2\Delta_{21}}+{\cal O}\left(\frac{v^2}{\Lambda^2}\right)\;.
\end{eqnarray}
It should be noted that the parameters in $C_6^{}$ do not contribute at the leading order. Using Eqs.~(\ref{eq:extract neutrino mass 2g}) and (\ref{eq:extract chargd-lepton mass 2g})-(\ref{eq:extract phi}) and taking into account ${\cal I}_{240}^{}\equiv {\rm Im}\,{\rm Tr}\left(X_l^{}X_5^{}G_{l5}^{}\right)$, one finally gets at the leading order
\begin{eqnarray}
	\label{eq:neutrino-antineutrino oscillation}
	{\cal A}_{\nu\bar{\nu}}^{e\mu}={\cal F}_{\nu\bar{\nu}}^{e\mu}\left[{\cal I}_{100}^{},{\cal I}_{200}^{},{\cal I}_{020}^{},{\cal I}_{120}^{},{\cal I}_{220}^{},{\cal I}_{040}^{}\right]\,{\cal I}_{240}^{}\;,
\end{eqnarray}
where 
\begin{eqnarray}
	&&{\cal F}_{\nu\bar{\nu}}^{e\mu}\left[{\cal I}_{100}^{},{\cal I}_{200}^{},{\cal I}_{020}^{},{\cal I}_{120}^{},{\cal I}_{220}^{},{\cal I}_{040}^{}\right]\nonumber\\
	&=&4\left(2{\cal I}_{040}^{}-{\cal I}_{020}^2\right)_{}^{1/2}\sin2\Delta_{21}^{}
	\left\{{\cal I}_{020}^{}\left[2{\cal I}_{120}^{}\left({\cal I}_{100}^{}{\cal I}_{020}^{}-{\cal I}_{120}^{}\right)-{\cal I}_{040}^{}\left({\cal I}_{100}^2-2{\cal I}_{200}^{}\right)-{\cal I}_{020}^2{\cal I}_{200}^{}\right]\right.\nonumber\\
	&&\left.+\cos2\Delta_{21}^{}\left[{\cal I}_{020}^{}\left({\cal I}_{020}^2{\cal I}_{100}^2+2{\cal I}_{120}^2-{\cal I}_{040}^{}{\cal I}_{100}^2\right)+4{\cal I}_{040}^{}\left({\cal I}_{100}^{}{\cal I}_{120}^{}-{\cal I}_{220}^{}\right)+2{\cal I}_{020}^2\right.\right.\nonumber\\
	&&\left.\left.\times\left({\cal I}_{220}^{}-2{\cal I}_{100}^{}{\cal I}_{120}^{}\right)\right]\right\}^{-1}\;,
\end{eqnarray}
with 
\begin{eqnarray*}
	\Delta_{21}^{}=\frac{Lv^4}{16E\Lambda^2}\left(2{\cal I}_{040}^{}-{\cal I}_{020}^2\right)_{}^{1/2}\;.
\end{eqnarray*}
Thus we have also written the CP asymmetries in neutrino-antineutrino oscillations for the two-generation case into the form of Eq.~(\ref{eq:observable decomposition}). They are linearly proportional to the CP-odd flavor invariant ${\cal I}_{240}^{}$ and the coefficient is the function of only CP-even primary flavor invariants. Contrary to the case of neutrino-neutrino oscillations, ${\cal A}_{\nu\bar{\nu}}^{e\mu}$ is not suppressed by ${\cal O}\left(v_{}^2/\Lambda_{}^2\right)$. This is because the Majorana-type CP phase in the two-generation flavor mixing matrix already enters the CP asymmetries in neutrino-antineutrino oscillations.

As has been mentioned in Sec.~\ref{subsec:conditions2g}, any CP-violating observable in the two-generation SEFT can be written as the linear combination of ${\cal I}_{121}^{(2)}$ and ${\cal I}_{240}^{}$ into the form of Eq.~(\ref{eq:observable decomposition}) with $j_{\rm max}^{}=2$. This can be realized by noticing that one can linearly extract $\tan\left(\phi/2\right)$ and $\tan\left(\beta/2\right)$ using ${\cal I}_{121}^{(2)}$ and ${\cal I}_{240}^{}$ with the coefficients being functions of CP-even invariants, and that any CP-violating observable in the two-generation SEFT must be proportional to $\tan\left(\phi/2\right)$ or $\tan\left(\beta/2\right)$. Since ${\cal I}_{121}^{(2)}$ and ${\cal I}_{240}^{}$ are respectively responsible for the CP asymmetries in neutrino-neutrino and neutrino-antineutrino oscillations, we draw the conclusion that ${\cal A}_{\nu\nu}^{e\mu}$ and ${\cal A}_{\nu\bar{\nu}}^{e\mu}$ already contain all the information about leptonic CP violation in the two-generation SEFT.

\subsection{Observables in the three-generation case}
In this subsection we discuss how to express CP-violating observables in terms of flavor invariants in the SEFT for the three-generation case. As has been explained above, for a theory with 6 independent phases, there are $2_{}^{6-1}=32$ linearly-independent monomials and one needs at least $32$ CP-odd basic flavor invariants to linearly expand all possible CP-violating observables in the most general case. However, since we are working in the effective theory, any physical observable suppressed more than ${\cal O}\left(1/\Lambda_{}^2\right)$ should be neglected. So we do not need to consider the observables that proportional to monomials with the power of $\tilde{x}_{ij}^{}$ higher than one, such as $x_{12}^{}\tilde{x}_{23}^{}\tilde{x}_{31}^{}$, $\tilde{x}_{12}^{}\tilde{x}_{23}^{}\tilde{x}_{31}^{}$ and so on,
where we have defined $x_{ij}^{}\equiv\tan\left(\alpha_{ij}^{}/2\right)$ and $\tilde{x}_{ij}^{}\equiv\tan\left(\beta_{ij}^{}/2\right)$.\footnote{A brief comment on the power counting of ${\cal O}(1/\Lambda)$ is helpful. One may wonder whether neutrino masses of ${\cal O}(v^2/\Lambda)$ could appear in the denominator and thus break the rule of power counting. For any CP-violating process, it is always possible to define a dimensionless working CP-violating observable, in which the power of $v^2_{}/\Lambda^{}_{}$ in neutrino mass matrix $M_\nu^{}$ in the numerator and that in the denominator cancel with each other [e.g., the definition of CP asymmetries in Eqs. (\ref{eq:CP asymmetry in neutrino oscillation def}) and (\ref{eq:CP asymmetry in neutrino-antineutrino oscillation def})]. This is because $M_\nu^{}$, in the mass basis that we have chosen, contains neutrino mass eigenvalues and does not account for CP violation, so its overall scale $v^2_{}/\Lambda$ can always be factorized out in describing CP-violating process. Therefore, the neutrino mass matrix cannot affect the power counting in CP-violating observables.} Therefore, we are left with only 16 possible monomials to the order of ${\cal O}\left(1/\Lambda_{}^2\right)$:
\begin{itemize}
	\item 4 monomials not suppressed: $x_{12}^{}$, $x_{23}^{}$, $x_{31}^{}$, $x_{12}^{}x_{23}^{}x_{31}^{}$\;;
	\item 12 monomials suppressed by ${\cal O}\left(1/\Lambda_{}^2\right)$: $\tilde{x}_{12}^{}$, $\tilde{x}_{23}^{}$, $\tilde{x}_{31}^{}$, $x_{12}^{}x_{23}^{}\tilde{x}_{12}^{}$, $x_{12}^{}x_{23}^{}\tilde{x}_{23}^{}$, $x_{12}^{}x_{23}^{}\tilde{x}_{31}^{}$,\\
	$x_{23}^{}x_{31}^{}\tilde{x}_{12}^{}$, $x_{23}^{}x_{31}^{}\tilde{x}_{23}^{}$, $x_{23}^{}x_{31}^{}\tilde{x}_{31}^{}$,
	$x_{31}^{}x_{12}^{}\tilde{x}_{12}^{}$, $x_{31}^{}x_{12}^{}\tilde{x}_{23}^{}$, $x_{31}^{}x_{12}^{}\tilde{x}_{31}^{}$\;.
\end{itemize}
Then we will demonstrate that they can indeed be linearly extracted using 16 CP-odd basic flavor invariants in the SEFT.

First, the 4 monomials not suppressed can be linearly extracted using 4 invariants not containing $C_6^{}$: $x_{ij}^{}$ (for $ij=12,23,31$) can be extracted using ${\cal J}_{240}^{(2)}$, ${\cal J}_{260}^{}$, ${\cal J}_{280}^{}$ which only involve $\sin2\alpha_{ij}^{}$ while $x_{12}^{}x_{23}^{}x_{31}^{}$ can be extracted using ${\cal J}_{360}^{}$ in which $\sin\left(\alpha_{12}^{}+\alpha_{23}^{}+\alpha_{31}^{}\right)$ is involved. Similarly, $\tilde{x}_{ij}^{}$ can be extracted using invariants only involving $\sin2\beta_{ij}^{}$, namely, ${\cal J}_{042}^{(2)}$, ${\cal J}_{062}^{}$ and ${\cal J}_{082}^{}$. Then the 3 cyclic monomials $x_{12}^{}x_{23}^{}\tilde{x}_{31}^{}$, $x_{23}^{}x_{31}^{}\tilde{x}_{12}^{}$ and $x_{31}^{}x_{12}^{}\tilde{x}_{23}^{}$ can be determined using the following 3 invariants involving $\sin\left(\alpha_{12}^{}+\alpha_{23}^{}+\beta_{31}^{}\right)$, $\sin\left(\alpha_{23}^{}+\alpha_{31}^{}+\beta_{12}^{}\right)$ and $\sin\left(\alpha_{31}^{}+\alpha_{12}^{}+\beta_{23}^{}\right)$:
\begin{eqnarray}
	{\cal J}_{221}^{}&\equiv& {\rm Im}\,{\rm Tr}\left(X_l^2X_5^{}C_6^{}\right)\;,\\
	{\cal J}_{241}^{}&\equiv& {\rm Im}\,{\rm Tr}\left(X_l^2X_5^{2}C_6^{}\right)\;,\\
	{\cal J}_{261}^{}&\equiv& {\rm Im}\,{\rm Tr}\left(X_l^2X_5^{2}C_6^{}X_5^{}\right)\;.
\end{eqnarray}
Finally, the remaining 6 non-cyclic monomials $x_{12}^{}x_{23}^{}\tilde{x}_{12}^{}$, $x_{12}^{}x_{23}^{}\tilde{x}_{23}^{}$,  $x_{23}^{}x_{31}^{}\tilde{x}_{23}^{}$, $x_{23}^{}x_{31}^{}\tilde{x}_{31}^{}$,
$x_{31}^{}x_{12}^{}\tilde{x}_{12}^{}$, and  $x_{31}^{}x_{12}^{}\tilde{x}_{31}^{}$
can be determined by using the following 6 invariants, where six sine functions of different phase combinations, i.e., $\sin\left(\alpha_{12}^{}+2\alpha_{23}^{}\pm\beta_{12}^{}\right)$, $\sin\left(2\alpha_{12}^{}+\alpha_{23}^{}\pm\beta_{23}^{}\right)$,
$\sin\left(\alpha_{23}^{}+2\alpha_{31}^{}\pm\beta_{23}^{}\right)$,
$\sin\left(2\alpha_{23}^{}+\alpha_{31}^{}\pm\beta_{31}^{}\right)$,
$\sin\left(2\alpha_{31}^{}+\alpha_{12}^{}\pm\beta_{12}^{}\right)$ and
$\sin\left(\alpha_{31}^{}+2\alpha_{12}^{}\pm\beta_{31}^{}\right)$, are present:
\begin{eqnarray}
	{\cal J}_{321}^{(1)}&\equiv&{\rm Im}\,{\rm Tr}\left(X_l^{}C_6^{}G_{l5}^{(2)}\right)\;,\\
	{\cal J}_{341}^{(1)}&\equiv&{\rm Im}\,{\rm Tr}\left(X_5^{}X_l^{}C_6^{}G_{l5}^{(2)}\right)\;,\\
	{\cal J}_{361}^{(1)}&\equiv&{\rm Im}\,{\rm Tr}\left(X_5^{2}X_l^{}C_6^{}G_{l5}^{(2)}\right)\;,\\
	{\cal J}_{321}^{(2)}&\equiv&{\rm Im}\,{\rm Tr}\left(X_l^2C_6^{}G_{l5}^{}\right)\;,\\
	{\cal J}_{341}^{(2)}&\equiv&{\rm Im}\,{\rm Tr}\left(X_5^{}X_l^2C_6^{}G_{l5}^{}\right)\;,\\
	{\cal J}_{361}^{(2)}&\equiv&{\rm Im}\,{\rm Tr}\left(X_5^2X_l^2C_6^{}G_{l5}^{}\right)\;,
\end{eqnarray}
where we have defined $G_{l5}^{(2)}\equiv C_5^{}\left(X_l^*\right)_{}^{2}C_5^\dagger$. 

To sum up, all the possible 16 monomials can be linearly extracted from the following 16 CP-odd basic flavor invariants 
$$\left\{{\cal J}_{360}^{},{\cal J}_{240}^{(2)},{\cal J}_{260}^{},{\cal J}_{280}^{},{\cal J}_{042}^{(2)},{\cal J}_{062}^{},{\cal J}_{082}^{},{\cal J}_{221}^{},{\cal J}_{241}^{},{\cal J}_{261}^{},{\cal J}_{321}^{(1)},{\cal J}_{341}^{(1)},{\cal J}_{361}^{(1)},{\cal J}_{321}^{(2)},{\cal J}_{341}^{(2)},{\cal J}_{361}^{(2)}
\right\}$$
and thus any CP-violating observable in the SEFT for three-generation case can be written as the linear combination of these 16 CP-odd flavor invariants into the form of Eq.~(\ref{eq:observable decomposition}) with $j_{\rm max}^{}=16$.

\section{Connection between the full theory and the effective theory}
\label{sec:matching}

In Secs.~\ref{sec:construction2g} and \ref{sec:construction3g} we have studied the algebraic structure of the invariant ring in the SEFT using the tool of invariant theory. In particular, we have explicitly constructed all the basic flavor invariants for the two-generation case and all the primary flavor invariants for the three-generation case. We have also shown that all the physical parameters in the theory can be extracted using primary flavor invariants. On the other hand, the algebraic structure of the invariant ring and the construction of flavor invariants in the full seesaw model have been partly studied in Refs.~\cite{Manohar:2009dy,Manohar:2010vu,Yu:2021cco}. Thus an intriguing question is what the connection between the invariant ring of the flavor space in the SEFT and that in the full theory is and how the two sets of flavor invariants match with each other.

\renewcommand\arraystretch{1.2}
\begin{table}[t!]
	\centering
	\begin{tabular}{l|c|c|c|c}
		\hline\hline
		{\bf Model} & {\bf Moduli} & {\bf Phases} & {\bf Physical parameters} & {\bf Primary invariants}\\
		\hline
		Two-generation SEFT & 8 & 2 & 10 & 10\\
		\hline
		Two-generation seesaw & 8 & 2 & 10 & 10\\
		\hline
		Three-generation SEFT & 15 & 6 & 21 & 21\\
		\hline
		Three-generation seesaw & 15 & 6 & 21 & 21\\
		\hline\hline
	\end{tabular}
	\vspace{0.5cm}
	\caption{\label{table:comparison} Comparison of the number of independent physical parameters in the theory and the number of primary invariants in the flavor space between the SEFT and the full seesaw model. Note that the moduli denote the parameters in the theory other than phases. It is obvious that the SEFT and the full seesaw model share exactly equal number of independent physical parameters and primary flavor invariants.}
\end{table}
\renewcommand\arraystretch{1.0}

In the full seesaw model introduced in Eq.~(\ref{eq:full lagrangian}), the building blocks for the construction of the flavor invariants are $Y_l^{}$, $Y_\nu^{}$ and $Y_{\rm R}^{}$ and they transform in the flavor space as Eq.~(\ref{eq:Yukawa trans}). Given the representations of the building blocks under the flavor-basis transformation, it is straightforward to calculate the HS, which encodes the information about the flavor structure in the full theory. The results of the HS for two-generation and three-generation seesaw are given in Eq.~(\ref{eq:HS seesaw 2g}) and Eq.~(\ref{eq:HS seesaw 3g}), respectively. The key observation is that the denominator of Eq.~(\ref{eq:HS seesaw 2g}) [or Eq.~(\ref{eq:HS seesaw 3g})] and that of Eq.~(\ref{eq:HS eff 2g main}) [or Eq.~(\ref{eq:HS eff 3g main})] have exactly the same number of factors, implying that there are equal number of algebraically-independent invariants in the flavor space of the full theory and that of the SEFT, i.e.,
$$
\boxed{
	\text{\# primary invariants in SEFT}=\text{\# primary invariants in seesaw}
}
$$
Moreover, the number of independent physical parameters in the full seesaw model also matches that in the SEFT~\cite{Broncano:2002rw}, as summarized in Table~\ref{table:comparison}. This nontrivial correspondence implies that for type-I seesaw model, which only extends the SM by adding gauge singlets, only one $d=5$ and one $d=6$ operator are already \emph{adequate} to incorporate all physical information about the UV theory, including the sources of CP violation~\cite{Broncano:2002rw,Broncano:2003fq}. 

\renewcommand\arraystretch{1.2}
\begin{table}[t!]
	\centering
	\begin{tabular}{l|c|c}
		\hline \hline
		flavor invariants &  degree & CP parity \\
		\hline \hline
		$I_{200}^{}\equiv {\rm Tr}\left(X_l^{}\right)$ &  2 & + \\
		\hline
		$I_{020}^{}\equiv {\rm Tr}\left(X_\nu^{}\right)$ &  2 & +\\
		\hline
		$I_{002}^{}\equiv {\rm Tr}\left(X_{\rm R}^{}\right)$ &  2 &+\\
		\hline
		$I_{400}^{}\equiv {\rm Tr}\left(X_l^2\right)$ & 4 &+\\
		\hline
		$I_{220}^{}\equiv {\rm Tr}\left(X_l^{}X_\nu^{}\right)$ & 4 &+\\
		\hline
		$I_{040}^{}\equiv {\rm Tr}\left(X_\nu^2\right)$ &  4 &+\\
		\hline
		$I_{022}^{}\equiv {\rm Tr}\left(\widetilde{X}_\nu^{}X_{\rm R}^{}\right)$ &  4 & $+$\\
		\hline
		$I_{004}^{}\equiv {\rm Tr}\left(X_{\rm R}^2\right)$ & 4 & $+$\\
		\hline
		$I_{222}^{}\equiv {\rm Tr}\left(X_{\rm R}^{}G_{l\nu}^{}\right)$ & 6 & $+$\\
		\hline
		$I_{042}^{}\equiv {\rm Tr}\left(\widetilde{X}_\nu^{}G_{\nu{\rm R}}^{}\right)$ & 6 & $+$\\
		\hline
		$I_{242}^{(1)}\equiv {\rm Tr}\left(G_{l\nu}^{}G_{\nu{\rm R}}^{}\right)$ & 8 & $+$\\
		\hline
		$I_{242}^{(2)}\equiv {\rm Im}\,{\rm Tr}\left(\widetilde{X}_\nu^{}X_{\rm R}^{}G_{l\nu}^{}\right)$ & 8 & $-$\\
		\hline
		$I_{044}^{}\equiv {\rm Im}\,{\rm Tr}\left(\widetilde{X}_\nu^{}X_{\rm R}^{}G_{\nu{\rm R}}^{}\right)$ & 8 & $-$\\
		\hline
		$I_{442}^{}\equiv {\rm Tr}\left(G_{l\nu}^{}G_{l\nu{\rm R}}^{}\right)$ & 10 & $+$\\
		\hline
		$I_{262}^{}\equiv {\rm Im}\,{\rm Tr}\left({\widetilde X}_\nu^{}G_{l\nu}^{}G_{\nu{\rm R}}^{}\right)$ & 10 & $-$\\
		\hline
		$I_{244}^{}\equiv {\rm Im}\,{\rm Tr}\left(X_{\rm R}^{}G_{l\nu}^{}G_{\nu{\rm R}}^{}\right)$ & 10 & $-$\\
		\hline
		$I_{462}^{}\equiv {\rm Im}\,{\rm Tr}\left(\widetilde{X}_\nu^{}G_{l\nu}^{}G_{l\nu{\rm R}}^{}\right)$ & 12 & $-$\\
		\hline
		$I_{444}^{}\equiv {\rm Im}\,{\rm Tr}\left(X_{\rm R}^{}G_{l\nu}^{}G_{l\nu{\rm R}}^{}\right)$ & 12 & $-$\\
		\hline
		\hline
	\end{tabular}
	\vspace{0.5cm}
	\caption{\label{table:2g seesaw}Summary of the basic flavor invariants along with their degrees and CP parities in the full seesaw model for two-generation case, where the subscripts of the invariants denote the degrees of $Y_l^{}$, $Y_\nu^{}$ and $Y_{\rm R}^{}$, respectively. Note that we have also defined some building blocks that transform adjointly under the flavor transformation: $X_l^{}\equiv Y_l^{}Y_l^\dagger$, $X_\nu^{}\equiv Y_\nu^{} Y_\nu^\dagger$, $\widetilde{X}_\nu^{}\equiv Y_\nu^\dagger Y_\nu^{}$, $X_{\rm R}^{}\equiv Y_{\rm R}^\dagger Y_{\rm R}^{}$, $G_{l\nu}^{}\equiv Y_\nu^\dagger X_l^{} Y_\nu^{}$, $G_{\nu{\rm R}}^{}\equiv Y_{\rm R}^{\dagger} \widetilde{X}_\nu^* Y_{\rm R}^{}$ and $G_{l\nu{\rm R}}^{}\equiv Y_{\rm R}^\dagger G_{l\nu}^*Y_{\rm R}^{}$. There are in total 12 CP-even and 6 CP-odd basic invariants in the invariant ring of the flavor space.}
\end{table}
\renewcommand\arraystretch{1}

This point can be seen more clearly from the basic invariants. We take the two-generation case for illustration. With the help of Eqs.~(\ref{eq:HS seesaw 2g}) and (\ref{eq:PL seesaw 2g}) one can explicitly construct all the basic flavor invariants in the full theory, as listed in Table~\ref{table:2g seesaw}. To one's surprise, there are exactly equal number of CP-odd and CP-even basic invariants in Table~\ref{table:2g eff} and Table~\ref{table:2g seesaw}, both are 6 and 12, i.e.,
$$
\boxed{
	\text{\# CP-odd (-even) basic invariants in SEFT}=\text{\# CP-odd (-even) basic invariants in seesaw}
}
$$
Recalling that the basic invariants serve as the generators of the invariant ring in the sense that any flavor invariant in the ring can be decomposed into the polynomial of the basic ones, we reach the conclusion that the ring of the invariants in the SEFT and that in the full seesaw model share an equal number of generators. One can then establish a direct connection between the two sets of generators by noticing that the building blocks $C_5^{}$ and $C_6^{}$ in the SEFT are related to the building blocks $Y_\nu^{}$ and $Y_{\rm R}^{}$ in the full theory by Eq.~(\ref{eq:wilson coe}). In Appendix~\ref{app:matching} we give the details of the matching procedure and the final conclusion is: \emph{All the basic invariants in the SEFT can be written as the rational functions of those in the full seesaw model.}\footnote{This result has been partly derived in Ref.~\cite{Yu:2021cco} for the minimal seesaw model, but the inclusion of the dimension-six operator as well as a complete matching is lacking therein.} 

For instance, the 18 basic flavor invariants in the SEFT have been explicitly expressed as the rational functions of the 18 basic flavor invariants in the full seesaw model in Eqs.~(\ref{eq:odd1 app})-(\ref{eq:even 12 app}) for the two-generation case. Moreover, one can establish a \emph{one-to-one} correspondence between the 6 CP-odd basic invariants in the SEFT and the 6 CP-odd basic invariants in the full theory (see Appendix~\ref{app:matching} for more details)
{\allowdisplaybreaks
	\begin{eqnarray}
		{\cal I}_{121}^{(2)}&=&\frac{2}{\left(I_{002}^2-I_{004}\right)^2}\left[I_{242}^{(2)}I_{022}^{}-I_{044}^{}I_{220}^{}+I_{262}^{}I_{002}^{}-I_{244}^{}I_{020}^{}\right]\;,\label{eq:odd1}\\
		{\cal I}_{221}^{}&=&\frac{2}{\left(I_{002}^2-I_{004}\right)^2}\left[I_{242}^{(2)}I_{222}^{}+I_{244}^{}I_{220}^{}+I_{462}^{}I_{002}^{}-I_{444}^{}I_{020}^{}\right]\;,\label{eq:odd2}\\
		{\cal I}_{122}^{}&=&\frac{2}{\left(I_{002}^2-I_{004}\right)^3}\left\{I_{242}^{(2)}\left[3I_{022}^2+2I_{040}^{}\left(I_{002}^2-I_{004}^{}\right)-4I_{020}^{}I_{002}^{}I_{022}^{}\right]\right.\nonumber\\
		&&\left.+I_{044}^{}\left(4I_{020}^{}I_{222}^{}-I_{220}^{}I_{022}^{}-2I_{242}^{(1)}\right)+I_{262}^{}\left[3I_{002}^{}I_{022}^{}-I_{020}^{}\left(I_{002}^2+3I_{004}^{}\right)\right]\right.\nonumber\\
		&&\left.+I_{244}^{}\left(3I_{020}^{}I_{022}^{}-2I_{042}^{}\right)\right\}\;,\label{eq:odd3}\\
		{\cal I}_{240}^{}&=&\frac{1}{\left(I_{002}^2-I_{004}\right)^2}\left[3I_{242}^{(2)}\left(I_{022}^{}I_{220}^{}-I_{020}^{}I_{222}^{}\right)-I_{044}^{}I_{220}^2+I_{262}^{}\left(3I_{002}^{}I_{220}^{}-2I_{222}^{}\right)\right.\nonumber\\
		&&\left.-2 I_{244}^{}I_{020}^{}I_{220}^{}+I_{462}^{}\left(2I_{022}^{}-3I_{002}^{}I_{020}\right)+I_{444}^{}I_{020}^2\right]\;,\label{eq:odd4}\\
		{\cal I}_{141}^{}&=&\frac{2}{\left(I_{002}^2-I_{004}\right)^3}\left\{I_{242}^{(2)}I_{020}^{}I_{022}^2+I_{044}^{}I_{020}^{}\left(I_{022}^{}I_{220}^{}-2I_{242}^{(1)}\right)\right.\nonumber\\
		&&\left.+I_{262}^{}\left[I_{002}^{}I_{020}^{}I_{022}^{}+I_{040}^{}\left(I_{004}^{}-I_{002}^2\right)\right]+I_{244}^{}I_{020}^{}\left(I_{020}^{}I_{022}-2I_{042}^{}\right)\right\}\;,\label{eq:odd5}\\
		{\cal I}_{042}^{}&=&\frac{2}{\left(I_{002}^2-I_{004}\right)^3}\,I_{044}^{}\left(I_{020}^2-I_{040}^{}\right)_{}^2\;.\label{eq:odd6}
	\end{eqnarray}
}
Note that Eqs.~(\ref{eq:odd1})-(\ref{eq:odd6}) form a system of \emph{linear} equations with respect to the CP-odd invariants and the determinant of the coefficient matrix in Eqs.~(\ref{eq:odd1})-(\ref{eq:odd6}) turns out to be
\begin{eqnarray}
	\label{eq:det}
	{\rm Det}&=&\frac{128}{\left(I_{002}^2-I_{004}^{}\right)^{14}}\,I_{020}^{}\left(I_{002}^{}I_{020}^{}-I_{022}^{}\right)\left(I_{020}^2-I_{040}^{}\right)_{}^2\nonumber\\
	&&\times\left\{I_{020}^2I_{022}^{}\left(3I_{020}^{}I_{022}^{}-4I_{002}^{}I_{040}^{}-3I_{042}^{}\right)-I_{022}^{}I_{040}^{}I_{042}^{}\right.\nonumber\\
	&&\left.+ I_{020}^{}I_{040}^{}\left[3I_{022}^2+2I_{002}^{}I_{042}^{}+I_{040}^{}\left(I_{002}^2-I_{004}^{}\right)\right]\right\}\;,
\end{eqnarray}
which is nonzero in general. This implies the vanishing of all CP-odd flavor invariants in the SEFT is equivalent to the vanishing of all CP-odd flavor invariants in the full seesaw model. Therefore, the absence of CP violation in the low-energy effective theory up to ${\cal O}\left(1/\Lambda_{}^2\right)$ is equivalent to the CP conservation in the full theory.

Eqs.~(\ref{eq:odd1})-(\ref{eq:odd6}) can be implemented to link the CP violation at low energies and that for leptogene-\\sis at high energies. For the purpose of illustration, we consider the (unflavored) CP asymmetries in the decays of RH Majorana neutrinos for the two-generation case, which are defined as
\begin{eqnarray}
	\epsilon_i^{}\equiv \frac{\sum_\alpha \left[\Gamma\left(N_i^{}\to \ell_\alpha+H\right)-\Gamma\left(N_i^{}\to \overline{\ell_\alpha}+\overline{H}\right)\right]}{\sum_\alpha \left[\Gamma\left(N_i^{}\to \ell_\alpha+H\right)+\Gamma\left(N_i^{}\to \overline{\ell_\alpha}+\overline{H}\right)\right]}\;,
\end{eqnarray}
where $\Gamma\left(N_i^{}\to \ell_\alpha+H\right)$ and $\Gamma\left(N_i^{}\to \overline{\ell_\alpha}+\overline{H}\right)$ denote the decay rate of $N_i^{}\to \ell_\alpha+H$ and $N_i^{}\to \overline{\ell_\alpha}+\overline{H}$ (for $i=1,2$ and $\alpha=e,\mu$), respectively. In the basis where the charged-lepton and the RH neutrino mass matrices are real and diagonal, $\epsilon_i^{}$ can be calculated as~\cite{Xing:2011zza}
\begin{eqnarray}
	\label{eq:epsiloni}
	\epsilon_i^{}=\frac{1}{8\pi\left(\widetilde{X}_\nu^{}\right)_{ii}^{}}\sum_{j\neq i}^{}{\rm Im}\,\left[\left(\widetilde{X}_\nu^{}\right)_{ij}^2\right]F\left(\frac{Y_j^2}{Y_i^2}\right)\;,
\end{eqnarray}
with
$$
F\left(x\right)\equiv \sqrt{x}\left[\frac{2-x}{1-x}+\left(1+x\right)\ln\left(\frac{x}{1+x}\right)\right]\;.
$$
In terms of the flavor invariants one can recast the CP asymmetries into the form of Eq.~(\ref{eq:observable decomposition}) with the unique CP-odd flavor invariant $I_{044}^{}$
\begin{eqnarray}
	\label{eq:epsiloni2}
	\epsilon_{1,2}^{}=\frac{\sqrt{2}\, I_{044}}{4\pi\sqrt{I_{002}^2-I_{004}}\left(I_{020}I_{002}-2I_{022}\pm I_{020}\sqrt{2I_{004}-I_{002}^2}\right)}F\left(\frac{I_{002}\pm\sqrt{2I_{004}-I_{002}^2}}{I_{002}\mp\sqrt{2I_{004}-I_{002}^2}}\right)\;,
\end{eqnarray}
where the upper and lower signs refer respectively to $\epsilon_1^{}$ and $\epsilon_2^{}$. Note that Eq.~(\ref{eq:epsiloni2}) is manifestly independent of the parametrization schemes and the flavor basis, though Eq.~(\ref{eq:epsiloni}) is calculated in a specific basis where $X_l^{}$ and $Y_{\rm R}^{}$ are real and diagonal. For the hierarchical mass spectrum $Y_2^{}\gg Y_1^{}$, only $\epsilon_1^{}$ from the decay of lighter Majorana neutrino is relevant for leptogenesis. In consideration of $F\left(x\right)=-3/\left(2\sqrt{x}\right)$ for $x\gg 1$ one obtains
\begin{eqnarray}
	\epsilon_1^{}=\frac{3}{16\pi}\frac{I_{044}}{I_{002}\left(I_{022}-I_{002}I_{020}\right)}\;.
\end{eqnarray}
Then using Eq.~(\ref{eq:odd6}) one can relate $\epsilon_1^{}$ to one CP-odd flavor invariant in the SEFT, namely, ${\cal I}_{042}^{}$, in a simple way
\begin{eqnarray}
	\label{eq:epsilon1}
	\epsilon_1^{}=\frac{3}{32\pi}\frac{\left(I_{002}^2-I_{004}\right)^3}{I_{002}\left(I_{022}-I_{002}I_{020}\right)\left(I_{020}^2-I_{040}\right)^2}\,{\cal I}_{042}^{}\;,
\end{eqnarray}
with the coefficient composed of all CP-even flavor invariants in the full theory. Furthermore, since there are only 2 independent phases in the SEFT for the two-generation case, only 2 of the 6 CP-odd basic invariants in Table~\ref{table:2g eff} are algebraically independent. With the help of the syzygies in Eqs.~(\ref{eq:syzygy1})-(\ref{eq:syzygy4}) one can express any four of them as the linear combinations of the other two, with the coefficients being rational functions of only CP-even invariants. To be explicit, one can express ${\cal I}_{042}^{}$ as the linear combination of ${\cal I}_{121}^{(2)}$ and ${\cal I}_{240}^{}$, which are responsible for the CP violation in neutrino-neutrino and neutrino-antineutrino oscillations, respectively [cf. Eqs.~(\ref{eq:neutrino oscillation}) and (\ref{eq:neutrino-antineutrino oscillation})]
\begin{eqnarray}
	{\cal I}_{042}^{}=\frac{{\cal P}_1\left[{\cal I}_{\rm even}\right]\,{\cal I}_{121}^{(2)}+{\cal P}_2\left[{\cal I}_{\rm even}\right]\,{\cal I}_{240}^{}}{\left[{\cal I}_{021}\left({\cal I}_{100}^2-2{\cal I}_{200}\right)+{\cal I}_{101}\left(2{\cal I}_{120}-{\cal I}_{020}{\cal I}_{100}\right)+{\cal I}_{001}\left({\cal I}_{020}{\cal I}_{200}-{\cal I}_{100}{\cal I}_{120}\right)\right]^2}\;,
\end{eqnarray}
where 
\begin{eqnarray}
	\label{eq:P1}
	{\cal P}_1^{}\left[{\cal I}_{\rm even}\right]
	&=&{\cal I}_{021}^3{\cal I}_{100}^{}\left(2{\cal I}_{200}^{}-{\cal I}_{100}^2\right)+{\cal I}_{021}^2\left[\left({\cal I}_{100}^2-2{\cal I}_{200}^{}\right)\left({\cal I}_{001}^{}{\cal I}_{120}^{}+2{\cal I}_{121}^{(1)}\right)\right.\nonumber\\
	&&\left. +{\cal I}_{020}^{}{\cal I}_{100}^{}\left({\cal I}_{100}^{}{\cal I}_{101}^{}-{\cal I}_{001}^{}{\cal I}_{200}^{}\right)+2{\cal I}_{220}^{}\left({\cal I}_{001}^{}{\cal I}_{100}^{}-2{\cal I}_{101}^{}\right)\right]\nonumber\\
	&&+{\cal I}_{021}^{}\left\{4{\cal I}_{121}^{(1)}\left[{\cal I}_{001}^{}\left({\cal I}_{020}^{}{\cal I}_{200}^{}-{\cal I}_{100}^{}{\cal I}_{120}^{}\right)+{\cal I}_{101}^{}\left(2{\cal I}_{120}^{}-{\cal I}_{020}^{}{\cal I}_{100}^{}\right)\right]\right.\nonumber\\
	&&\left.+{\cal I}_{001}^2\left[{\cal I}_{020}^{}\left({\cal I}_{120}^{}{\cal I}_{200}^{}-{\cal I}_{100}^{}{\cal I}_{220}^{}\right)+{\cal I}_{120}^{}\left({\cal I}_{100}^{}{\cal I}_{120}^{}-2{\cal I}_{220}^{}\right)\right]\right.\nonumber\\
	&&\left.+2{\cal I}_{001}^{}{\cal I}_{020}^{}{\cal I}_{101}^{}\left(2{\cal I}_{220}^{}-{\cal I}_{100}^{}{\cal I}_{120}^{}\right)+{\cal I}_{100}^3{\cal I}_{020}^{}{\cal I}_{022}^{}+{\cal I}_{100}^2{\cal I}_{120}^{}\left({\cal I}_{002}^{}{\cal I}_{020}^{}-2{\cal I}_{022}^{}\right)\right.\nonumber\\
	&&\left.-2{\cal I}_{100}^{}\left[{\cal I}_{002}^{}\left({\cal I}_{120}^2+{\cal I}_{020}^{}{\cal I}_{220}^{}\right)+{\cal I}_{020}^{}{\cal I}_{022}^{}{\cal I}_{200}^{}\right]+4{\cal I}_{120}^{}\left({\cal I}_{022}^{}{\cal I}_{200}^{}+{\cal I}_{002}^{}{\cal I}_{220}^{}\right)\right\}\nonumber\\
	&&+{\cal I}_{120}^3{\cal I}_{001}^{}\left(2{\cal I}_{002}^{}-{\cal I}_{001}^2\right)+{\cal I}_{120}^2\left[{\cal I}_{001}^2\left({\cal I}_{020}^{}{\cal I}_{101}^{}+2{\cal I}_{121}^{(1)}\right)+{\cal I}_{001}^{}{\cal I}_{100}^{}\left(2{\cal I}_{022}^{}-{\cal I}_{002}^{}{\cal I}_{020}^{}\right)\right.\nonumber\\
	&&\left.-4\left({\cal I}_{022}^{}{\cal I}_{101}^{}+{\cal I}_{002}^{}{\cal I}_{121}^{(1)}\right)\right]+{\cal I}_{120}^{}{\cal I}_{020}^{}\left[4{\cal I}_{121}^{(1)}\left({\cal I}_{002}^{}{\cal I}_{100}^{}-{\cal I}_{001}^{}{\cal I}_{101}^{}\right)+{\cal I}_{001}^3{\cal I}_{220}^{}\right.\nonumber\\
	&&\left.-{\cal I}_{001}^{}{\cal I}_{022}^{}\left({\cal I}_{100}^2+2{\cal I}_{200}^{}\right)+2\left(2{\cal I}_{022}^{}{\cal I}_{100}^{}{\cal I}_{101}^{}-{\cal I}_{001}^{}{\cal I}_{002}^{}{\cal I}_{220}^{}\right)
	\right]\nonumber\\
	&&-{\cal I}_{001}^2{\cal I}_{020}^2\left({\cal I}_{121}^{(1)}{\cal I}_{200}^{}+{\cal I}_{101}^{}{\cal I}_{220}^{}\right)+{\cal I}_{001}^{}{\cal I}_{020}^2{\cal I}_{100}^{}\left(2{\cal I}_{101}^{}{\cal I}_{121}^{(1)}+{\cal I}_{022}^{}{\cal I}_{200}^{}+{\cal I}_{002}^{}{\cal I}_{220}^{}\right)\nonumber\\
	&&-{\cal I}_{020}^2{\cal I}_{100}^2\left({\cal I}_{022}^{}{\cal I}_{101}^{}+{\cal I}_{002}^{}{\cal I}_{121}^{(1)}\right)\;,
\end{eqnarray}
and
\begin{eqnarray}
	\label{eq:P2}
	{\cal P}_2^{}\left[{\cal I}_{\rm even}\right]=\left[{\cal I}_{001}^2{\cal I}_{120}^{}-{\cal I}_{001}^{}\left({\cal I}_{021}^{}{\cal I}_{100}^{}+{\cal I}_{020}^{}{\cal I}_{101}^{}\right)+{\cal I}_{002}^{}\left({\cal I}_{020}^{}{\cal I}_{100}^{}-2{\cal I}_{120}^{}\right)+2{\cal I}_{021}^{}{\cal I}_{101}^{}\right]_{}^2\;,
\end{eqnarray}
are polynomials of CP-even basic flavor invariants in the SEFT. 
Since any CP-even basic flavor invariant in the SEFT can be written as the rational function of those in the full theory using Eqs.~(\ref{eq:even 1 app})-(\ref{eq:even 12 app}), one finally arrives at
\begin{eqnarray}
	\label{eq:connection}
	\epsilon_1^{}={\cal R}_1^{}\left[I_{\rm even}^{}\right]\,{\cal I}_{121}^{(2)}+{\cal R}_2^{} \left[I_{\rm even}^{}\right]\,{\cal I}_{240}^{}\;,
\end{eqnarray} 
where ${\cal R}_1^{}\left[I_{\rm even}^{}\right]$ and ${\cal R}_2^{}\left[I_{\rm even}^{}\right]$ are rational functions of CP-even basic flavor invariants in the full theory. The complete expressions of ${\cal R}_1^{}\left[I_{\rm even}^{}\right]$ and ${\cal R}_2^{}\left[I_{\rm even}^{}\right]$ are too lengthy to be listed here, though they can be straightforwardly obtained by substituting Eqs.~(\ref{eq:even 1 app})-(\ref{eq:even 12 app}) into Eqs.~(\ref{eq:P1}) and (\ref{eq:P2}) and combining them with Eq.~(\ref{eq:epsilon1}). In Eq.~(\ref{eq:connection}) we have expressed the CP asymmetries in the decays of RH neutrinos as the linear combination of two CP-odd flavor invariants in the low-energy effective theory, which respectively measure the CP violation in neutrino-neutrino and neutrino-antineutrino oscillations. This establishes a direct link between CP-violating observables at high- and low-energy scales in a basis- and parametrization-independent way.\footnote{The connection between CP violation at low and high energies has also been discussed in some previous works~\cite{Broncano:2003fq,Branco:2001pq,Branco:2003rt,Branco:2004hu,Branco:2006ce,Pascoli:2006ie,Pascoli:2006ci,Antusch:2009gn}, but without the full language of invariant theory, and the independence of flavor bases and parametrization schemes is not manifest therein.} Particularly, if CP violation is absent in both neutrino-neutrino and neutrino-antineutrino oscillations, i.e., ${\cal A}_{\nu\nu}^{e\mu}={\cal A}_{\nu\bar{\nu}}^{e\mu}=0$ implying ${\cal I}_{121}^{(2)}={\cal I}_{240}^{}=0$, then the CP asymmetries in RH neutrino decays also vanish. Conversely, if CP violation is measured at low energies either in neutrino-neutrino or neutrino-antineutrino oscillations, indicating either ${\cal I}_{121}^{(2)}$ or ${\cal I}_{240}^{}$ is nonvanishing, then CP asymmetries may exist in the decays of RH neutrinos. This result is consistent with the conclusion drawn from Eqs.~(\ref{eq:odd1})-(\ref{eq:det}) that the absence of CP violation in the SEFT also implies the CP conservation in the full seesaw model.

The above analysis about the basic invariants can be directly generalized to the three-generation case. Since there are more than 200 basic invariants in both the SEFT and the full theory, we shall not attempt to write down the complete matching conditions among these two sets of basic invariants for the three-generation case. However, all the basic invariants in the SEFT can still be written as the rational functions of those in the full theory using the matching procedure in Appendix~\ref{app:matching} and taking advantage of Eq.~(\ref{eq:inverse 3g}). In this case, it can be shown that the absence of CP violation in the SEFT is equivalent to the CP conservation in the full seesaw model~\cite{Broncano:2003fq}. 

In Ref.~\cite{Branco:2001pq} the authors constructed the following 6 CP-odd flavor invariants\footnote{It should be noticed that the notations of building blocks and flavor invariants in Ref.~\cite{Branco:2001pq} are different from those in the present paper.}
\begin{eqnarray}
	J_{044}^{}&\equiv&{\rm Im}\,{\rm Tr}\left(\widetilde{X}_\nu^{}X_{\rm R}^{}G_{\nu{\rm R}}^{}\right)\;,\\
	J_{046}^{}&\equiv&{\rm Im}\,{\rm Tr}\left(\widetilde{X}_\nu^{}X_{\rm R}^{2}G_{\nu{\rm R}}^{}\right)\;,\\
	J_{048}^{}&\equiv&{\rm Im}\,{\rm Tr}\left(\widetilde{X}_\nu^{}X_{\rm R}^{2}G_{\nu{\rm R}}^{}X_{\rm R}^{}\right)\;,\\
	J_{444}^{}&\equiv&{\rm Im}\,{\rm Tr}\left(G_{l\nu}^{}X_{\rm R}^{}G_{l\nu{\rm R}}^{}\right)\;,\\
	J_{446}^{}&\equiv&{\rm Im}\,{\rm Tr}\left(G_{l\nu}^{}X_{\rm R}^{2}G_{l\nu{\rm R}}^{}\right)\;,\\
	J_{448}^{}&\equiv&{\rm Im}\,{\rm Tr}\left(G_{l\nu}^{}X_{\rm R}^{2}G_{l\nu{\rm R}}^{}X_{\rm R}^{}\right)\;,
\end{eqnarray}
and mentioned that the vanishing of these 6 invariants serves as the sufficient and necessary conditions for CP conservation in the full seesaw model for the three-generation case. However, as it is emphasized in Refs.~\cite{Yu:2019ihs,Yu:2020gre}, these are not \emph{linear} equations with respect to the sine functions of the phases in $Y_\nu^{}$ so there may exist some parameter space where all these equations are satisfied but the phases can take some nontrivial values. Therefore, without
any information about the physical parameters at high-energy scales, these equations can only be understood to guarantee CP conservation in \emph{some} particular parameter space. This shortcoming can be overcome by taking advantage of the CP-odd flavor invariants in the SEFT rather than in the full theory. First, it has been proved in Ref.~\cite{Yu:2019ihs} that Eqs.~(\ref{eq:cp conservation condition 1})-(\ref{eq:cp conservation condition 3}) are sufficient to guarantee CP conservation in \emph{all} experimentally allowed parameter space to the order of ${\cal O}\left(1/\Lambda\right)$. Then Eqs.~(\ref{eq:j121})-(\ref{eq:j161}) supply three linear equations and enforce $\alpha_{ij}^{}=\beta_{ij}^{}+k\pi$ without any nontrivial solutions of the phases. Thus the vanishing of $\left\{{\cal J}_{360}^{},{\cal J}_{240}^{(2)},{\cal J}_{260}^{},{\cal J}_{121}^{},{\cal J}_{141}^{},{\cal J}_{161}^{}\right\}$ are able to guarantee CP conservation in the SEFT in all experimentally allowed parameter space. Finally, since the CP conservation in the SEFT is sufficient for CP conservation in the full theory, the vanishing of these 6 flavor invariants in the SEFT also serves as the sufficient and necessary conditions for CP conservation in the full seesaw model in all experimentally allowed parameter space.

To sum up, the connection between the full theory and its low-energy effective theory can be established through the matching of flavor invariants. The matching conditions, such as those in Eqs.~(\ref{eq:odd1 app})-(\ref{eq:even 12 app}), serve as a bridge to link the observables at high energies and those at low energies in a basis- and parametrization-independent way. In addition, the matching conditions are also necessary to determine the initial values of the renormalization-group running of the flavor invariants in the effective theory~\cite{Wang:2021wdq}.

Finally, it is worthwhile to make some brief comments on the practical determination of the physical parameters in the full theory via low-energy measurements. Although the SEFT with only $C^{}_5$ and $C^{}_6$ already contains the same number of physical parameters as the full seesaw model does, the precision of the SEFT itself is limited to the order of ${\cal O}(1/\Lambda^2)$. More precise determination of the physical parameters in $C^{}_5$ and $C^{}_6$, and thus those in the full theory, may require the inclusion of the effective operators of higher mass dimensions at the treel level or even the loop-level matching.

\section{Summary}
\label{sec:summary}
In the language of invariant theory, we have systematically investigated the algebraic structure of the ring of the flavor invariants in the SEFT, which includes one dimension-five and one dimension-six non-renormalizable operator. Particular attention has been paid to the sources of CP violation and the connection between the full seesaw model and the SEFT.

For the first time, we calculate the HS of the flavor space in the SEFT and explicitly construct all the basic (primary) flavor invariants in the invariant ring for the two- (three-) generation case. We have shown that all the physical parameters in the theory can be extracted using the primary flavor invariants, so that any physical observable can be recast into the function of flavor invariants. Furthermore, we prove that any CP-violating observable in the SEFT can be expressed as the linear combination of CP-odd flavor invariants. The minimal sufficient and necessary conditions for leptonic CP conservation in both the SEFT and the full seesaw model have been  clarified.

Based on the observation that there is an equal number of independent physical parameters in the SEFT and in the full seesaw model, we reveal the intimate connection between their rings of flavor invariants. With the HS, we show that the invariant ring of the SEFT shares equal number of primary invariants with that of the full theory, indicating that the inclusion of only one dimension-five and one dimension-six operator in the SEFT is adequate to incorporate all physical information about the full seesaw model. Through a proper matching procedure, we establish a direct link between the flavor invariants in the SEFT and those in the full theory: The former can be expressed as the rational functions of the latter. The matching of the flavor invariants can be used to build a bridge between the CP asymmetries in leptogenesis and those in low-energy neutrino oscillation experiments in a basis- and parametrization-independent way.

The physical observables, which can be measured directly in experiments, should depend on neither the flavor basis nor the specific parametrization of Yukawa matrices. This is exactly the feature of flavor invariants. Therefore, it will be helpful (and more natural) to describe physical observables in terms of flavor invariants. The previous efforts~\cite{Jarlskog:1985ht,Jarlskog:1985cw,Branco:1986quark,Branco:1986lepton,Manohar:2009dy,Manohar:2010vu,Yu:2019ihs,Yu:2020gre,Wang:2021wdq,Yu:2021cco,Bonnefoy:2021tbt,Yu:2022nxj} and the results in this work have demonstrated the great power of the invariant theory in studying CP violation in the quark and leptonic sector, and call for more applications of the invariant theory in other aspects of particle physics.

\section*{Acknowledgements}
This work was supported by the National Natural Science Foundation of China under grant No.~11835013 and the Key Research Program of the Chinese Academy of Sciences under grant No. XDPB15.

\begin{appendix}
\section{Calculation of the Hilbert series}
\label{app:HS}
In this appendix, we present the computational details of the HS in the SEFT.\footnote{A concise and pedagogical introduction to the invariant theory and the HS can be found in Appendix B of Ref.~\cite{Wang:2021wdq}.} The HS plays an important role in the invariant theory and supplies a powerful tool for studying algebraic structure of the invariant ring and the polynomial identities among invariants. The HS is defined as the generating function of the invariants
\begin{eqnarray}
	\label{eq:HS def}
	{\mathscr H}\left(q\right)\equiv \sum_{k=0}^{\infty}c_k^{}q_{}^k\;,
\end{eqnarray} 
where $c_k^{}$ (with $c_0^{}\equiv 1$) denote the number of linearly-independent invariants at degree $k$ while $q$ is an arbitrary complex number that satisfies $\left|q\right|<1$ and labels the degree of the building blocks.

The HS encodes all the information about the invariant ring. A general HS can always be written as the ratio of two polynomial functions~\cite{sturmfels2008algorithms,derksen2015computational}
\begin{eqnarray}
	{\mathscr H}\left(q\right)=\frac{1+a_1q+\cdots a_{l-1}q^{l-1}+q^l}{\prod_{k=1}^r\left(1-q^{d_k}\right)}\;,
\end{eqnarray} 
where the numerator has the palindromic structure (i.e., $a_k^{}=a_{l-k}^{}$) and the denominator exhibits the standard Euler product form. A highly nontrivial result is that the total number of the denominator factors $r$ equals the number of the \emph{primary} invariants in the ring, which also matches the number of independent physical parameters in the theory. Here primary invariants refer to those invariants that are algebraically independent, which means there does not exist any polynomial function of them that is identically equal to zero. 

It can be proved that as long as the symmetry group is reductive (including the $N$-dimensional unitary group ${\rm U}(N)$ in the flavor space that we consider throughout this paper), the ring is finitely generated~\cite{sturmfels2008algorithms, derksen2015computational}. This implies that there exist a finite number of \emph{basic} invariants such that any invariant in the ring can be decomposed into the polynomial of these basic invariants. One should keep in mind that the number of basic invariants is in general no smaller than that of primary invariants. This is because the basic invariants may not be algebraically independent, i.e., there may exist nontrivial polynomial relations among basic invariants that are identically equal to zero, known as syzygies. In order to obtain the information of basic invariants, one can calculate the plethystic logarithm (PL) function of the HS
\begin{eqnarray}
	\label{eq:PL def}
	{\rm PL}\left[{\mathscr H}(q)\right]\equiv 
	\sum_{k=1}^{\infty}\frac{\mu(k)}{k}{\rm ln}\left[{\mathscr H}(q_{}^k)\right]\;,
\end{eqnarray}
where $\mu(k)$ is the M\"obius function. The great power of the PL function is that from it one can directly read off the number and degrees of the basic invariants and syzygies: The leading positive terms of PL correspond to the basic invariants while the leading negative terms correspond to the syzygies among them~\cite{Hanany:2006qr}. We will see how this principle is applied from the examples below.

To calculate the HS from the definition in Eq.~(\ref{eq:HS def}) is very difficult in most cases. A systematic approach is to utilize the Molien-Weyl (MW) formula, which reduces the calculation of the HS to contour integrals in the complex plane~\cite{molien1897invarianten, weyl1926darstellungstheorie}
\begin{eqnarray}
	\label{eq:MW formula}
	{\mathscr H}(q)=\int \left[{\rm d}\mu\right]_{G}^{} {\rm PE}\left(z_1^{},...,z_{r_0}^{};q\right)\;,
\end{eqnarray}
where $\left[{\rm d}\mu\right]_{G}^{}$ stands for the Haar measure of the symmetry group $G$ while the integrand is the  plethystic exponential (PE) function defined as
\begin{eqnarray}
	{\rm PE}\left(z_1^{},...z_{r_0}^{};q\right)\equiv {\rm exp}\left[\sum_{k=1}^{\infty}\sum_{i=1}^{n}\frac{\chi_{R_i}\left(z_1^k,...,z_{r_0}^k\right)q^k}{k}\right]\;,
\end{eqnarray}
where $z_1^{},...,z_{r_0}^{}$ are coordinates on the maximum torus of $G$ with $r_0^{}$ the rank of $G$ and $\chi_{R_i}^{}$ (for $i=1,...n$) is the character function of the $i$-th building block that is in the $R_i^{}$ representation of $G$. Below we will use the MW formula to calculate the HS in the flavor space of the SEFT for two- and three-generation cases.	

\subsection{Two-generation SEFT}
In the two-generation scenario, the building blocks in the SEFT to construct flavor invariants ($X_l^{}$, $C_5^{}$ and $C_6^{}$) transform under the symmetry group ${\rm U}(2)$ in the flavor space as 
\begin{eqnarray}
	X_l^{}: {\bf 2}\otimes {\bf 2}_{}^*\;,\quad
	C_5^{}: \left({\bf 2}\otimes {\bf 2}\right)_{\rm s}^{}\;,\quad
	C_5^{\dagger}: \left({\bf 2}_{}^*\otimes {\bf 2}_{}^*\right)_{\rm s}^{}\;,\quad
	C_6^{}: {\bf 2}\otimes {\bf 2}_{}^*\;,
\end{eqnarray}
where ${\bf 2}$ and ${\bf 2}_{}^*$ stand for the fundamental and anti-fundamental representation of ${\rm U}(2)$ respectively while the subscript ``s? denotes the symmetric part. The character function of ${\bf 2}$ and ${\bf 2}_{}^*$ are $z_1^{}+z_2^{}$ and $z_1^{-1}+z_2^{-1}$ respectively, which lead to the character functions of the building blocks
\begin{eqnarray}
	\chi_l^{}\left(z_1^{},z_2^{}\right)&=&\left(z_1^{}+z_2^{}\right)\left(z_1^{-1}+z_2^{-1}\right)\;,\nonumber\\
	\chi_5^{}\left(z_1^{},z_2^{}\right)&=&z_1^2+z_2^2+z_1^{}z_2^{}+z_1^{-1}+z_2^{-1}+z_1^{-1}z_2^{-1}\;,\nonumber\\
	\chi_6^{}\left(z_1^{},z_2^{}\right)&=&\left(z_1^{}+z_2^{}\right)\left(z_1^{-1}+z_2^{-1}\right)\;,
\end{eqnarray}
where $z_1^{}$ and $z_2^{}$ denote the coordinates on the maximum torus of ${\rm U}(2)$ group. Then one can calculate the PE function
\begin{eqnarray}
	\label{eq:PE eff 2g}
	{\rm PE}\left(z_1^{},z_2^{};q\right)&=& {\rm exp}\left(\sum_{k=1}^\infty\frac{\chi_l\left(z_1^k,z_2^k\right)q^k+\chi_5\left(z_1^k,z_2^k\right)q^k+\chi_6\left(z_1^k,z_2^k\right)q^k}{k}\right)\nonumber\\
	&=&\left[\left(1-q\right)_{}^4\left(1-qz_1^{}z_2^{-1}\right)_{}^2\left(1-qz_2^{}z_1^{-1}\right)_{}^2\left(1-qz_1^2\right)\left(1-qz_2^2\right)\left(1-qz_1^{}z_2^{}\right)\right.\nonumber\\
	&&\left.\times\left(1-qz_1^{-2}\right)\left(1-qz_2^{-2}\right)\left(1-qz_1^{-1}z_2^{-1}\right)\right]_{}^{-1}\;,
\end{eqnarray}
where the identity $\sum_{k=1}^{\infty}(x_{}^k/k)=-{\rm ln}(1-x)$ (for $\left|x\right|<1$) has been used. Note that the degrees of $X_l^{}$, $C_5^{}$ and $C_6^{}$ are all labeled by $q$. Substituting the PE function in Eq.~(\ref{eq:PE eff 2g}) into the MW formula in Eq.~(\ref{eq:MW formula}) and taking into account the Haar measure of the ${\rm U}(2)$ group, one obtains the HS in the SEFT for the two-generation case
\begin{eqnarray}
	\label{eq:HS eff 2g}
	{\mathscr H}_{\rm SEFT}^{(2\rm g)}(q)&=&\int \left[{\rm d}\mu\right]_{\rm U (2)}^{} {\rm PE}\left(z_1^{},z_2^{};q\right)\nonumber\\
	&=&\frac{1}{2}\frac{1}{\left(2\pi {\rm i}\right)^2}\oint_{\left|z_1\right|=1}\oint_{\left|z_2\right|=1}\left(2-\frac{z_1}{z_2}-\frac{z_2}{z_1}\right) {\rm PE}\left(z_1^{},z_2^{};q\right)\nonumber\\
	&=&\frac{1+3q^4+2q^5+3q^6+q^{10}}{\left(1-q\right)^2\left(1-q^2\right)^4\left(1-q^3\right)^2\left(1-q^4\right)^2}\;,
\end{eqnarray}
where in the second line of Eq.~(\ref{eq:HS eff 2g}) the integrals are performed on the unit circle and in the final step the contour integrals are accomplished via the residue theorem. From Eq.~(\ref{eq:HS eff 2g}) one finds that the numerator of HS exhibits the palindromic structure as expected while the denominator owns totally 10 factors. The latter implies that there are 10 primary flavor invariants, corresponding to the 10 physical parameters in the theory. The number of basic invariants can be obtained by substituting Eq.~(\ref{eq:HS eff 2g}) into Eq.~(\ref{eq:PL def}) and calculate the PL function
\begin{eqnarray}
	\label{eq:PL eff 2g}
	{\rm PL}\left[{\mathscr H}_{\rm SEFT}^{(2\rm g)}(q)\right]=2q+4q^2+2q^3+5q^4+2q^5+3q^6-{\cal O}\left(q^8\right)\;,
\end{eqnarray}
from which one can read off that there are in total 18 basic invariants (two of degree 1, four of degree 2, two of degree 3, five of degree 4, two of degree 5 and three of degree 6), and the syzygies begin to appear at degree 8. With the help of the leading positive terms in Eq.~(\ref{eq:PL eff 2g}), one can explicitly construct all the basic invariants, as listed in Table~\ref{table:2g eff}. 

\subsection{Three-generation SEFT}
We then proceed to calculate the HS in the SEFT for the three-generation case. The representations of the building blocks under the ${\rm U}(3)$ group turn out to be
\begin{eqnarray}
	X_l^{}: {\bf 3}\otimes {\bf 3}_{}^*\;,\quad
	C_5^{}: \left({\bf 3}\otimes {\bf 3}\right)_{\rm s}^{}\;,\quad
	C_5^{\dagger}: \left({\bf 3}_{}^*\otimes {\bf 3}_{}^*\right)_{\rm s}^{}\;,\quad
	C_6^{}: {\bf 3}\otimes {\bf 3}_{}^*\;,
\end{eqnarray}
where ${\bf 3}$ and ${\bf 3}_{}^*$ denote the fundamental and anti-fundamental representation of ${\rm U}(3)$. Recalling that the character functions of ${\bf 3}$ and ${\bf 3}_{}^*$ are $z_1^{}+z_2^{}+z_3^{}$ and $z_1^{-1}+z_2^{-1}+z_3^{-1}$ respectively, one can calculate the character functions of the building blocks
\begin{eqnarray}
	\chi_l^{}\left(z_1^{},z_2^{},z_3^{}\right)&=&\left(z_1^{}+z_2^{}+z_3^{}\right)\left(z_1^{-1}+z_2^{-1}+z_3^{-1}\right)\;,\nonumber\\
	\chi_5^{}\left(z_1^{},z_2^{},z_3^{}\right)&=&z_1^2+z_2^2+z_3^2+z_1^{}z_2^{}+z_1^{}z_3^{}+z_2^{}z_3^{}\nonumber\\
	&&+z_1^{-2}+z_2^{-2}+z_3^{-2}+z_1^{-1}z_2^{-1}+z_1^{-1}z_3^{-1}+z_2^{-1}z_3^{-1}\;,\nonumber\\
	\chi_6^{}\left(z_1^{},z_2^{},z_3^{}\right)&=&\left(z_1^{}+z_2^{}+z_3^{}\right)\left(z_1^{-1}+z_2^{-1}+z_3^{-1}\right)\;,
\end{eqnarray}
where $z_i^{}$ (for $i=1,2,3$) denote the coordinates on the maximum torus of ${\rm U}(3)$ group. Then the PE function can be written as 
\begin{eqnarray}
	\label{eq:PE eff 3g}
	{\rm PE}\left(z_1^{},z_2^{},z_3^{};q\right)&=& {\rm exp}\left(\sum_{k=1}^\infty\frac{\chi_l\left(z_1^k,z_2^k,z_3^k\right)q^k+\chi_5\left(z_1^k,z_2^k,z_3^k\right)q^k+\chi_6\left(z_1^k,z_2^k,z_3^k\right)q^k}{k}\right)\nonumber\\
	&=&\left[\left(1-q\right)_{}^6\left(1-q z_1^{} z_2^{-1}\right)_{}^2\left(1-q z_2^{} z_1^{-1}\right)_{}^2\left(1-q z_1^{} z_3^{-1}\right)_{}^2\left(1-q z_3^{} z_1^{-1}\right)_{}^2\right.\nonumber\\
	&&\left.\times \left(1-q z_2^{} z_3^{-1}\right)_{}^2 \left(1-q z_3^{} z_2^{-1}\right)_{}^2 \left(1-q z_1^2\right)\left(1-q z_2^2\right)\left(1-q z_3^2\right)\left(1-q z_1^{}z_2^{}\right)\right.\nonumber\\
	&&\left.\times \left(1-q z_1^{}z_3^{}\right)\left(1-q z_2^{}z_3^{}\right)
	\left(1-q z_1^{-2}\right)\left(1-q z_2^{-2}\right)\left(1-q z_3^{-2}\right)\left(1-q z_1^{-1}z_2^{-1}\right)\right.\nonumber\\
	&& \left.\times \left(1-q z_1^{-1}z_3^{-1}\right)\left(1-q z_2^{-1}z_3^{-1}\right)\right]_{}^{-1}\;.
\end{eqnarray}
Using the MW formula in Eq.~(\ref{eq:MW formula}), one obtains the HS in the SEFT for the three-generation case
\begin{eqnarray}
	\label{eq:HS eff 3g1}
	{\mathscr H}_{\rm SEFT}^{(3\rm g)}(q)&=&\int \left[{\rm d}\mu\right]_{\rm U (3)}^{} {\rm PE}\left(z_1^{},z_2^{},z_3^{};q\right)\nonumber\\
	&=&\frac{1}{6}\frac{1}{\left(2\pi {\rm i}\right)^3}\oint_{\left|z_1\right|=1}\oint_{\left|z_2\right|=1}\oint_{\left|z_3\right|=1}\left[-\frac{\left(z_2-z_1\right)^2\left(z_3-z_1\right)^2\left(z_3-z_2\right)^2}{z_1^2z_2^2z_3^2}\right]{\rm PE}\left(z_1^{},z_2^{},z_3^{};q\right)\;,\nonumber\\
\end{eqnarray}
where in the second line of Eq.~(\ref{eq:HS eff 3g1}) the Haar measure of ${\rm U}(3)$ group has been substituted and the integrals should be performed on the unit circle. Taking advantage of the residue theorem to calculate the contour integrals and after some tedious algebra one finally obtains
\begin{eqnarray}
	\label{eq:HS eff 3g}
	{\mathscr H}_{\rm SEFT}^{(3\rm g)}(q)=\frac{{\mathscr N}_{\rm SEFT}^{(3\rm g)}(q)}{{\mathscr D}_{\rm SEFT}^{(3\rm g)}(q)}\;,
\end{eqnarray}
where 
\begin{eqnarray}
	\label{eq:numerator eff 3g}
	{\mathscr N}_{\rm SEFT}^{(3\rm g)}(q)&=&q^{65}+2 q^{64}+4 q^{63}+11 q^{62}+23 q^{61}+48 q^{60}+120 q^{59}+269 q^{58}+587 q^{57}+1258 q^{56}\nonumber\\
	&&+2543 q^{55}+4895 q^{54}+9124 q^{53}+16281 q^{52}+27963 q^{51}+46490 q^{50}+74644 q^{49}\nonumber\\
	&&+115871q^{48}+174433 q^{47}+254494 q^{46}+360055 q^{45}+494873 q^{44}+660820 q^{43}\nonumber\\
	&&+857677 q^{42}+1083226 q^{41}+1331628 q^{40}+1593650 q^{39}+1858178 q^{38}+2111158 q^{37}\nonumber\\
	&&+2337226 q^{36}+2522435
	q^{35}+2654026 q^{34}+2721987 q^{33}+2721987 q^{32}+2654026q^{31}\nonumber\\
	&&+2522435 q^{30}+2337226 q^{29}+2111158 q^{28}+1858178 q^{27}+1593650 q^{26}+1331628 q^{25}\nonumber\\
	&&+1083226 q^{24}+857677 q^{23}+660820
	q^{22}+494873 q^{21}+360055 q^{20}+254494 q^{19}\nonumber\\
	&&+174433 q^{18}+115871 q^{17}+74644 q^{16}+46490 q^{15}+27963 q^{14}+16281 q^{13}+9124 q^{12}\nonumber\\
	&&+4895 q^{11}+2543 q^{10}+1258 q^9+587 q^8+269 q^7+120
	q^6+48 q^5+23 q^4+11 q^3+4 q^2\nonumber\\
	&&+2 q+1\;,
\end{eqnarray}
and
\begin{eqnarray}
	\label{eq:denominator eff 3g}
	{\mathscr D}_{\rm SEFT}^{(3\rm g)}(q)=\left(1-q^2\right)^3 \left(1-q^3\right) \left(1-q^4\right)^5 \left(1-q^5\right)^6 \left(1-q^6\right)^6\;.
\end{eqnarray}
It can be seen that the HS in the three-generation SEFT is much more complicated than that in the two-generation case, reflecting the richness of the leptonic flavor structure and the complexity of the invariant ring. As a nontrivial cross-check, the numerator in Eq.~(\ref{eq:numerator eff 3g}) does exhibit the palindromic structure, and more importantly, the denominator in Eq.~(\ref{eq:denominator eff 3g}) has 21 factors, which correctly matches the number of independent physical parameters in the SEFT for the three-generation case.

\subsection{Full theory}
For completeness, we also calculate the HS in the full seesaw model using the MW formula, although the results have been given in the literature~\cite{Manohar:2009dy,Manohar:2010vu}. In the full theory, the building blocks to construct flavor invariants are $Y_l^{}$, $Y_\nu^{}$ and $Y_{\rm R}^{}$, which transform under the flavor group ${\rm U}(m)\otimes {\rm U}(n)$ as in Eq.~(\ref{eq:Yukawa trans}). Their representations are assigned as
\begin{eqnarray}
	X_l^{}\equiv Y_l^{}Y_l^\dagger: \textbf{\emph{m}}\otimes \textbf{\emph{m}}_{}^*\;,\quad
	Y_\nu^{}: \textbf{\emph{m}}\otimes \textbf{\emph{n}}_{}^{ *}\;,\quad
	Y_\nu^{\dagger}: \textbf{\emph{n}}\otimes \textbf{\emph{m}}_{}^{*}\;,\quad
	Y_{\rm R}^{}: \left(\textbf{\emph{n}}_{}^{*}\otimes \textbf{\emph{n}}_{}^{*}\right)_{\rm s}^{}\;,\quad
	Y_{\rm R}^{\dagger}: \left(\textbf{\emph{n}}\otimes \textbf{\emph{n}}\right)_{\rm s}^{}\;,\qquad
\end{eqnarray}
where $m$ (or $n$) is the number of the generations of active  (or RH) neutrinos, $\textbf{\emph{m}}$ (or $\textbf{\emph{n}}$) and $\textbf{\emph{m}}_{}^{*}$ (or $\textbf{\emph{n}}_{}^{*}$) denote respectively the fundamental and anti-fundamental representation of ${\rm U}(m)$ [or ${\rm U}(n)$] group. In order to compare with the HS in the SEFT, we consider two special scenarios: $m=n=2$ and $m=n=3$.

For the case of $m=n=2$, the character functions of the building blocks read
\begin{eqnarray}
	\chi_l^{}\left(z_1^{},z_2^{}\right)&=&\left(z_1^{}+z_2^{}\right)\left(z_1^{-1}+z_2^{-1}\right)\,\nonumber\\
	\chi_\nu^{}\left(z_1^{},z_2^{},z_3^{},z_4{}\right)&=&\left(z_1^{}+z_2^{}\right)\left(z_3^{-1}+z_4^{-1}\right)+\left(z_3^{}+z_4^{}\right)\left(z_1^{-1}+z_2^{-1}\right)\;,\nonumber\\
	\chi_{\rm R}\left(z_3^{},z_4^{}\right)&=&z_3^2+z_4^2+z_3^{}z_4^{}+z_3^{-2}+z_4^{-2}+z_3^{-1}z_4^{-1}\;,
\end{eqnarray}
where $z_1^{}$ and $z_2$ (or $z_3^{}$ and $z_4^{}$) denote the coordinates on the maximum torus of the ${\rm U}(2)$ group that corresponds to the flavor-basis transformation in the active neutrino (or RH neutrino) sector. The PE function turns out to be
\begin{eqnarray}
	\label{eq:PE seesaw 2g}
	&&{\rm PE}\left(z_1^{},z_2^{},z_3^{},z_4^{};q\right)\nonumber\\
	&=&{\rm exp}\left(\sum_{k=1}^\infty\frac{\chi_l\left(z_1^k,z_2^k\right)q^{2k}+\chi_\nu\left(z_1^k,z_2^k,z_3^k,z_4^k\right)q^{k}+\chi_{\rm R}\left(z_3^k,z_4^k\right)q^k}{k}\right)\nonumber\\
	&=&\left[\left(1-q_{}^2\right)_{}^2\left(1-q_{}^2z_1^{}z_2^{-1}\right)\left(1-q_{}^2z_2^{}z_1^{-1}\right)
	\left(1-qz_1^{}z_3^{-1}\right)\left(1-qz_3^{}z_1^{-1}\right)\left(1-qz_1^{}z_4^{-1}\right)\left(1-qz_4^{}z_1^{-1}\right)\right.\nonumber\\
	&&\left.\times \left(1-qz_2^{}z_3^{-1}\right)\left(1-qz_3^{}z_2^{-1}\right)\left(1-qz_2^{}z_4^{-1}\right)\left(1-qz_4^{}z_2^{-1}\right)\left(1-qz_4^2\right)\left(1-qz_5^2\right)\right.\nonumber\\
	&&\left.\times
	\left(1-qz_4^{}z_5^{}\right)\left(1-qz_4^{-2}\right)\left(1-qz_5^{-2}\right)\left(1-qz_4^{-1}z_5^{-1}\right)
	\right]_{}^{-1}\;.
\end{eqnarray}
Note that we have counted the degrees of $Y_l^{}$, $Y_\nu^{}$ and $Y_{\rm R}^{}$ by $q$, such that the degree of $X_l^{}\equiv Y_l^{}Y_l^\dagger$ is labeled by $q_{}^2$, which is different from the convention in the scenario of the SEFT.\footnote{Different conventions under which the degrees of building blocks are labeled will change the form of the HS. However, they have no influence on the algebraic structure of the invariant ring. Namely, the construction of primary invariants, basic invariants as well as the syzygies are not affected by different conventions.} Substituting Eq.~(\ref{eq:PE seesaw 2g}) into Eq.~(\ref{eq:MW formula}) one obtains
\begin{eqnarray}
	\label{eq:HS seesaw 2g}
	{\mathscr H}_{\rm SS}^{(2\rm g)}(q)
	&=&\int \left[{\rm d}\mu\right]_{{\rm U}(2)\otimes{\rm U}(2)}^{} {\rm PE}\left(z_1^{},z_2^{},z_3^{},z_4^{};q\right)\nonumber\\
	&=&\frac{1}{4}\frac{1}{\left(2\pi {\rm i}\right)^4}\oint_{\left|z_1\right|=1}\oint_{\left|z_2\right|=1}\oint_{\left|z_3\right|=1}\oint_{\left|z_4\right|=1}\left(2-\frac{z_1}{z_2}-\frac{z_2}{z_1}\right)\left(2-\frac{z_3}{z_4}-\frac{z_4}{z_3}\right) {\rm PE}\left(z_1^{},z_2^{},z_3^{},z_4^{};q\right)\nonumber\\
	&=&\frac{1+q^6+3q^8+2q^{10}+3q^{12}+q^{14}+q^{20}}{\left(1-q^2\right)^3\left(1-q^4\right)^5\left(1-q^6\right)\left(1-q^{10}\right)}\;,
\end{eqnarray}
which agrees with the result obtained in Ref.~\cite{Manohar:2009dy}. The denominator in Eq.~(\ref{eq:HS seesaw 2g}) has 10 factors, corresponding to the 10 physical parameters in the two-generation seesaw model. The PL function of the HS is given by
\begin{eqnarray}
	\label{eq:PL seesaw 2g}
	{\rm PL}\left[{\mathscr H}_{\rm SS}^{(2\rm g)}(q)\right]=3q^2+5q^4+2q^6+3q^8+3q^{10}+2q^{12}-{\cal O}\left(q^{14}\right)\;,
\end{eqnarray}
from which one can read off there are in total 18 basic invariants (three of degree 2, five of degree 4, two of degree 6, three of degree 8, three of degree 10 and two of degree 12) and the syzygies begin to appear at the degree 14. With the help of Eq.~(\ref{eq:PL seesaw 2g}) one can explicitly construct all the basic invariants, as shown in Table~\ref{table:2g seesaw}.

For the case of $m=n=3$, the character functions of the building blocks read
\begin{eqnarray}
	\chi_l^{}\left(z_1^{},z_2^{},z_3^{}\right)&=&\left(z_1^{}+z_2^{}+z_3^{}\right)\left(z_1^{-1}+z_2^{-1}+z_3^{-1}\right)\,\nonumber\\
	\chi_\nu^{}\left(z_1^{},z_2^{},z_3^{},z_4{},z_5^{},z_6^{}\right)&=&\left(z_1^{}+z_2^{}+z_3^{}\right)\left(z_4^{-1}+z_5^{-1}+z_6^{-1}\right)+\left(z_4^{}+z_5^{}+z_6^{}\right)\left(z_1^{-1}+z_2^{-1}+z_3^{-1}\right)\;,\nonumber\\
	\chi_{\rm R}\left(z_4^{},z_5^{},z_6^{}\right)&=&z_4^2+z_5^2+z_6^2+z_4^{}z_5^{}+z_4^{}z_6^{}+z_5^{}z_6^{}+z_4^{-2}+z_5^{-2}+z_6^{-2}\nonumber\\
	&&+z_4^{-1}z_5^{-1}+z_4^{-1}z_6^{-1}+z_5^{-1}z_6^{-1}\;,
\end{eqnarray}
where $z_i^{}$, for $i=1,2,3$ (or for $i=4,5,6$) denote the coordinates on the maximum torus of the ${\rm U}(3)$ group that corresponds to the flavor-basis transformation in the active neutrino (or RH neutrino) sector. Labeling the degrees of $Y_l^{}$, $Y_\nu^{}$ and $Y_{\rm R}^{}$ by $q$, one can calculate the PE function
\begin{eqnarray}
	\label{eq:PE seesaw 3g}
	&&{\rm PE}\left(z_1^{},z_2^{},z_3^{},z_4^{},z_5^{},z_6^{};q\right)\nonumber\\
	&=&\left[\left(1-q_{}^2\right)_{}^3\left(1-q_{}^2z_1^{}z_2^{-1}\right)\left(1-q_{}^2z_2^{}z_1^{-1}\right)\left(1-q_{}^2z_1^{}z_3^{-1}\right)\left(1-q_{}^2z_3^{}z_1^{-1}\right)\left(1-q_{}^2z_2^{}z_3^{-1}\right)\right.\nonumber\\
	&&\left.\times\left(1-q_{}^2z_3^{}z_2^{-1}\right)\left(1-qz_1^{}z_4^{-1}\right)\left(1-qz_4^{}z_1^{-1}\right)\left(1-qz_1^{}z_5^{-1}\right)\left(1-qz_5^{}z_1^{-1}\right)\left(1-qz_1^{}z_6^{-1}\right)\right.\nonumber\\
	&&\left.\times\left(1-qz_6^{}z_1^{-1}\right)\left(1-qz_2^{}z_4^{-1}\right)\left(1-qz_4^{}z_2^{-1}\right)\left(1-qz_2^{}z_5^{-1}\right)\left(1-qz_5^{}z_2^{-1}\right)\left(1-qz_2^{}z_6^{-1}\right)\right.\nonumber\\
	&&\left.\times\left(1-qz_6^{}z_2^{-1}\right)\left(1-qz_3^{}z_4^{-1}\right)\left(1-qz_4^{}z_3^{-1}\right)\left(1-qz_3^{}z_5^{-1}\right)\left(1-qz_5^{}z_3^{-1}\right)\left(1-qz_3^{}z_6^{-1}\right)\right.\nonumber\\
	&&\left.\times\left(1-qz_6^{}z_3^{-1}\right)\left(1-qz_4^2\right)\left(1-qz_5^2\right)\left(1-qz_6^2\right)\left(1-qz_4^{}z_5^{}\right)\left(1-qz_4^{}z_6^{}\right)\left(1-qz_5^{}z_6^{}\right)\right.\nonumber\\
	&&\left.\times \left(1-qz_4^{-2}\right)\left(1-qz_5^{-2}\right)\left(1-qz_6^{-2}\right)\left(1-qz_4^{-1}z_5^{-1}\right)\left(1-qz_4^{-1}z_6^{-1}\right)\left(1-qz_5^{-1}z_6^{-1}\right)
	\right]_{}^{-1}\;.
\end{eqnarray}
Inserting Eq.~(\ref{eq:PE seesaw 3g}) into Eq.~(\ref{eq:MW formula}) and performing the complex integrals by virtue of the residue theorem, one gets  
\begin{eqnarray}
	\label{eq:HS seesaw 3g}
	{\mathscr H}_{\rm SS}^{(3\rm g)}(q)&=&\int \left[{\rm d}\mu\right]_{{\rm U} (3)\otimes{\rm U}(3)}^{} {\rm PE}\left(z_1^{},z_2^{},z_3^{},z_4^{},z_5^{},z_6^{};q\right)\nonumber\\
	&=&\frac{1}{36}\frac{1}{\left(2\pi {\rm i}\right)^6}\oint_{\left|z_1\right|=1}\oint_{\left|z_2\right|=1}\oint_{\left|z_3\right|=1}\oint_{\left|z_4\right|=1}\oint_{\left|z_5\right|=1}\oint_{\left|z_6\right|=1}\left[-\frac{\left(z_2-z_1\right)^2\left(z_3-z_1\right)^2\left(z_3-z_2\right)^2}{z_1^2z_2^2z_3^2}\right]\nonumber\\
	&&\times\left[-\frac{\left(z_5-z_4\right)^2\left(z_6-z_4\right)^2\left(z_6-z_5\right)^2}{z_4^2z_5^2z_6^2}\right]{\rm PE}\left(z_1^{},z_2^{},z_3^{},z_4^{},z_5^{},z_6^{};q\right)\;,\nonumber\\
	&=&\frac{{\mathscr N}_{\rm SS}^{(3\rm g)}(q)}{{\mathscr D}_{\rm SS}^{(3\rm g)}(q)}\;,
\end{eqnarray}
where 
\begin{eqnarray}
	\label{eq:numerator seesaw 3g}
	{\mathscr N}_{\rm SS}^{(3\rm g)}(q)&=&1+q^4+5q^6+9q^8+22q^{10}+61q^{12}+126q^{14}+273q^{16}+552q^{18}+1038q^{20}+1880q^{22}\nonumber\\
	&&+3293q^{24}+5441q^{26}+8712q^{28}+13417q^{30}+19867q^{32}+28414q^{34}+39351q^{36}\nonumber\\
	&&+52604q^{38}+68220q^{40}+85783q^{42}+104588q^{44}+123852q^{46}+142559q^{48}+159328q^{50}\nonumber\\
	&&+173201q^{52}+183138q^{54}+188232q^{56}+188232q^{58}+183138q^{60}+173201q^{62}\nonumber\\
	&&+159328q^{64}+142559q^{66}+123852q^{68}+104588q^{70}+85783q^{72}+68220q^{74}+52604q^{76}\nonumber\\
	&&+39351q^{78}+28414q^{80}+19867q^{82}+13417q^{84}+8712q^{86}+5441q^{88}+3293q^{90}\nonumber\\
	&&+1880q^{92}+1038q^{94}+552q^{96}+273q^{98}+126q^{100}+61q^{102}+22q^{104}+9q^{106}+5q^{108}\nonumber\\
	&&+q^{110}+q^{114}\;,
\end{eqnarray}
and
\begin{eqnarray}
	\label{eq:denominator seesaw 3g}
	{\mathscr D}_{\rm SS}^{(3\rm g)}(q)=\left(1-q^2\right)^3\left(1-q^4\right)^4(1-q^6)^4\left(1-q^8\right)^2\left(1-q^{10}\right)^2\left(1-q^{12}\right)^3\left(1-q^{14}\right)^2\left(1-q^{16}\right)\;, \quad
\end{eqnarray}
which is in agreement with the result in Ref.~\cite{Manohar:2010vu}. Note that the numerator in Eq.~(\ref{eq:numerator seesaw 3g}) also exhibits the palindromic structure and the denominator in Eq.~(\ref{eq:denominator seesaw 3g}) has totally 21 factors, exactly corresponding to the 21 independent physical parameters in the three-generation seesaw model.

\section{Matching of flavor invariants}
\label{app:matching}
In this appendix, we explain how to relate the flavor invariants in the SEFT to those in the full seesaw model. In fact, all the basic invariants in the SEFT can be expressed as the rational functions of the basic invariants in the full theory. 

This matching can be realized by noticing that the building blocks $C_5^{}$ and $C_6^{}$ in the SEFT are related to the building blocks $Y_\nu^{}$ and $Y_{\rm R}^{}$ in the full theory by Eq.~(\ref{eq:wilson coe}), and $Y_l^{}$ is the building block both in the SEFT and the full theory. Below we will explicitly show how to express the 18 basic invariants in the SEFT (i.e., the invariants in Table~\ref{table:2g eff}) as the rational functions of those in the full theory (i.e., the invariants in Table~\ref{table:2g seesaw}) in the two-generation case. The generalization to the three-generation case is straightforward.

We take ${\cal I}_{121}^{(2)}\equiv {\rm Im}\,{\rm Tr}\left(X_l^{}X_5^{}C_6^{}\right)$, the first CP-odd basic invariant in Table~\ref{table:2g eff}, as a concrete example. The first step is to replace $C_5^{}$ and $C_6^{}$ with $Y_\nu^{}$ and $Y_{\rm R}^{}$ using Eq.~(\ref{eq:wilson coe})
\begin{eqnarray}
	\label{eq:inverse matrix ex1}
	{\cal I}_{121}^{(2)}={\rm Im}\,{\rm Tr}\left[X_l^{}Y_\nu Y_{\rm R}^{-1}Y_{\nu}^{\rm T}Y_\nu^*\left(Y_{\rm R}^{\dagger }\right)_{}^{-1}Y_\nu^{\dagger}Y_\nu^{}\left(Y_{\rm R}^{\dagger}Y_{\rm R}^{}\right)_{}^{-1}Y_\nu^\dagger\right]\;.
\end{eqnarray}
In order to deal with the inverse matrix in Eq.~(\ref{eq:inverse matrix ex1}), we can utilize the following identity
\begin{eqnarray}
	\label{eq:inverse matrix 2g}
	A_{}^{-1}=\frac{2\left[\,{\rm Tr}\left(A\right) {\bf 1}_2-A\right]}{{\rm Tr}\left(A\right)^2-{\rm Tr}\left(A^2\right)}\;,
\end{eqnarray}
where $A$ is any $2\times 2$ non-singular matrix and ${\bf 1}_2^{}$ is the 2-dimensional identity matrix. Note that ${\cal I}_{121}^{(2)}$ is unchanged under the transformation in the flavor space, so is the right-hand side of Eq.~(\ref{eq:inverse matrix ex1}). Therefore one cannot substitute $Y_{\rm R}^{}$ directly into Eq.~(\ref{eq:inverse matrix 2g}), because $Y_{\rm R}^{}$ does not possess a bi-unitary transformation in the flavor space and thus ${\rm Tr}\left(Y_{\rm R}^{}\right)$ is not invariant under the flavor-basis transformati-\\on [recalling that $Y_{\rm R}^{}\to Y_{\rm R}^{\prime}=U_{\rm R}^{*} Y_{\rm R}^{} U_{\rm R}^\dagger$ and ${\rm Tr}\left(Y_{\rm R}^{\prime}\right)\neq{\rm Tr}\left(Y_{\rm R}^{}\right)$]. So it is necessary to rearrange the matrices on the right-hand side of Eq.~(\ref{eq:inverse matrix ex1}) into the form that transform \emph{adjointly} in the flavor space
\begin{eqnarray}
	\label{eq:inverse matrix ex2}
	{\cal I}_{121}^{(2)}&=&{\rm Im}\,{\rm Tr}\left\{\left(Y_\nu^\dagger X_l^{}Y_\nu\right)\left[Y_{\rm R}^\dagger\left(Y_\nu^{\rm T}Y_\nu^*\right)_{}^{-1}Y_{\rm R}^{}\right]_{}^{-1}\left(Y_\nu^\dagger Y_\nu^{}\right)\left(Y_{\rm R}^{\dagger}Y_{\rm R}^{}\right)_{}^{-1}\right\}\nonumber\\
	&=&{\rm Im}\,{\rm Tr}\left(G_{l\nu}^{}G_{\tilde{\nu}{\rm R}}^{-1}\widetilde{X}_\nu^{}X_{\rm R}^{-1}\right)\;,
\end{eqnarray}
where $G_{\tilde{\nu}{\rm R}}^{}\equiv Y_{\rm R}^\dagger(\widetilde{X}_{\nu}^{*})_{}^{-1}Y_{\rm R}^{}$, while $G_{l\nu}^{}$, $\widetilde{X}_\nu^{}$ and $X_{\rm R}^{}$ have been defined in the caption of Table~\ref{table:2g seesaw}. Note that all the matrices on the right-hand side of Eq.~(\ref{eq:inverse matrix ex2}) transform as the bi-unitary representation in the flavor space
\begin{eqnarray}
	G_{l\nu}^{}\to U_{\rm R}^{}G_{l\nu}^{} U_{\rm R}^\dagger\;,\quad
	G_{\tilde{\nu}{\rm R}}^{}\to U_{\rm R}^{}G_{\tilde{\nu}{\rm R}}^{} U_{\rm R}^\dagger\;,\quad
	\widetilde{X}_{\nu}^{}\to U_{\rm R}^{}\widetilde{X}_{\nu}^{} U_{\rm R}^\dagger\;,\quad
	X_{{\rm R}}^{}\to U_{\rm R}^{}X_{{\rm R}}^{} U_{\rm R}^\dagger\;,\quad
\end{eqnarray}
and thus their traces are all invariant under the flavor-basis transformation. Then one can substitute $\widetilde{X_\nu}^{}$, $G_{\tilde{\nu}{\rm R}}$ and $X_{\rm R}^{}$ into Eq.~(\ref{eq:inverse matrix 2g}) to obtain
\begin{eqnarray}
	&&G_{\tilde{\nu}{\rm R}}^{}=\frac{2}{\left(I_{020}^2-I_{040}\right)}\left(I_{020}^{}X_{\rm R}^{}-G_{\nu{\rm R}}^{}\right)\;,\\
	\label{eq:inverse matrix ex3}
	&&G_{\tilde{\nu}{\rm R}}^{-1}=\frac{-2}{\left(I_{002}^2-I_{004}\right)}\left[I_{020}^{}X_{\rm R}^{}-G_{\nu {\rm R}}^{}-\left(I_{020}^{}I_{002}^{}-I_{022}^{}\right){\bf 1}_2^{}\right]\;,\\
	\label{eq:inverse matrix ex4}
	&&X_{\rm R}^{-1}=\frac{2}{\left(I_{002}^2-I_{004}\right)}\left(I_{002}^{}{\bf 1}_2^{}-X_{\rm R}^{}\right)\;.
\end{eqnarray}
Inserting Eqs.~(\ref{eq:inverse matrix ex3})-(\ref{eq:inverse matrix ex4}) back into Eq.~(\ref{eq:inverse matrix ex2}) and after some algebra one obtains
\begin{eqnarray}
	\label{eq:inverse matrix ex5}
	{\cal I}_{121}^{(2)}=\frac{4}{\left(I_{002}^2-I_{004}\right)^2}\left[I_{242}^{(2)}I_{022}^{}+I_{262}^{}I_{002}^{}-{\rm Im}\,{\rm Tr}\left(G_{l\nu}^{}G_{\nu{\rm R}}^{}\widetilde{X}_\nu^{}X_{\rm R}^{}\right)\right]\;.
\end{eqnarray}
The final step is to decompose all the flavor invariants on the right-hand side of Eq.~(\ref{eq:inverse matrix ex5}) into the polynomials of the basic invariants in Table~\ref{table:2g seesaw}. Using the algorithm of decomposition developed in Appendix C of Ref.~\cite{Wang:2021wdq}, we have
\begin{eqnarray}
	\label{eq:inverse matrix ex6}
	{\rm Im}\, {\rm Tr}\left(G_{l\nu}^{}G_{\nu{\rm R}}^{}\widetilde{X}_\nu^{}X_{\rm R}^{}\right)=\frac{1}{2}\left(I_{242}^{(2)}I_{022}^{}+I_{044}^{}I_{220}^{}+I_{262}^{}I_{002}^{}+I_{244}^{}I_{020}^{}\right)\;.
\end{eqnarray}
Substituting Eq.~(\ref{eq:inverse matrix ex6}) into Eq.~(\ref{eq:inverse matrix ex5}) we finally get the expression of ${\cal I}_{121}^{(2)}$ in terms of the rational function of the basic invariants in the full theory
\begin{eqnarray}
	\label{eq:odd1 app}
	{\cal I}_{121}^{(2)}=\frac{2}{\left(I_{002}^2-I_{004}\right)^2}\left[I_{242}^{(2)}I_{022}^{}-I_{044}^{}I_{220}^{}+I_{262}^{}I_{002}^{}-I_{244}^{}I_{020}^{}\right]\;,
\end{eqnarray}
which is exactly Eq.~(\ref{eq:odd1}). The remaining 5 CP-odd basic invariants in Table~\ref{table:2g eff} can be handled in the same manner as ${\cal I}_{121}^{(2)}$, and thus we ultimately obtain
{\allowdisplaybreaks
	\begin{eqnarray}
		{\cal I}_{221}^{}&=&\frac{2}{\left(I_{002}^2-I_{004}\right)^2}\left[I_{242}^{(2)}I_{222}^{}+I_{244}^{}I_{220}^{}+I_{462}^{}I_{002}^{}-I_{444}^{}I_{020}^{}\right]\;,\label{eq:odd2 app}\\
		{\cal I}_{122}^{}&=&\frac{2}{\left(I_{002}^2-I_{004}\right)^3}\left\{I_{242}^{(2)}\left[3I_{022}^2+2I_{040}^{}\left(I_{002}^2-I_{004}^{}\right)-4I_{020}^{}I_{002}^{}I_{022}^{}\right]\right.\nonumber\\
		&&\left.+I_{044}^{}\left(4I_{020}^{}I_{222}^{}-I_{220}^{}I_{022}^{}-2I_{242}^{(1)}\right)+I_{262}^{}\left[3I_{002}^{}I_{022}^{}-I_{020}^{}\left(I_{002}^2+3I_{004}^{}\right)\right]\right.\nonumber\\
		&&\left.+I_{244}^{}\left(3I_{020}^{}I_{022}^{}-2I_{042}^{}\right)\right\}\;,\label{eq:odd3 app}\\
		{\cal I}_{240}^{}&=&\frac{1}{\left(I_{002}^2-I_{004}\right)^2}\left[3I_{242}^{(2)}\left(I_{022}^{}I_{220}^{}-I_{020}^{}I_{222}^{}\right)-I_{044}^{}I_{220}^2+I_{262}^{}\left(3I_{002}^{}I_{220}^{}-2I_{222}^{}\right)\right.\nonumber\\
		&&\left.-2 I_{244}^{}I_{020}^{}I_{220}^{}+I_{462}^{}\left(2I_{022}^{}-3I_{002}^{}I_{020}\right)+I_{444}^{}I_{020}^2\right]\;,\label{eq:odd4 app}\\
		{\cal I}_{141}^{}&=&\frac{2}{\left(I_{002}^2-I_{004}\right)^3}\left\{I_{242}^{(2)}I_{020}^{}I_{022}^2+I_{044}^{}I_{020}^{}\left(I_{022}^{}I_{220}^{}-2I_{242}^{(1)}\right)\right.\nonumber\\
		&&\left.+I_{262}^{}\left[I_{002}^{}I_{020}^{}I_{022}^{}+I_{040}^{}\left(I_{004}^{}-I_{002}^2\right)\right]+I_{244}^{}I_{020}^{}\left(I_{020}^{}I_{022}-2I_{042}^{}\right)\right\}\;,\label{eq:odd5 app}\\
		{\cal I}_{042}^{}&=&\frac{2}{\left(I_{002}^2-I_{004}\right)^3}\,I_{044}^{}\left(I_{020}^2-I_{040}^{}\right)_{}^2\;.\label{eq:odd6 app}
	\end{eqnarray}
}
Therefore, the 6 CP-odd basic invariants in the SEFT have been written as the linear combinations of the 6 CP-odd basic invariants in the full theory, with the coefficients being rational functions of CP-even basic invariants.

For completeness, we also list below the matching conditions of the CP-even basic invariants and all of them can be deduced in the same manner as ${\cal I}_{121}^{(2)}$, i.e.,
{\allowdisplaybreaks
	\begin{eqnarray}
		\label{eq:even 1 app}
		{\cal I}_{100}^{}&=&I_{200}^{}\;,\\
		{\cal I}_{001}^{}&=&\frac{2}{\left(I_{002}^2-I_{004}\right)}\left(I_{002}^{}I_{020}^{}-I_{022}^{}\right)\;,\\
		{\cal I}_{200}^{}&=&I_{400}^{}\;,\\
		{\cal I}_{101}^{}&=&\frac{2}{\left(I_{002}^2-I_{004}\right)}\left(I_{002}^{}I_{220}^{}-I_{222}^{}\right)\;,\\
		{\cal I}_{020}^{}&=&\frac{2}{\left(I_{002}^2-I_{004}\right)}\left(I_{042}^{}-2I_{022}^{}I_{020}^{}+I_{020}^2I_{002}^{}\right)\;,\\
		{\cal I}_{002}^{}&=&\frac{2}{\left(I_{002}^2-I_{004}\right)^2}\left[2I_{022}^{}\left(I_{022}^{}-2I_{002}^{}I_{020}^{}\right)+I_{004}^{}\left(I_{020}^2-I_{040}^{}\right)+I_{002}^2\left(I_{020}^2+I_{040}^{}\right)\right]\;,\\
		{\cal I}_{120}^{}&=&\frac{2}{\left(I_{002}^2-I_{004}\right)}\left[I_{220}^{}\left(I_{020}^{}I_{002}^{}-I_{022}^{}\right)+I_{242}^{(1)}-I_{020}^{}I_{222}^{}\right]\;,\\
		{\cal I}_{021}^{}&=&\frac{1}{\left(I_{002}^2-I_{004}\right)^2}\left[I_{004}^{}I_{020}^{}\left(I_{020}^2-I_{040}^{}\right)+I_{002}^2I_{020}^{}\left(3I_{020}^2+I_{040}^{}\right)-4I_{022}^{}\left(I_{042}^{}-2I_{020}^{}I_{022}^{}\right)\right.\nonumber\\
		&&\left.+4I_{002}^{}I_{020}^{}\left(I_{042}^{}-3I_{020}^{}I_{022}^{}\right)\right]\;,\\
		{\cal I}_{220}^{}&=&\frac{2}{\left(I_{002}^2-I_{004}\right)}\left[I_{220}^{}\left(I_{220}^{}I_{002}^{}-2I_{222}^{}\right)+I_{442}^{}\right]\;,\\
		{\cal I}_{121}^{(1)}&=&\frac{1}{\left(I_{002}^2-I_{004}\right)^2}\left[I_{004}^{}I_{220}^{}\left(I_{020}^2-I_{040}^{}\right)+I_{002}^2I_{220}^{}\left(3I_{020}^2+I_{040}^{}\right)\right.\nonumber\\
		&&\left.+4I_{022}^{}\left(I_{022}^{}I_{220}^{}+I_{020}^{}I_{222}^{}-I_{242}^{(1)}\right)-4I_{002}^{}I_{020}^{}\left(2I_{022}^{}I_{220}^{}+I_{020}^{}I_{222}^{}-I_{242}^{(1)}\right)\right]\;,\\
		{\cal I}_{040}^{}&=&\frac{1}{\left(I_{002}^2-I_{004}\right)^2}\left[I_{004}^{}\left(I_{020}^2-I_{040}^{}\right)_{}^2+I_{002}^2\left(3I_{020}^2-I_{040}^{}\right)\left(I_{020}^2+I_{040}^{}\right)\right.\nonumber\\
		&&\left.-4\left(2I_{020}^{}I_{022}^{}-I_{042}^{}\right)\left(2I_{002}^{}I_{020}^2-2I_{020}^{}I_{022}^{}+I_{042}^{}\right)
		\right]\;,\\
		{\cal I}_{022}^{}&=&\frac{1}{\left(I_{002}^2-I_{004}\right)^3}\left[I_{002}^3\left(5I_{020}^4+2I_{020}^2I_{040}+I_{040}^2\right)+8I_{020}^{}\left(I_{002}^2I_{020}^{}I_{042}^{}-2I_{022}^3\right)\right.\nonumber\\
		&&\left.+8I_{022}^2\left(5I_{002}^{}I_{020}^2+I_{042}\right)-4I_{022}^{}I_{002}^{}I_{020}^{}\left(7I_{002}^{}I_{020}^2+I_{002}^{}I_{040}^{}+4I_{042}^{}\right)\right.\nonumber\\
		\label{eq:even 12 app}
		&&\left.+I_{004}^{}\left(I_{020}^2-I_{040}^{}\right)\left(3I_{002}^{}I_{020}^2+I_{002}^{}I_{040}^{}-4I_{020}^{}I_{022}^{}\right)\right]\;.
	\end{eqnarray}
}
From Eqs.~(\ref{eq:even 1 app})-(\ref{eq:even 12 app}) we can observe that all the 12 CP-even basic invariants in the SEFT can be expressed as the rational functions of those 12 CP-even basic invariants in the full theory, which however are independent of any CP-odd invariants.

The generalization to the three-generation case is straightforward. One just needs to replace Eq.~(\ref{eq:inverse matrix 2g}) by
\begin{eqnarray}
	\label{eq:inverse 3g}
	A_{}^{-1}=\frac{6A^2-6\,{\rm Tr}\left(A\right)A+3\left[{\rm Tr}\left(A\right)^2-{\rm Tr}\left(A^2\right)\right]{\bf 1}_3}{{\rm Tr}\left(A\right)^3-3\,{\rm Tr}\left(A^2\right){\rm Tr}\left(A\right)+2\,{\rm Tr}\left(A^3\right)}\;,
\end{eqnarray}
where $A$ is a $3\times 3$ non-singular matrix and ${\bf 1}_3^{}$ is the 3-dimensional identity matrix. Following the same strategy as before, one can also express all the basic invariants in the SEFT as the rational functions of those in the full theory for the three-generation case.

Conversely, one can also express all the basic invariants in the full theory as the functions of those in the SEFT (but not rational functions) as described in the following procedure. First, the number of independent physical parameters in the full seesaw model exactly matches that in $C_5$ and $C_6$, both of which are 10 (or 21) for two- (or three-) generation case.  Moreover, one can prove that all the physical parameters in the full seesaw model can be expressed in terms of those in the SEFT (see Refs.~\cite{Broncano:2002rw,Broncano:2003fq} for details). Second, we have shown in Sec.~\ref{subsec:extract2g} and Sec.~\ref{subsec:extract3g} that all the physical parameters in the SEFT can be extracted using primary invariants (which are a subset of basic invariants). Therefore, all the basic invariants in the full seesaw model, which are composed of the physical parameters in the full theory, can be expressed as the functions of the basic invariants in the SEFT. Since any flavor invariant in the ring can be decomposed as the polynomial of basic ones, we conclude that the matching of two sets of flavor invariants can be inverted. Such an inversion may be accomplished up to some possible discrete degeneracies due to the non-linearity of polynomial functions. However, as some complication arises at each step, we shall not explicitly invert the complete invariant matching.

\end{appendix}

\bibliographystyle{JHEP}
\bibliography{SEFT_ref}
\end{document}